%% file: main.tex
\documentclass[12pt]{article}
\usepackage[utf8]{inputenc}
\usepackage{titling} 
\usepackage{lipsum}  
\usepackage{geometry}
\usepackage{indentfirst}
\usepackage{array}
\usepackage{nccmath}
\usepackage{caption}
\usepackage{colortbl}
\usepackage{textcomp}
\usepackage{stfloats}
\usepackage{url}
\usepackage{verbatim}
\usepackage{graphicx}
\usepackage{xcolor}
\usepackage{subcaption}
\usepackage{bm}
\usepackage{url}
\usepackage{tabularx}
\usepackage[normalem]{ulem}
\usepackage{multicol,lipsum}
\usepackage{cuted}
\usepackage{authblk}
\usepackage{listings}
\usepackage{hyperref}
\usepackage{amsmath, amssymb}
\usepackage{tikz}   
\usepackage[backend=biber,style=numeric,sorting=none]{biblatex}

\newcommand{\realnumbers}{\mathbb{R}}

\newcommand{\interpolatefunc}{p}
\newcommand{\pointx}{\mathbf{x}}
\newcommand{\gradiant}{\mathbf{g}}
\newcommand{\hessian}{\mathbf{H}}
\newcommand{\eigenvec}{\mathbf{q}}
\newcommand{\velocity}{\mathbf{v}}
\newcommand{\acceleration}{\mathbf{a}}
\newcommand{\curvature}{c}
\newcommand{\bigQ}{\mathbf{Q}}
\newcommand{\tensor}{\mathbf{T}}
\newcommand{\bigLambda}{\bf{\Lambda}}
\newcommand{\bigL}{\mathbf{L}}

\newcommand{\pblurryring}{p_{\text{blurry\_ring}}}
\newcommand{\pring}{p_{\text{ring}}}

\newcommand{\psinlf}{p_{\text{sin,lf}}}
\newcommand{\psinhf}{p_{\text{sin,hf}}}

\newcommand{\functionf}{f}

\title{Tubular Curvature Filter: Pointwise Curvature Calculation for Tubular Objects in Images}

\author[1]{\small Elifnur Sunger\thanks{These authors contributed equally to this work.}}
\author[1]{\small Beyza Kalkanli$^*$}
\author[2]{\small Veysi Yildiz}
\author[3]{\small Tales Imbiriba}
\author[4]{\small Giovanna Guidoboni}
\author[5]{\small Peter Campbell}
\author[1]{\small Deniz Erdogmus}
\date{}
\affil[1]{\small Department of Electrical and Computer Engineering, Northeastern University, Boston, USA}
\affil[2]{\small Identifeye Health, Guilford, USA}
\affil[3]{\small Department of Computer Science, University of Massachusetts Boston, Boston, USA}
\affil[4]{\small Department of Electrical and Computer Engineering, University of Maine, Orono, USA}
\affil[5]{\small Casey Eye Institute, Oregon Health and Sciences University, Portland, USA}

\lstdefinestyle{latex}{
  language=[LaTeX]TeX,
  escapeinside={\%*}{*)},
}
\lstdefinelanguage{bibtex}
{keywords={%
        @article,@book,@collectedbook,@conference,@electronic,@ieeetranbstctl,%
        @inbook,@incollectedbook,@incollection,@injournal,@inproceedings,%
        @manual,@mastersthesis,@misc,@patent,@periodical,@phdthesis,@preamble,%
        @proceedings,@standard,@string,@techreport,@unpublished%
    },
    comment=[l][shape]{@comment},
    sensitive=false,
}

\lstnewenvironment{lstlistingtex}{\lstset{language=[LaTeX]TeX}}{}

\makeatletter
\renewcommand\footnoterule{%
  \kern -3pt
  \hrule width \textwidth height 0.4pt
  \kern 2.6pt
}
\makeatother
\begin{document}

\maketitle
\vspace{-2cm}
\begin{abstract}
Purpose: Accurate estimation of blood vessel tortuosity from medical images is an extremely important and challenging task. It is particularly relevant in the context of retinopathy of prematurity (ROP), where the staging of disease severity and consequent therapeutic approaches are heavily informed by the presence and prominence of vessel tortuosity.
Existing methods based on centerline or skeleton curvature fail to capture curvature gradients across a rotating tubular structure, thereby limiting their effectiveness in the case of ROP. 

Methods: This paper defines local tubular curvature and presents the Tubular Curvature Filter (TCF) method, which locally calculates the acceleration of curve bundles traversing a tubular object parallel to its centerline. This is achieved by examining the directional rate of change in the eigenvectors of the Hessian matrix of a tubular intensity function in space. TCF implicitly calculates the local tubular curvature without the need to explicitly segment or extracting the centerline of the tubular object.        

Results: Experimental results demonstrate that TCF provides accurate estimates of local curvature at any point inside tubular structures. Results on 2D and 3D images show that TCF discerns curvature differences between the inner and outer sides of curved tubular objects, while centerline-based approaches cannot.

Conclusion: Our findings highlight that TCF's ability to discern between the inner and outer sides of curved tubular objects is particularly useful in medical fields that require vasculature curvature analysis from images, especially where vascular structures often have non-uniform diameters, such as in ROP.
\end{abstract}
\section{Introduction}
\input{JMO/tex/intro}

\section{Method} \label{sec:method}
\input{JMO/tex/method}

\section{Results}

\input{JMO/tex/results}
\section{Conclusions and future perspectives}
\input{JMO/tex/conclusion}

\appendix
\section{Derivation of Definition 1}
\input{JMO/tex/appendix_derivation}
\section{Generalization to Functions of n-Dimensional Input}
\input{JMO/tex/appendix_n_dimensions}
\section*{Acknowledgements}
This work was supported by grants R01 EY019474, R01 EY034718, and P30 EY10572 from the National Institutes of Health, by unrestricted departmental funding from Research to Prevent Blindness, and by the Malcolm Marquis Innovation Fund and National Science Foundation grant DMS 2108711/2327640.

\printbibliography
\end{document}

%% file: JMO/tex/intro.tex
\label{sec:intro}

Curvature calculations in images have utility in many application domains, such as object recognition \cite{RegionDetector, thanh2020melanoma}, corner detection \cite{735812, zhang2019discrete, he2008corner}, face recognition \cite{10.1007/s10851-017-0728-2}, image registration \cite{ImageDescriptors}, image reconstruction \cite{ImageReconstructionbyMinimizingCurvatures}, image denoising \cite{zhong2020minimizing, lee2005noise, kim2006pde} and segmentation \cite{Nc2019OptimizedMP, 9223670, goldluecke2011introducing, chen2016new, el2016contrast}. 
Curvature calculations are particularly important in the context of retinopathy of prematurity (ROP), a vasoproliferative disease that poses high risks for retinal detachment and irreversible vision loss in babies who are born prematurely~\cite{gilbert2001childhood}. ROP treatment options and outcomes strongly depend on the severity of the disease stage, which is classified based on the appearance of blood vessels in the retina~\cite{chiang2021international}. In particular, the presence and prominence of vessel tortuosity is a major indicator of disease severity~\cite{chiang2021international}, hence the importance of accurately computing curvature.

The image processing literature explores multiple forms of curvature. Motivated by ROP vessel analysis, this paper introduces a method for calculating curvature locally along a bundle of parallel curves that extend through a tubular object in the image, without extracting the centerline.

In medical applications, centerline-based curvature methods are commonly used to design medical devices~\cite{choi2008methods}; assess disease severity~\cite{fabrydisease, bullitt2003measuring}; segment vascular structures~\cite{lorigo2001curves, law2009efficient}; model and visualize vascular or bony structures~\cite{twistedbloodvessels, hathout2012vascular, gerig2004analysis, ge2024automatic, zhou2015assessment, vogel2022robust}; analyze fluid– structure interactions in arterial flow~\cite{ALETTI201677} and blood flow in the retina~\cite{koogler2023analysis}; and understand how bone curvature influences the response to mechanical stimuli~\cite{javaheri2020lasting}. There are also deep learning methods to estimate the centerline curvature of bony or vascular structures~\cite{ernst2019cnn, liu2020spinal, wang2024b}. Note that centerline-based curvature assessments do not accurately capture the varying thickness of tubular structures or the differences in curvature between the inner and outer sides of tubular objects during turns. Since biological vascular structures often vary in diameter, this difference is crucial for modeling purposes, such as the relationship between curvature and local wall shear stress, with local wall shear stress being known to differ between the inner and outer bends of curved vessels~\cite{santamarina1998computational}.

\begin{figure}[t]
    \centering
    \includegraphics[trim={0 1.6cm 0 0.1cm },clip, width=0.6\textwidth]{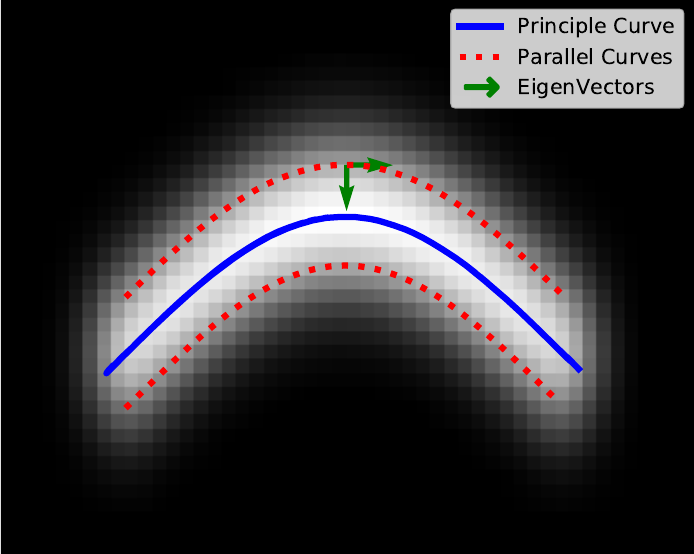}
    \caption{Example function, its principal curve~\cite{ozertem2011locally}, parallel curves and eigenvectors of Hessian at a sample point.}
    \label{fig:pdf_parallel}
\end{figure}

This paper focuses on local tubular curvature and proposes the \textit{Tubular Curvature Filter (TCF)} method, which is based on the local acceleration of tangent vectors to a bundle of curves in the tubular object. These curves are parallel to the centerline, which is traced by the ridge of a suitable intensity function that highlights the tubular object. In prior work, these tangent vectors are identified as specific eigenvectors of the Hessian of the intensity function~\cite{ozertem2011locally}. Fig.~\ref{fig:pdf_parallel} is an illustration of a function and its principal curve, along with parallel curves and the eigenvectors of the Hessian at a sample point. The method assumes that the image is appropriately preprocessed to yield an intensity image that highlights the tube-shaped objects and is based on interpolating this intensity image to obtain an intensity function over spatial coordinates. 

To the best of our knowledge, TCF is \emph{the first pointwise curvature calculation method for tubular objects}. TCF can be applied pointwise to any thrice continuously differentiable interpolated vessel/tube intensity image, thereby \emph{eliminating the need to segment or extract centerlines of tubular objects}, unlike centerline-based methods. Experiments with 2D and 3D images show that TCF can discern curvature differences between the inner and outer sides of a curved tubular object, achieving a level of spatial precision that centerline-based approaches cannot. With appropriate local image interpolation techniques, TCF is highly suitable for parallel processing.

%% file: JMO/tex/method.tex
\label{sec:methods}

In this section, TCF will be described and a method to compute local tubular curvature at any point for a given thrice continuously differentiable function of Cartesian spatial coordinates in an Euclidean space will be developed. Application to intensity images of tubular shapes will also be discussed.

\subsection{Tubular Curvature}\label{sec:tub_curv}
To compute the curvature of tubular shapes over space we exploit the geometrical interpretation of the eigenvalues of the Hessian provided by Ozertem et al.~\cite{ozertem2011locally}. In Ozertem et al.~\cite{ozertem2011locally}, the authors redefined principal curves and surfaces in terms of first and second derivatives of probability density functions. This definition of the principal curves corresponds to the \textit{ridges} of said probability density function. We motivate our discussion using the example in Fig~\ref{fig:pdf_parallel}, which presents a discretized image of a function that forms a tubular shape over 2-dimensional (2-D) space; function values are pixel intensities. The \textit{principal curve} that forms the ridge of this function and \textit{parallel} curves traced by select eigenvectors of the Hessian matrix, as defined by Ozertem et al.~\cite{ozertem2011locally}, are represented by the blue-solid and red-dotted lines, respectively. The eigenvectors of the function's Hessian evaluated at a given point are represented by green arrows. At any point over the 2-D space, such eigenvectors point toward the principal and to a direction tangential to the parallel curves.

The curvature of parallel curves at any point in space can be defined based on the rate of change of eigenvectors tangent to the parallel curves; for unit-length eigenvectors, this rate of change becomes the acceleration vector. Curvature of these curves at any point is the norm of the local acceleration vector. When computing the curvature of tubular structures in an image, an interpolation method provides a thrice-continuously differentiable intensity function, from which necessary gradient and Hessian values is computed for any given point.
    
Let $ \pointx (s) \in \realnumbers^2 $ be a point on a curve with arc length parameter $s$~\cite{montiel2009curves}; $\interpolatefunc(\cdot): \realnumbers^{2} \rightarrow \realnumbers^*_+=\{\interpolatefunc(\pointx(s)) \in  \realnumbers\,|\,\interpolatefunc(\pointx(s))>0\}$ be a positive, at least thrice differentiable function, for example an interpolated image as explained in Section~\ref{sec:tcf_images}. $\hessian(\pointx(s))$ denotes the Hessian of the logarithm of the function (see Appendix~\ref{appendix:derivation_of_definition1}), with $\{\left((\lambda_1 (\pointx(s)), \eigenvec_1(\pointx(s))\right),\left((\lambda_2(\pointx(s)), \eigenvec_2(\pointx(s))\right)\}$ eigenvalue-vector pairs of $\hessian(\pointx(s))$, where $|\lambda_1 (\pointx(s))| < |\lambda_2(\pointx(s))|$.

\noindent{\textbf{Definition 1.}} Acceleration $\acceleration(\pointx(s)) = \partial \velocity(\pointx(s)) / \partial s$, at any point $\pointx(s)$, is the rate of change of velocity $\velocity = \eigenvec_1$ with respect to the arc length parameter $s$, which is the unit-length tangent to the curve. Using chain rule:
    \begin{align}
        \label{eq:acceleration_eq}
        \frac{\partial \velocity(\pointx(s))}{ \partial s} 
        = \frac{\partial \eigenvec_1(\pointx(s))}{ \partial s} 
        = \frac{\partial \eigenvec_1(\pointx(s))}{\partial \pointx(s)} \frac{\partial \pointx(s)}{\partial s}\cdot
    \end{align}

Since $ \eigenvec_1(\pointx (s)) $ is the unit length velocity vector, we write $\partial \pointx(s) / \partial s = \eigenvec_1(\pointx (s))$ as $\partial \pointx(s) = \eigenvec_1(\pointx (s))\partial s$.

\noindent{\textbf{Definition 2.}} Tubular curvature of the intensity function $\interpolatefunc(\pointx (s))$ at $\pointx(s)$ is the Euclidean norm of its acceleration vector: curvature $\curvature(\pointx(s)) = || \acceleration(\pointx(s)) ||$.

In the remainder of this paper, we will not explicitly indicate dependency on $s$ for notation simplicity.

\noindent{\textbf{Derivation 1.}} To calculate $\acceleration$ at $\pointx$, the directional derivatives of $\eigenvec_1$, i.e., ${\partial \textbf{q}_{1}}/{\partial x_1}$ and $ {\partial \textbf{q}_{1}}/{\partial x_2}$, are required; which can be computed using the eigendecomposition of $ \hessian(\pointx) = \bigQ(\pointx)\bigLambda(\pointx)\bigQ(\pointx)^T $ as:

\begin{align}
    \label{eq:hessian}
    \hessian(\pointx) 
    =\begin{bmatrix}\eigenvec_1 &\eigenvec_2\end{bmatrix}
    \begin{bmatrix}\lambda_1 &0\\ 0 &\lambda_2\end{bmatrix}
    \begin{bmatrix}\eigenvec_1 \\ \eigenvec_2\end{bmatrix},
\end{align}
where $\bigQ$ is the $2\times2$ orthonormal matrix whose $i^{th}$ column is the eigenvector $\eigenvec_i$, and $\bigLambda$ is diagonal with elements as corresponding eigenvalues. In \eqref{eq:hessian} $\hessian(\pointx)$ is assumed to be diagonalizable with two linearly independent eigenvectors. If this assumption is invalid, ambiguity in eigenvectors leads to inability to uniquely trace a curve passing through point $\pointx$, resulting in local curvature ambiguity. 

Calculating ${\partial \eigenvec_1(\pointx)}/{\partial \pointx} $ requires taking derivatives of this Hessian decomposition. Let $\tensor_i(\pointx)$, $\textbf{V}_i(\pointx)$ and $\bigL_i(\pointx)$ be the $i^{th}$ directional derivatives of $\hessian(\pointx)$, $\bigQ(\pointx)$ and $\bigLambda(\pointx)$ respectively:
    \begin{equation}
          \label{eq:tensor_i_derivation_2d}
            \tensor_i(\pointx) = \frac{\partial \hessian(\pointx)}{ \partial x_i} = \begin{bmatrix}
        \text{t}_\text{i11} & \text{t}_\text{i12} \\ 
        \text{t}_\text{i21} &\text{t}_\text{i22}
        \end{bmatrix},
    \end{equation}
    \begin{equation}
        \textbf{V}_i(\pointx) = \frac{\partial \bigQ(\pointx)}{\partial x_i} =
        \begin{bmatrix}
        \frac{\partial \eigenvec_1}{\partial x_i} &\frac{\partial \eigenvec_2}{\partial x_i}
        \end{bmatrix}
        =
        \begin{bmatrix}
        \text{v}_\text{i11} & \text{v}_\text{i12} \\ 
        \text{v}_\text{i21} &\text{v}_\text{i22}
        \end{bmatrix},
    \end{equation}
    \begin{equation}
        \label{eq:l_i_derivation_2d}
        \bigL_{i}(\pointx)=
        \begin{bmatrix}
        \frac{\partial \lambda_{1}}{\partial x_i} &0\\
        0 &\frac{\partial \lambda_{2}}{\partial x_i}
        \end{bmatrix}=
        \begin{bmatrix}
        l_\text{i1} &0 \\ 
        0 &l_\text{i2}
        \end{bmatrix}.
    \end{equation}
Omitting dependencies on $(\pointx)$ for brevity, $\tensor_i(\pointx)$ is:
    \begin{equation}
        \label{eq:hessianderivative}
        \begin{aligned}
        \tensor_i = & \textbf{V}_i \bigLambda \bigQ^T +\bigQ\bigL_i \bigQ^T +\bigQ\bigLambda\textbf{V}_i^T.
        \end{aligned}
    \end{equation}
Here, the derivatives of eigenvectors and eigenvalues are unknown. Since $\tensor_i$ matrices are symmetric, we get 3 equations from $\tensor_i$. Using two properties of eigenvectors, we obtain additional equations: 1) eigenvectors are unit length; hence, derivatives of their magnitude are zero, and 2) they are perpendicular to each other. With these, we have:
    \begin{equation}
        \label{eq:eig_property1}
        \frac{{\partial || \eigenvec_1(\pointx)||}^2}{\partial x_i}= 2\eigenvec_1^T\frac{\partial \eigenvec_1}{\partial x_i} = 0,
    \end{equation}
    \begin{equation}
        \label{eq:eig_property2}
        \frac{{\partial || \eigenvec_2(\pointx)||}^2}{\partial x_i}= 2\eigenvec_2^T\frac{\partial \eigenvec_2}{\partial x_i} = 0,
    \end{equation}
    \begin{equation}
        \label{eq:last_prependicular}
        \frac{\partial \eigenvec_1^T\eigenvec_2}{\partial x_i} =\eigenvec_1^T \frac{\partial \eigenvec_2}{\partial x_i} +
        \frac{\partial \eigenvec_1^T}{\partial x_i} \eigenvec_2 =0.
    \end{equation}
From $\tensor_i$ and the properties of eigenvectors, we obtain 6 equations with 12 unknowns. These yield the following sets of linear equations from \eqref{eq:hessianderivative}-\eqref{eq:last_prependicular}, where $\eigenvec_i = \begin{bmatrix}\text{q}_{i1} &\text{q}_{i2}\end{bmatrix}^T$:
    \begin{equation}
        \label{eq:matrix_form_both}
        \begin{aligned} \begin{bmatrix}
        \text{t}_{111}\\ \text{t}_{112}\\ \text{t}_{122}\\ 0\\ 0\\ 0\\
        \end{bmatrix}
        \!= \textbf{M}
        \begin{bmatrix}
        \text{v}_{111}\\ \text{v}_{112}\\ l_{11}\\ l_{12}\\
        \text{v}_{121}\\ \text{v}_{122}\\
        \end{bmatrix} 
        \!\textrm{and}\!
        \begin{bmatrix}
        \text{t}_{211}\\ \text{t}_{212}\\ \text{t}_{222}\\ 0\\ 0\\ 0\\
        \end{bmatrix}
        \!=
        \textbf{M}
        \begin{bmatrix}
        \text{v}_{211}\\ \text{v}_{212}\\ l_{21}\\ l_{22}\\
        \text{v}_{221}\\ \text{v}_{222}\\
        \end{bmatrix},
        \end{aligned}
    \end{equation}

    \begin{equation}
        \label{eq:matrix_m}
        \!\textbf{M}\!=\!\begin{bmatrix}
        2\lambda_1\text{q}_{11} &2\lambda_2\text{q}_{21} &\text{q}_{11}^2 &\text{q}_{21}^2  &0 &0\\
        \lambda_1\text{q}_{12} & \lambda_2\text{q}_{22} &\text{q}_{11}\text{q}_{12} &\text{q}_{21}\text{q}_{22} &\lambda_1\text{q}_{11} & \lambda_2\text{q}_{21}\\
        0 &0 &\text{q}_{12}^2 &\text{q}_{22}^2 &2\lambda_1\text{q}_{12} &2\lambda_2\text{q}_{22}\\
        \text{q}_{11} & &0 &0&\text{q}_{12}&0\\
        0 &\text{q}_{21} &0 &0 &0 &\text{q}_{22}\\
        \text{q}_{21} &\text{q}_{11} &0 &0 &\text{q}_{22} &\text{q}_{12}\\
        \end{bmatrix}.
        \end{equation}

We provide the complete derivation of linear equations that leads to $\textbf{M}$ in Appendix ~\ref{appendix:derivation_of_definition1}. Please see Appendix \ref{appendix:N-dimensional} for derivations of linear equations solved in order to obtain the acceleration vector for functions with n-dimensional arguments, including the 3-D special case (for volumetric images).

By solving the linear equations in Appendix~\ref{appendix:derivation_of_definition1}, we find $\textbf{V}(\pointx)$ and $\bigL(\pointx)$. 
Then, we calculate $\partial \eigenvec_1(\pointx) / \partial \pointx$ from $\textbf{V}(\pointx)$. Finally, we obtain the acceleration vector at $\pointx$ following (\ref{eq:acceleration_eq}).

\subsection{TCF on Images}
\label{sec:tcf_images}
The methodology discussed above is designed to operate over continuous thrice differentiable functions defined over $\mathbb{R}^2$ (and in general $\mathbb{R}^n$). To apply the TCF to images, we need to smoothly interpolate intensity images to obtain intensity functions with necessary continuity and differentiability properties. This paper does not advocate a particular interpolation methodology. For illustration, we consider a kernel regression based approach. Note that to operate on continuous thrice-differentiable functions defined over $\mathbb{R}^2$/$\mathbb{R}^3$, tubular shapes should have a width of no less than 3 or 4 pixels/voxel lengths.

We use point location to calculate curvature; thus, a pixel can be converted to physical units as:
\begin{equation}
    \pointx = \begin{bmatrix}
    x_1\\
    x_2 
    \end{bmatrix} = \begin{bmatrix}
    (p_1 - c_1 + 0.5) \times \nu_1 \\
    (p_2 - c_2 + 0.5) \times \nu_2
    \end{bmatrix},
\end{equation}
where $\mathbf{p} \in \mathbb{Z}_+^2$ corresponds to a pixel in the image, $\pointx \in \mathbb{R}^2$ represents the physical location in physical units of pixel $\mathbf{p}$, $\mathbf{c} \in \mathbb{Z}_+^2$ is the pixel corresponding to the physical origin in pixel coordinates, and $\mathbf{\nu} \in \mathbb{R}^2$ represents the physical lengths per pixel in the horizontal and vertical directions. A similar conversion can be done for three-dimensional images (or $n$-dimensional). $\tilde{\pointx}_j\in \mathbb{R}^2$ and $w_j\in\mathbb{R}_+$, $j\in\{1,\ldots, n\}$ are the location and intensity of the $j$-th pixel on a image, and $K_{S_j}(\cdot): \mathbb{R}^2 \rightarrow \realnumbers$ is the interpolation kernel with scale matrix $\textbf{S}_j$. The kernel-based interpolation is:
    \begin{equation}
        \label{eq:kde}
        \interpolatefunc(\pointx)= \sum_{j=1}^{n} w_{j}K_{\textbf{S}_j} (\pointx-\tilde{\pointx}_j).
    \end{equation}

Variable-scale (width) kernels may be employed in interpolation to achieve robustness against outlier pixel intensities that may arise due to noise or other reasons. The approach outlined in Ozertem et al.~\cite{ozertem2007nonparametric} proposes to select $k$-nearest pixels to $\pointx$. For each $\pointx$, a sequence $\{\tilde{\pointx}_{\pi(1)},\tilde{\pointx}_{\pi(2)},\ldots,\tilde{\pointx}_{\pi(n)}\}$ is defined, where $\pi(i) \in \{1,2,\ldots,n\}$ represents the sorted set of pixel indexes such that $\lVert \pointx - \tilde{\pointx}_{\pi(i)} \rVert_{2} \leq \lVert \pointx - \tilde{\pointx}_{\pi(i+1)} \rVert_{2}$, and the $k$-nearest pixels are selected based on this sequence, denoted as $\{\tilde{\pointx}_{\pi(1)},\tilde{\pointx}_{\pi(2)},...,\tilde{\pointx}_{\pi(k)}\}$, with $k\leq n$. These $k$-nearest pixels can then be used for kernel-based interpolation:
    \begin{equation}
        \label{eq:kde_knearest}
        \interpolatefunc_{\pi}(\pointx)= \sum_{j=1}^{k} w_{\pi(j)}K_{\textbf{S}_j} (\pointx-\tilde{\pointx}_{\pi(j)}).
    \end{equation}

Given $\interpolatefunc(\pointx)$, considering that the tubular structure of interest is represented with large function values (high-intensity pixels over a dark background), the application of the TCF is straightforward. 
The methodology presented in Section~\ref{sec:tub_curv} was derived, without loss of generality, for images with tubular shapes represented by light pixels over dark backgrounds and used $\eigenvec_1$ for curvature calculations. 
If tubular objects appear as dark pixels over a light background, the intensity function is negated to achieve tubular objects with high intensity values.

%% file: JMO/tex/results.tex
\label{sec:experiments}
TCF is applied to artificial and medical images in both two and three dimensions. Within this section, we have categorized the results based on the number of dimensions and image type (artificial or medical). We also included additional analyses to see the limits and capabilities of TCF. If not specified otherwise, Gaussian kernels are used to obtain these results, but other unimodal kernels can be used as well. For medical images, one should carefully consider structures in the image and choose interpolation parameters accordingly, as selecting appropriate Gaussian scales is important when adjacent tubular structures are present in the image~\cite{law2008three, 
 law2010oriented} or in noisy images~\cite{law2013gradient}.

\subsection{Two-Dimensional Results}
\label{sec_experiments_2d}
This subsection has three parts: 
(1) curvature results for artificial images with various curved structures, 
(2) analyses of medical images, and 
(3) additional analyses on TCF and its comparison with centerline-based curvature.

\begin{figure}[t]
    \centering
    \begin{subfigure}[b]{0.90\textwidth}
        \centering    
        \includegraphics[trim={0 0 0 0.75cm },clip,width=0.32\textwidth]{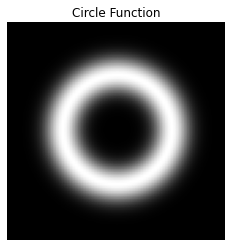}
        \includegraphics[trim={0 0 0 0.75cm },clip,width=0.32\textwidth]{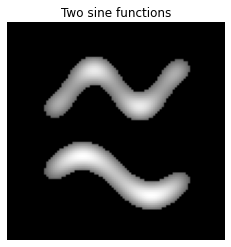}
        \includegraphics[trim={0 0 0 0.75cm },clip,width=0.32\textwidth]{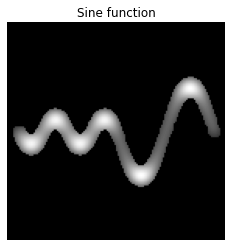}        
    \end{subfigure} 
    \begin{subfigure}[b]{0.90\textwidth}
        \centering    
        \includegraphics[trim={0 0 0 0.75cm },clip,width=0.32\textwidth]{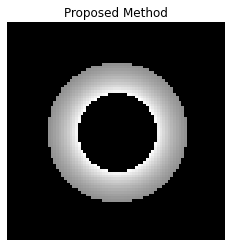}
        \includegraphics[trim={0 0 0 0.75cm },clip,width=0.32\textwidth]{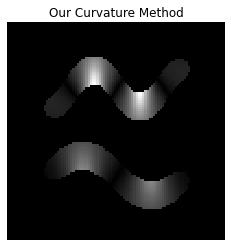}
        \includegraphics[trim={0 0 0 0.75cm },clip,width=0.32\textwidth]{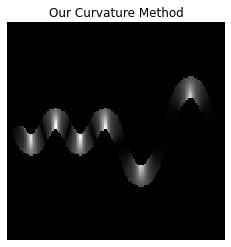} 
    \end{subfigure} 
    
    \caption{First row: Blurry ring function $\pring(\pointx)$, two sine functions with different frequencies $p_{sin,f}(\pointx)$, and a sine function with changing amplitude and frequency $p_{sin,A,f}(\pointx)$ are presented, respectively. Second row: Curvature values with TCF, where brighter pixels indicate higher curvature values.}
    \label{fig:artificial_results}
\end{figure}
\subsubsection{Artificial Images}
\label{sec:artificial_images}
This section presents two-dimensional artificial images, including a blurry ring, two sine waves, and a sine wave with changing amplitude and frequency. Fig.~\ref{fig:artificial_results} shows these images with their TCF curvature results, where brighter pixels indicate higher curvature values.

To generate images of the two sine waves, and a sine wave with changing amplitude and frequency, we employ Gaussian kernels as described in \eqref{eq:kde}, with the scale matrix $\mathbf{S}_j = 2\mathbf{I}_2$, where $\mathbf{I}_2$ represents the two-dimensional identity matrix, and $w_j = 1$ for $j \in \{1, \ldots, n\}$.  All functions are defined by placing evenly spaced isotropic Gaussian kernels, followed by the creation of images using uniform grids $\mathcal{G}(l,a,b)$ where $\mathcal{G}$ represents a uniform grid with $l$ samples in the interval $[a,b]$.

\paragraph{Image of a  Blurry Ring}
\label{sec:blurry_ring}
$\pblurryring(\pointx) = e^{-{{(\sqrt{\pointx^T \pointx}-r)}^2}/m}$ is a blurry ring function
where $r=1$ is the radius of the ring and $m=0.1$ is the thickness parameter.

The first column in Fig.~\ref{fig:artificial_results} displays $\pblurryring(\pointx)$ along with its curvature results. At every point in the ring function, the curvature is reciprocal to the radius of the ring. Consequently, points closer to the center of the ring are expected to exhibit higher curvature values, as demonstrated by TCF.

\paragraph{Image of Two Sine Waves}
We present the difference in curvature values for two sine functions with different frequencies. We place equally spaced isotropic Gaussian kernels on $y=\sin{(f\pointx)}$ function. The sine function is defined as $p_{sine,f}(\pointx)= \sum_{j=1}^{n} K_{I} (\pointx-\pointx_{j,f}),$ where $\pointx_{j,f} \in \{\pointx = [m,\sin{(fm)}]^T \,|\, m \in \mathcal{G}(50,-\pi,\pi)\}$. $f$ is the frequency of the sine wave. We created an image using a higher-frequency sine function $\psinhf(\pointx)$ with $f=1.5$ and a lower-frequency sine function $\psinlf(\pointx)$ with $f=1.0$. 

The third column of Fig.~\ref{fig:artificial_results} present the two sine waves with their TCF results. As expected, TCF values at local minima and maxima in $\psinhf(\pointx)$ are higher than those in $\psinlf(\pointx)$. 

\paragraph{Image of A Sine Wave with Changing Amplitude and Frequency}
We show the change of the curvature values along a single sine function by modifying its amplitude and frequency. Equally spaced isotropic Gaussian kernels are placed on the $y=A\sin{(f\pointx)}$ function, defined as $p_{sine,A,f}(\pointx)= \sum_{j=1}^{n} K_{I} (\pointx-\pointx_{j,A,f}),$
where $\pointx_{j,A,f} \in \{\pointx = [m,A\sin{(fm)}]^T \,|\, m \in \mathcal{G}(50,-2\pi,2\pi)\}$. $f$ is the frequency and $A$ is the amplitude of the sine wave. We generated an image by varying the values of $A$ and $f$ along the function. Specifically, for the first 25 samples (in the interval $[-2\pi, 0]$), we set $A = 1$ and $f = 2$, while for the remaining samples (in the interval $[0, 2\pi]$), we used $A = 3$ and $f = 1$.

The last column of Fig.~\ref{fig:artificial_results} presents $p_{sin,A,f}(\pointx)$ with its curvature results. Along the function, the amplitude of the sine wave increases while the frequency decreases. TCF gives higher values at the local minima and maxima in $p_{sin,A,f}(\pointx)$ at the first half of the function since the frequency in the first half is higher than the second half. Also, the amplitude of the sine wave increases along the function, and this creates flat function parts where the curvature values are almost zero.

\begin{figure}[t]
    \centering
    \begin{subfigure}[b]{0.49\textwidth}
        \centering
        \includegraphics[trim={0 0 0 0.75cm },clip,width=0.48\textwidth]{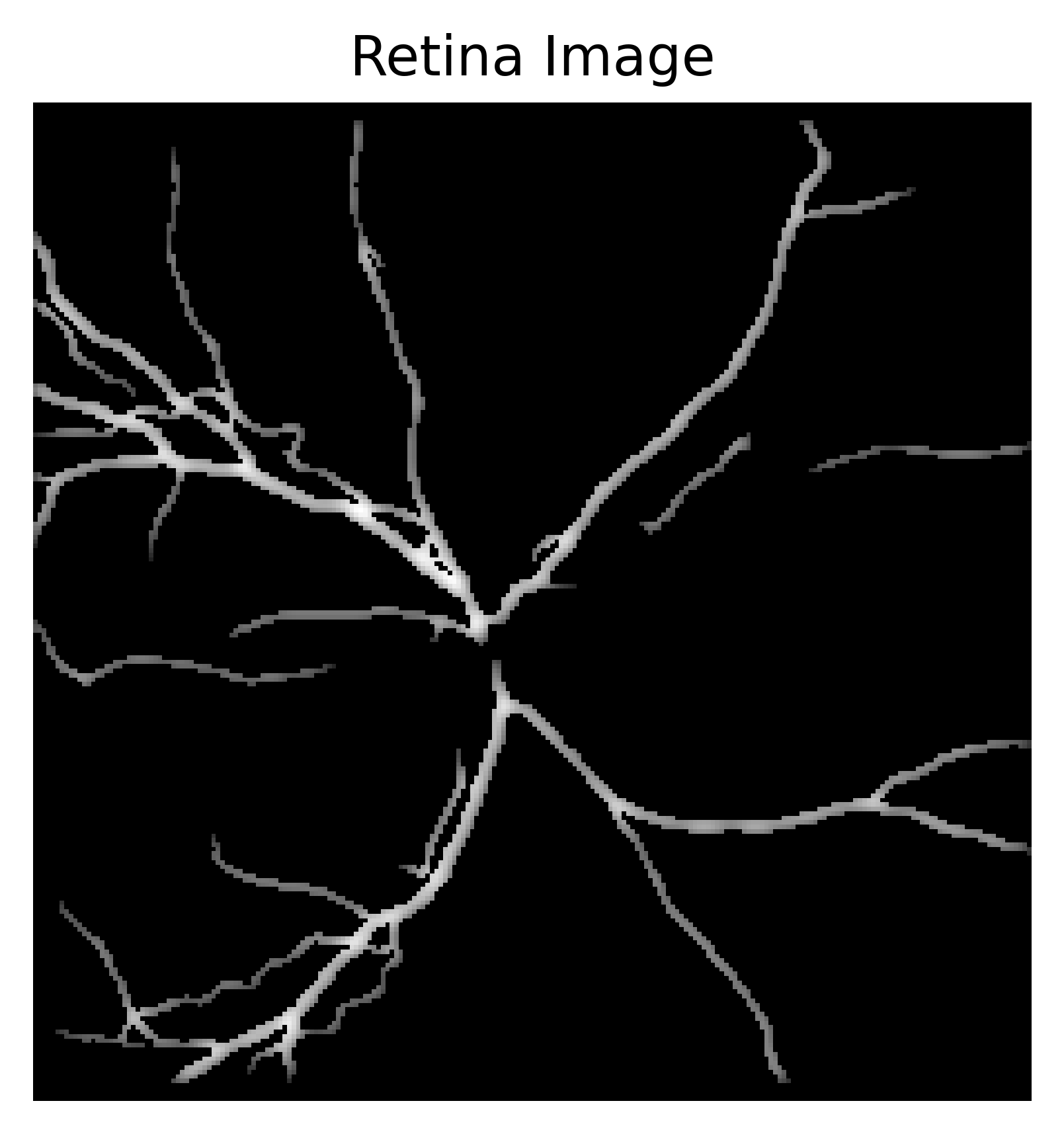}
        \includegraphics[trim={0 0 0 0.75cm },clip,width=0.48\textwidth]{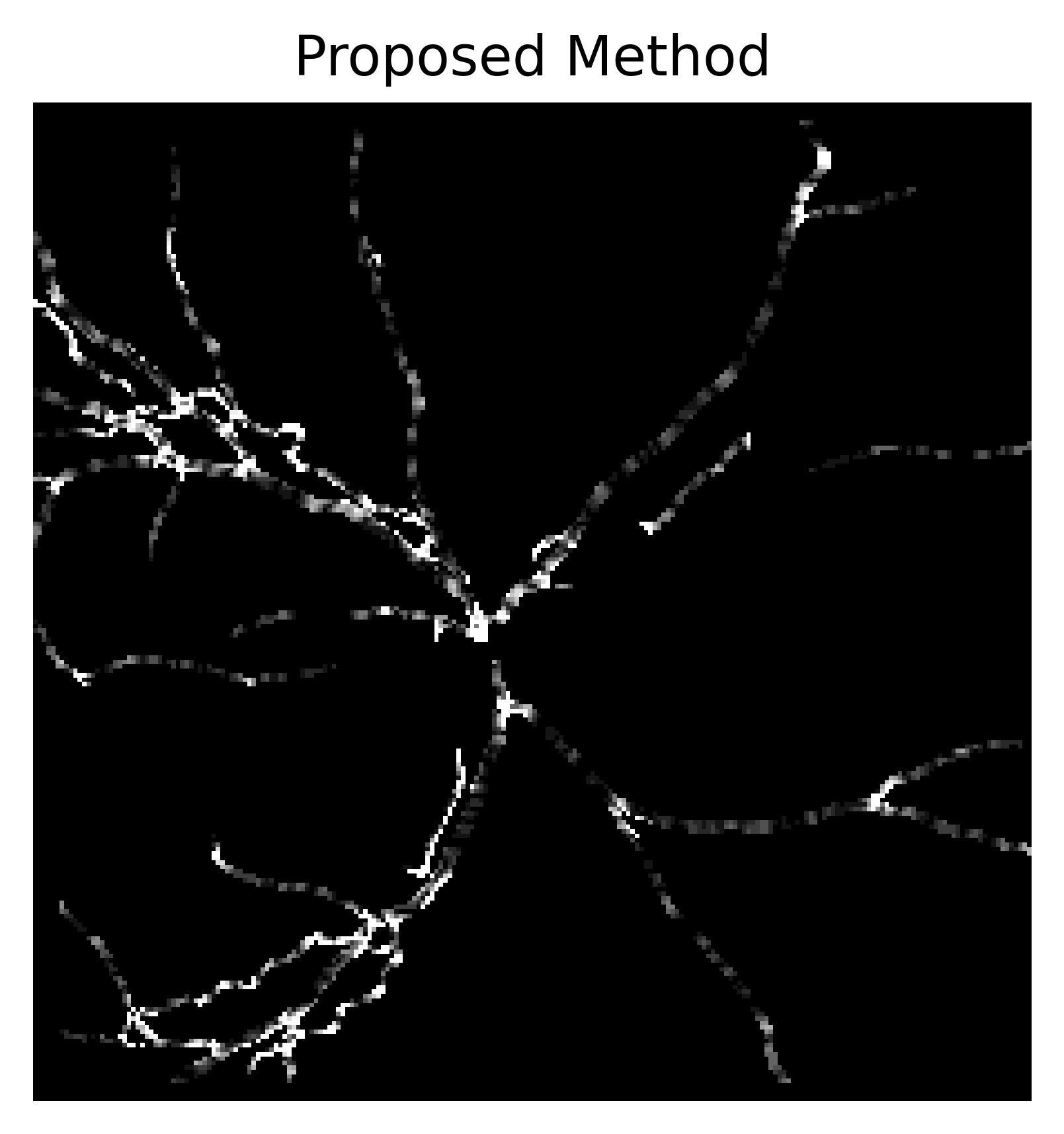}
    \end{subfigure} 
    \begin{subfigure}[b]{0.49\textwidth}
        \centering
        \includegraphics[trim={0 0 0 0.75cm },clip,width=0.48\textwidth]{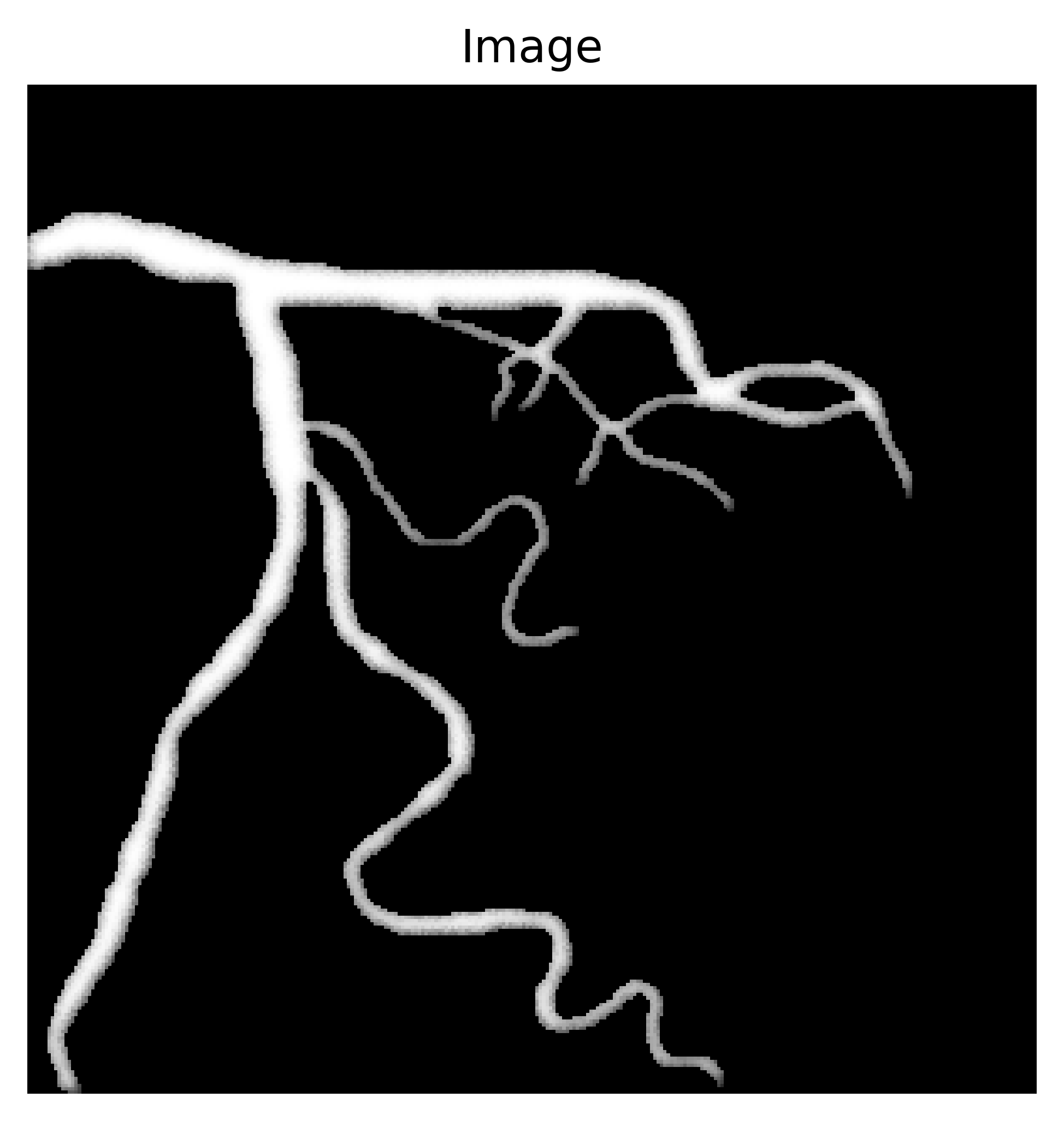}
        \includegraphics[trim={0 0 0 0.75cm },clip,width=0.48\textwidth]{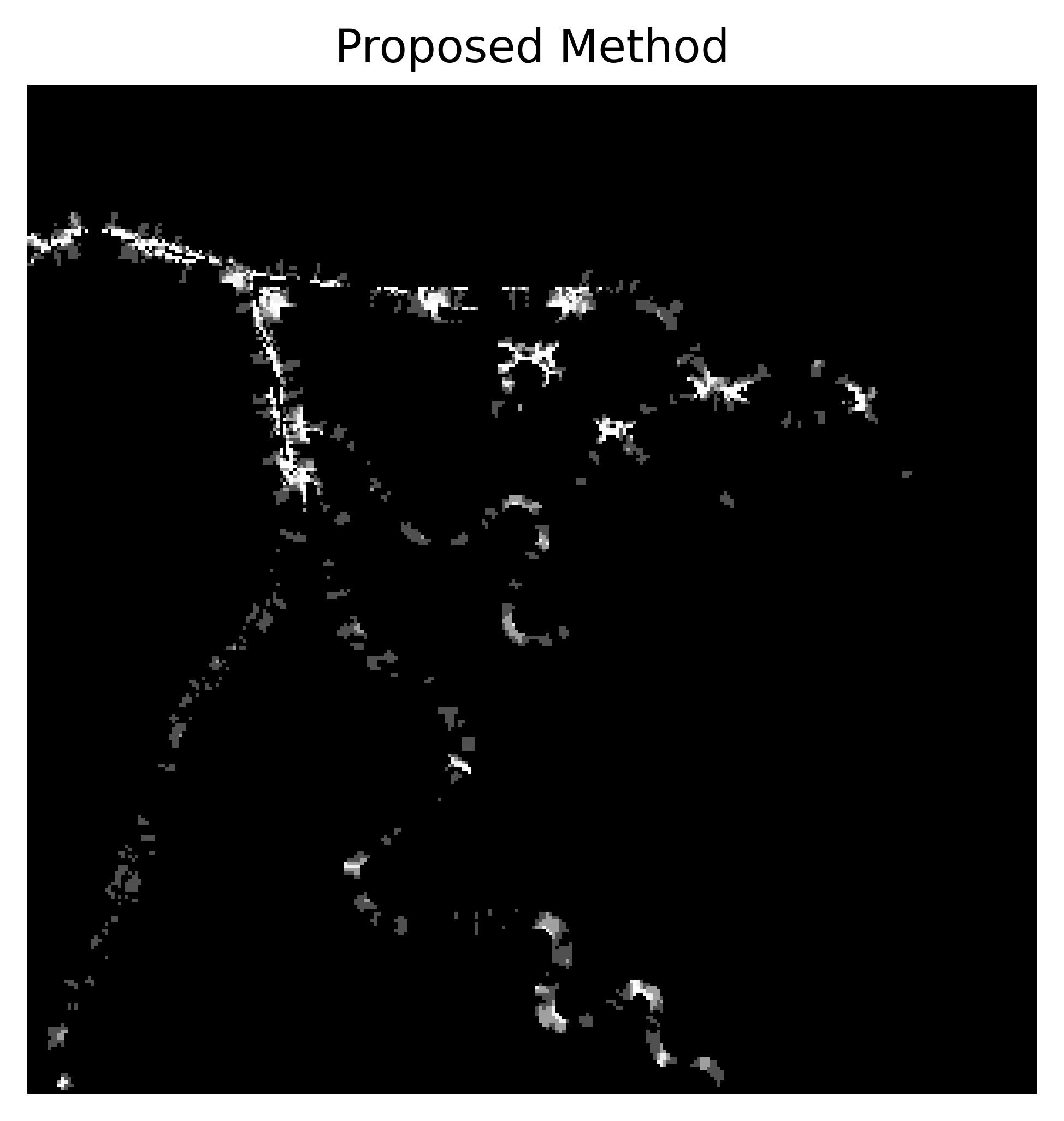}
    \end{subfigure}
        \caption{On the left: An Interpolated Retina image and its curvature calculated using TCF. On the right: An interpolated Angiography image and its curvature calculated using TCF. Contrast enhancement is applied to the curvature values for visualization. In the curvature plots, intensity values represent curvature, with brighter pixels indicating higher values.}
    \label{fig:real_results}
\end{figure}
\subsubsection{Medical Images}

\label{sec:real_images}
We examine two-dimensional medical images from the Retina \cite{brown2018automated} and X-ray coronary angiogram \cite{app9245507} datasets. Fig.~\ref{fig:real_results} presents these examples with their TCF curvature results. In the curvature images, intensity values indicate curvature, with brighter pixels representing higher values. Also, contrast enhancement is applied for visualization. Gaussian kernels, as explained in \eqref{eq:kde_knearest}, are applied to the medical examples using the scale matrix $\textbf{S}_j = 0.5 \times \mathbf{I}_2$, where $\mathbf{I}_2$ is the two-dimensional identity matrix, and $k = 440$.



\paragraph{Retina Images}
\label{sec:real_images_retina}
The left of Fig.~\ref{fig:real_results} shows a sample retinal vessel segmentation image and its curvature results. The segmented retina image includes branching points and vessel segments with different degrees of tortuosity.
Branching points occur where a vessel splits into smaller vessels or where multiple vessel segments overlap. Branching points are naturally high-curvature regions because vessels create a cross-shaped structure at these points. We observe this in TCF. Since retinal vessels originate around the optic disc, many branching points appear in this region, where we observe high curvature values. The vessel segments on the left of the image have high tortuosity, whereas the segments on the right are almost flat. The curvature image shows higher values for tortuous vessel segments compared to other segments, demonstrating that TCF captures the tortuosity of retinal blood vessels.

See Section~\ref{sec:ROP} for a detailed analysis of retina images with ROP categories and an explanation of why centerline methods fail to capture vessel structures.

\paragraph{Coronary Angiography Images}
\label{sec:real_images_angiography}
Fig.~\ref{fig:real_results} shows the interpolation function of a segmented coronary angiography and its TCF curvature values on the right. Similar to the retina example in Section \ref{sec:real_images_retina}, high curvature values appear at branching points and in highly tortuous regions. The vessel segment at the bottom left is nearly flat, while the right segment is highly tortuous. TCF captures this difference, assigning low curvature values to the flat segment and high values to the tortuous regions.

\subsubsection{Additional Analyses}
This section analyzes TCF to assess centerline limitations, kernel choices, intensity variations, and image resolution. In all cases where we define a ring function, using an inner radius of 0.5 and an outer radius of 1.0, except in Section~\ref{sec:same_centerline}. We also include an analysis of the numerical stability of TCF on a retinal image. Additionally, we analyze retinal images to qualitatively highlight regions where the method fails for different stages of ROP.

We define the ground-truth curvature for ring images as $1/r$, where $r$ is the distance of pixels from the ring's center. For ring examples with an inner radius of $0.5$ and the outer radius is $1.0$, the curvature values range from $1.0$ to $2.0$. We evaluate performance using the mean absolute difference between the ground truth and TCF.

\paragraph{Limitations of the Centerline}
\label{sec:same_centerline}
\begin{figure}[t]
    \centering
    \begin{subfigure}[b]{0.90\textwidth}
        \centering    
        \includegraphics[trim={0 0 0 0.75cm },clip,width=0.32\textwidth]{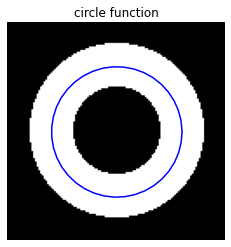}
        \includegraphics[trim={0 0 0 0.75cm },clip,width=0.32\textwidth]{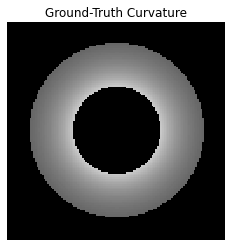}
        \includegraphics[trim={0 0 0 0.75cm },clip,width=0.32\textwidth]{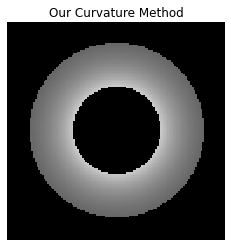}       
    \end{subfigure} 
    \begin{subfigure}[b]{0.90\textwidth}
        \centering    
        \includegraphics[trim={0 0 0 0.75cm },clip,width=0.32\textwidth]{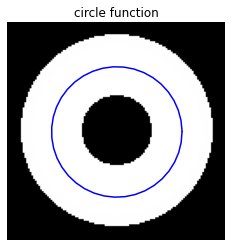}
        \includegraphics[trim={0 0 0 0.75cm },clip,width=0.32\textwidth]{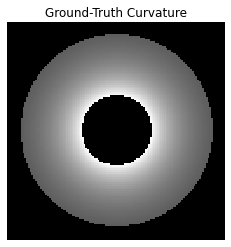}
        \includegraphics[trim={0 0 0 0.75cm },clip,width=0.32\textwidth]{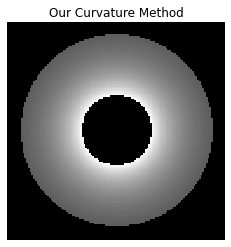}       
    \end{subfigure} 
    \caption{First row: From left to right, the thin ring function (inner radius 0.5, outer radius 1.0), ground-truth curvature, and TCF result. Second row: From left to right, the thick ring function (inner radius 0.4, outer radius 1.1), ground-truth curvature, and TCF result. Intensity values represent curvature, with brighter pixels indicating higher values. Both rings have the same centerline (in blue), illustrating that centerline-based methods fail to capture the shape.}
    \label{fig:same_cent}
\end{figure}

\begin{figure}[t]
    \centering
    \includegraphics[trim={0 3.0cm 0 3.75cm },clip,width=0.49\textwidth]{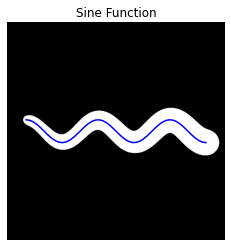}
    \includegraphics[trim={0 3.0cm 0 3.75cm },clip,width=0.49\textwidth]{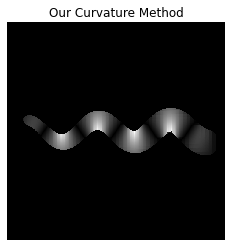}  
    \caption{From left to right: a tube with changing thickness and its centerline $y=sin(x)$, followed by the curvature result with TCF, where intensity values represent curvature, with brighter pixels indicating higher values. The centerline alone cannot capture the varying thickness, highlighting that methods relying solely on the centerline fail to represent the tube's shape.}
    \label{fig:thick tube}
\end{figure}

\begin{figure}[t]
    \centering
    \begin{subfigure}[b]{0.49\textwidth}
    \centering   
        \includegraphics[trim={0 0 0 0.75cm },clip,width=0.48\textwidth]{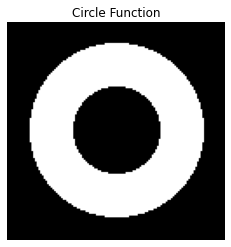}   
        \includegraphics[trim={0 0 0 0.75cm },clip,width=0.48\textwidth]{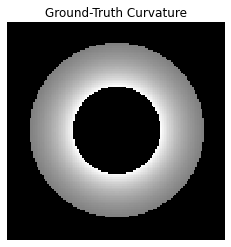}
        \caption{}
    \end{subfigure} 
    \begin{subfigure}[b]{0.49\textwidth}
        \centering 
        \includegraphics[trim={0 0 0 0.75cm },clip,width=0.48\textwidth]{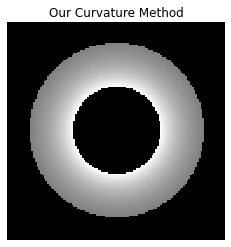}
        \includegraphics[trim={0 0 0 0.75cm },clip,width=0.48\textwidth]{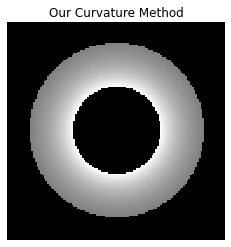}
        \caption{}
    \end{subfigure} 
    \caption{(a) Ring function $\pring(\pointx)$ and its ground-truth curvature. (b) Curvature values with TCF, from left to right, using the Gaussian and Epanechnikov kernels, with intensity values representing curvature, where brighter pixels indicate higher values. We calculated the mean absolute difference between the ground-truth and TCF. For the Gaussian kernel, it is $1.09 \times 10^{-8}$, and for the Epanechnikov kernel, it is $5.98 \times 10^{-15}$, respectively.}
    \label{fig:kernel}
\end{figure}
In this section, we focus on why centerline methods cannot capture curvature changes across the image. We present two examples: one compares two different ring examples with the same centerline, and the other examines a sine wave with increasing thickness. We generate an artificial ring example by defining two circles to generate a ring image  $\pring(\pointx)= \sum_{j=1}^{n} K (\pointx-\pointx_j),$ where $\pointx_j \in \{\pointx_j = \left[ r_j \cos\left(\theta)\right), r_j \sin\left( \theta \right) \right]^T \,|\, r_j  \in [0.5,1], \,|\, \theta \in [0,2\pi)\,|\, \pointx_j \in \mathcal{G}(125,-1.25,1.25)\}$. $K$ is either the Gaussian kernel. The pixel intensity is set to $1$ for pixels between the rings ($w_j = 1$ for $j \in {1, \ldots, n}$) and the scale matrix is set to $\mathbf{S}_j = 0.1 \times \mathbf{I}_2$.

Fig.~\ref{fig:same_cent} shows two ring images with the same centerline: one with radii 0.5 and 1.0 (first row) and another with radii 0.4 and 1.1 (second row). From left to right, it displays the ring function, ground-truth curvature, and TCF results. The ring functions differ but their centerline remains the same. Centerline-based approaches compute curvature along this line but cannot capture curvature variations across the ring’s radii.

Fig.~\ref{fig:thick tube} presents a tube example where the centerline is defined as $y=sin(x)$ and the thickness varies along the tube. From left to right, it shows the tube with its centerline and the TCF results. Similar to the ring example, the centerline alone cannot capture the thickness variations across the tube.

\paragraph{Results with Different Kernels}
\label{sec:kernels}
In this paper, we primarily use the Gaussian kernel; however, in this section, we also employ the Epanechnikov kernel, defined as, $K_e(\mathbf{x}) = 1-\pointx^{T}\pointx$ for $|\mathbf{x}| \leq 1$, for a given point $\pointx \in \realnumbers^2$.

We generate a ring example as in Section~\ref{sec:same_centerline}. Here, $K$ is either the Gaussian kernel or the Epanechnikov kernel. 

Fig.~\ref{fig:kernel}(a) shows the image of the ring function $\pring(\pointx)$ and the ground-truth curvature. Fig.~\ref{fig:kernel}(b) shows TCF results from left to right with Gaussian and Epanechnikov kernels. Both kernels perform well in calculating the pixel-wise curvature, with mean absolute differences between the ground-truth and TCF of $1.09 \times 10^{-8}$ and $5.98 \times 10^{-15}$, respectively.

\paragraph{Intensity Profiles}
\begin{figure}[t]
    \centering
    \begin{subfigure}[b]{0.90\textwidth}
        \centering    
        \includegraphics[trim={0 0 0 0.75cm },clip,width=0.32\textwidth]{JMO/figures/Appendix_results/circle_func_meshstep0.02_ones.png}
        \includegraphics[trim={0 0 0 0.75cm },clip,width=0.32\textwidth]{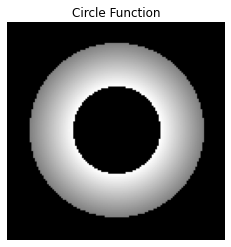}
        \includegraphics[trim={0 0 0 0.75cm },clip,width=0.32\textwidth]{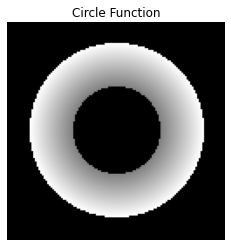}     
    \end{subfigure} 
    \begin{subfigure}[b]{0.90\textwidth}
        \centering    
        \includegraphics[trim={0 0 0 0.75cm },clip,width=0.32\textwidth]{JMO/figures/Appendix_results/curv_circle_meshstep0.02_ones.png}
        \includegraphics[trim={0 0 0 0.75cm },clip,width=0.32\textwidth]{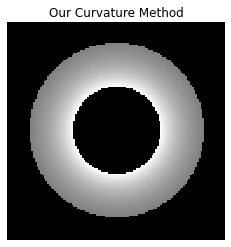}
        \includegraphics[trim={0 0 0 0.75cm },clip,width=0.32\textwidth]{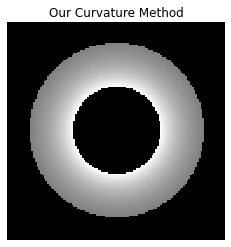}
    \end{subfigure} 
    
    \caption{First row: Ring function $\pring(\pointx)$ with different intensity profiles, uniform inside the inner and outer rings, higher near the inner ring, and lower near the outer ring. Second row: Curvature values with TCF, intensity represents curvature, with brighter pixels indicating higher values. We computed the mean absolute difference between the ground truth and TCF. From left to right, the values are: $1.09 \times 10^{-8}$, $1.56 \times 10^{-8}$, and $0.94 \times 10^{-8}$.}
    \label{fig:intensity}
\end{figure}
\begin{figure}[t]
    \centering
    \begin{subfigure}[b]{0.90\textwidth}
        \centering    
        \includegraphics[trim={0 0 0 0.75cm },clip,width=0.32\textwidth]{JMO/figures/Appendix_results/circle_func_meshstep0.02_ones.png}
        \includegraphics[trim={0 0 0 0.75cm },clip,width=0.32\textwidth]{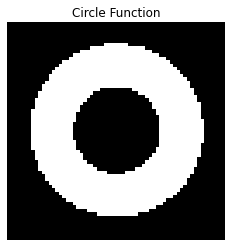}
        \includegraphics[trim={0 0 0 0.75cm },clip,width=0.32\textwidth]{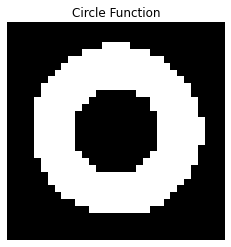}       
    \end{subfigure} 
    \begin{subfigure}[b]{0.90\textwidth}
        \centering    
        \includegraphics[trim={0 0 0 0.75cm },clip,width=0.32\textwidth]{JMO/figures/Appendix_results/curv_circle_meshstep0.02_ones.png}
        \includegraphics[trim={0 0 0 0.75cm },clip,width=0.32\textwidth]{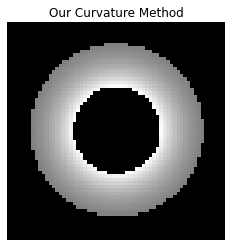}
        \includegraphics[trim={0 0 0 0.75cm },clip,width=0.32\textwidth]{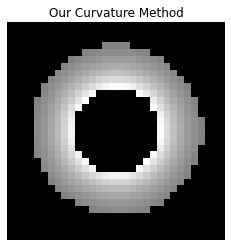}
    \end{subfigure} 
    \caption{First row: Ring function $\pring(\pointx)$ with resolutions of $125 \times 125$, $62.5 \times 62.5$, and $31.25 \times 31.25$. Second row: Curvature values with TCF, intensity represents curvature, with brighter pixels indicating higher values. We computed the mean absolute difference between the ground truth and TCF. From left to right, the values are $1.09 \times 10^{-8}$, $6.75 \times 10^{-8}$, and $1.27 \times 10^{-3}$.}
    \label{fig:resolution}
\end{figure}

We use the ring function $\pring(\pointx)$ as described in Section~\ref{sec:same_centerline}. We evaluate the sensitivity of TCF to different intensity profiles, such as ``uniform'', where all of the pixels in the region have a value of $1$; ``higher inside'', where $w_j = 1.5 - r_j$; ``higher outside'', where $w_j = r_j$, with $r_j$ being the distance of the $j$-th pixel from the center.

Fig.~\ref{fig:intensity} demonstrates the ring functions with different intensity values in the first row and their respective TCF results with a Gaussian kernel in the second row. All of these cases perform well in estimating the curvature, achieving mean absolute differences from the ground-truth results of $1.09 \times 10^{-8}$, $1.56 \times 10^{-8}$, and $0.94 \times 10^{-8}$, respectively.

\paragraph{Resolution}
We use ring images with pixel widths and heights of $0.02$, $0.04$, and $0.08$, maintaining uniform intensity as in Section~\ref{sec:same_centerline}. These correspond to resolutions of $125 \times 125$, $62.5 \times 62.5$, and $31.25 \times 31.25$. Fig.~\ref{fig:resolution} displays these resolutions from left to right in the first row and their TCF results with a Gaussian kernel in the second row. As the resolution decreases, the estimation quality worsens. The mean absolute differences from the ground-truth are $1.09 \times 10^{-8}$, $6.75 \times 10^{-8}$, and $1.27 \times 10^{-3}$, respectively.

\paragraph{Numerical Stability}
\label{sec:stability}
We focus on the numerical stability of the inversion of the linear system given in~\eqref{eq:matrix_form_both} for medical images, such as the retina image. To assess stability, we calculate the condition number of $\mathbf{M}$. A large condition number indicates the matrix is ill-conditioned, meaning that small changes in the input can lead to large changes in the solution, while a smaller condition number suggests greater stability. We calculate the condition number as $\kappa(\mathbf{M}) = \|\mathbf{M}\|_2 \cdot \|\mathbf{M}^{-1}\|_2$.

\begin{figure}[t]
    \centering
    \includegraphics[trim={0 0 0 0.75cm },clip,width=0.33\textwidth]{JMO/figures/Retina.png}
    \includegraphics[trim={0 0 0 0.75cm },clip,width=0.33\textwidth]{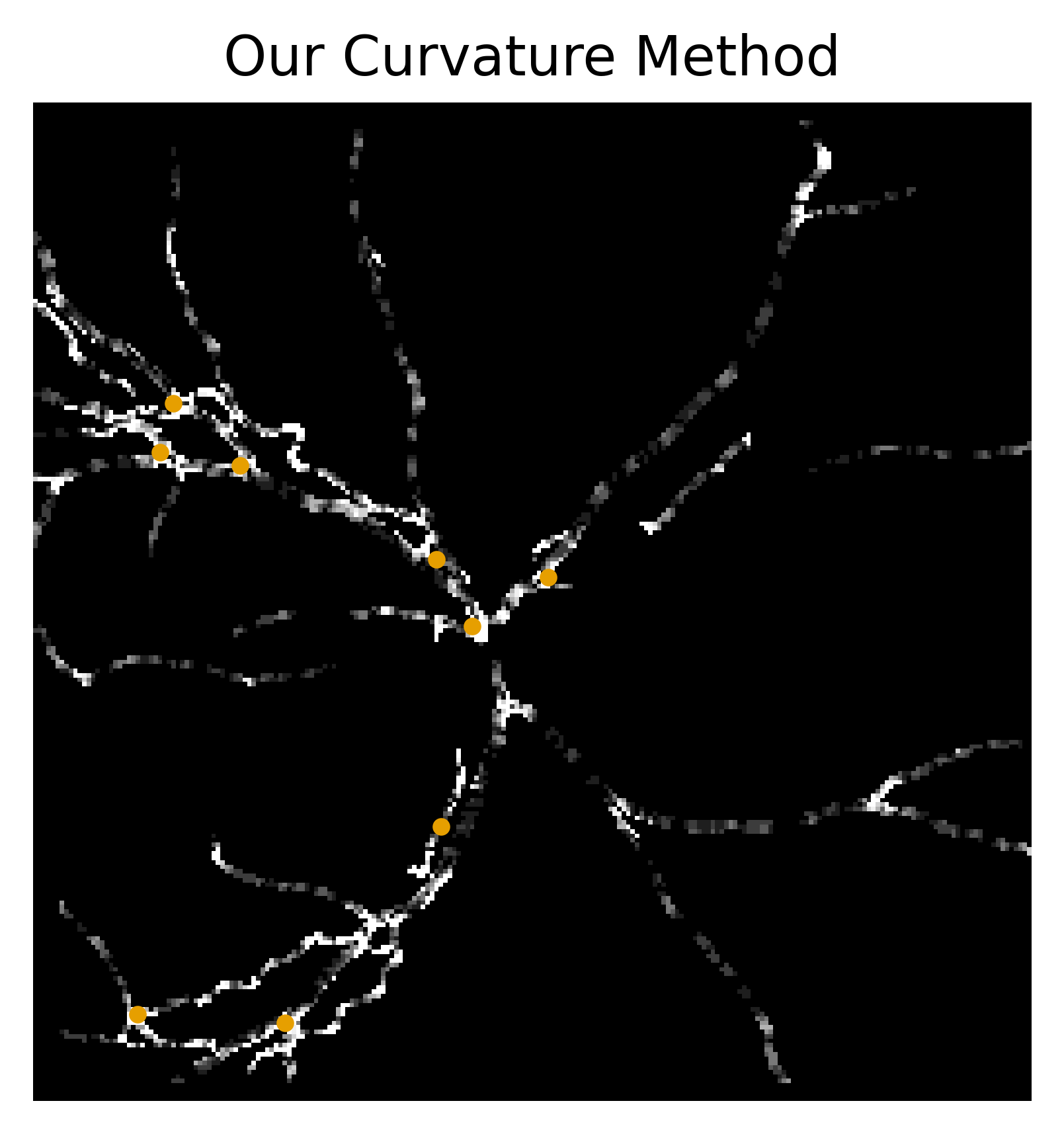}    
    \caption{From left to right: a Retina image and its curvature with TCF, where pixels with a condition number for matrix M greater than 100 are marked in orange. The condition number of $\mathbf{M}$ ranges from $5.1$ to $234.1$ for this image.}
    \label{fig:stability_retina}
\end{figure}

Fig.~\ref{fig:stability_retina} presents the retina image and its corresponding curvature image, where pixels with a condition number larger than 100 are marked in orange. The condition number of $\mathbf{M}$ ranges from $5.1$ to $234.1$. Eq.~\eqref{eq:hessian} assumes the Hessian matrix is diagonalizable. Therefore, we cannot calculate curvature values with TCF when this occurs, which is expected in the middle of vessel branching regions. Additionally, as shown in Fig.~\ref{fig:stability_retina}, even if the Hessian is diagonalizable in the middle of these regions, the condition number of $\mathbf{M}$ is large, indicating that TCF calculations are not numerically stable in these areas.

\begin{figure}[t]
    \centering
    \begin{subfigure}[b]{0.90\textwidth}
        \centering    
        \begin{tikzpicture}
            \node[anchor=south west,inner sep=0] (image) at (0,0) 
                {\includegraphics[trim={0 0 0 0.75cm },clip,width=0.32\textwidth]{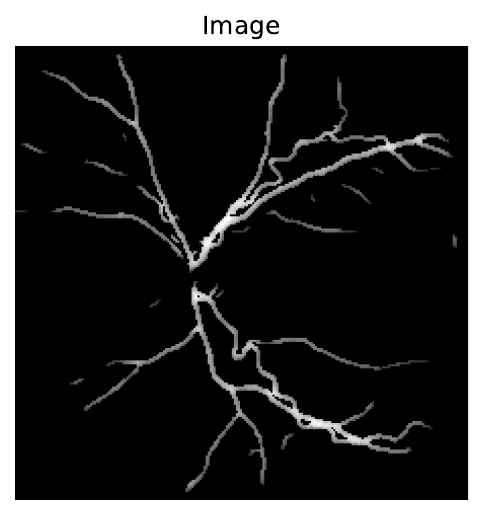}};
            \begin{scope}[x={(image.south east)},y={(image.north west)}]
                \draw[red, thick] (0.42,0.57) rectangle (0.52,0.67);
            \end{scope}
        \end{tikzpicture}
        \begin{tikzpicture}
            \node[anchor=south west,inner sep=0] (image) at (0,0) 
                {\includegraphics[width=0.32\textwidth]{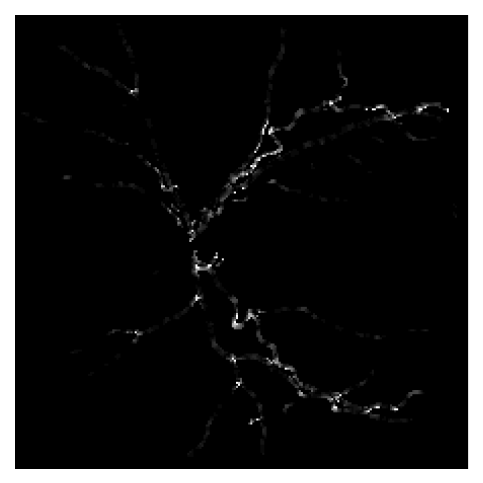}};
            \begin{scope}[x={(image.south east)},y={(image.north west)}]
                \draw[red, thick] (0.42,0.55) rectangle (0.52,0.65);
            \end{scope}
        \end{tikzpicture}
        \begin{tikzpicture}
            \node[anchor=south west,inner sep=0] (image) at (0,0) 
                {\includegraphics[width=0.32\textwidth]{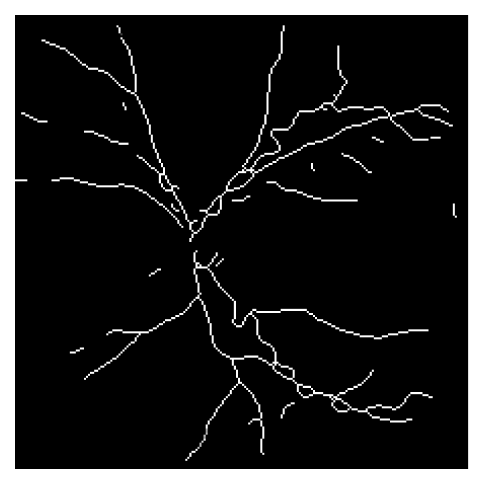}};
            \begin{scope}[x={(image.south east)},y={(image.north west)}]
                \draw[red, thick] (0.42,0.55) rectangle (0.52,0.65);
            \end{scope}
        \end{tikzpicture}   
    \end{subfigure} 
    \begin{subfigure}[b]{0.90\textwidth}
        \centering    
        \begin{tikzpicture}
            \node[anchor=south west,inner sep=0] (image) at (0,0) 
                {\includegraphics[trim={0 0 0 0.75cm },clip,width=0.32\textwidth]{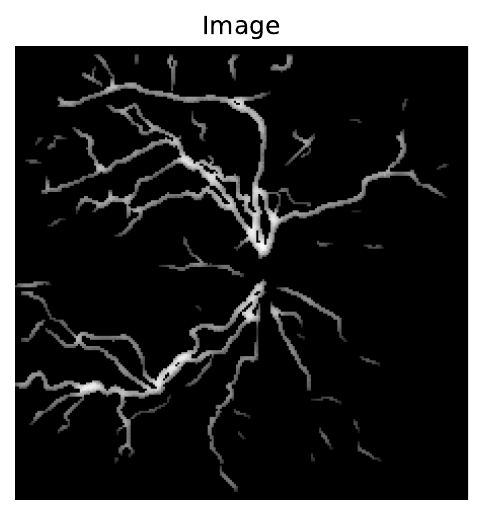}};
            \begin{scope}[x={(image.south east)},y={(image.north west)}]
                \draw[red, thick] (0.31,0.28) rectangle (0.41,0.38);
            \end{scope}
        \end{tikzpicture}
        \begin{tikzpicture}
            \node[anchor=south west,inner sep=0] (image) at (0,0) 
                {\includegraphics[width=0.32\textwidth]{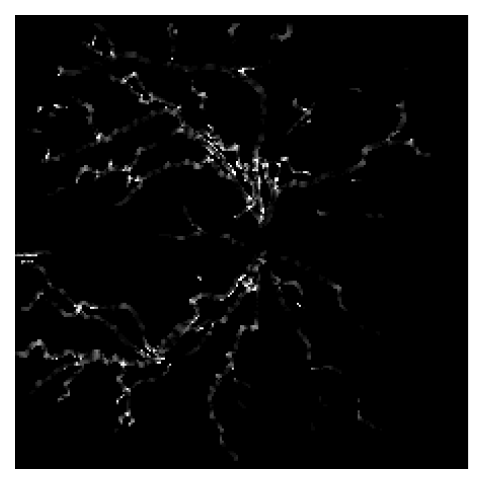}};
            \begin{scope}[x={(image.south east)},y={(image.north west)}]
                 \draw[red, thick] (0.31,0.26) rectangle (0.41,0.36);
            \end{scope}
        \end{tikzpicture}
        \begin{tikzpicture}
            \node[anchor=south west,inner sep=0] (image) at (0,0) 
                {\includegraphics[width=0.32\textwidth]{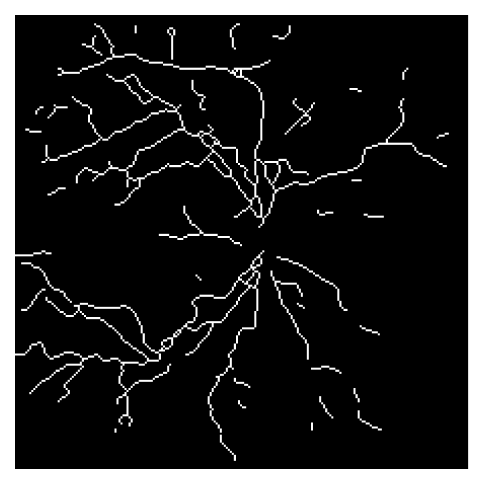}};
            \begin{scope}[x={(image.south east)},y={(image.north west)}]
                 \draw[red, thick] (0.31,0.26) rectangle (0.41,0.36);
            \end{scope}
        \end{tikzpicture}   
    \end{subfigure} 
    
    \caption{From left to right: images from the Retina dataset (with Pre-Plus, and Plus ROP from top to bottom), the corresponding curvature with TCF, and curvature with the centerline. The centerline is calculated using the ``thin'' function from the morphological operations in scikit-image~\cite{van2014scikit}. We compute the centerline curvature using the implicit planar curvature calculation~\cite{GOLDMAN2005632}. Contrast enhancement is applied to the curvature values. In the curvature plots, intensity represents curvature, with brighter pixels indicating higher values. The red squares highlight examples where vessels are dilated, and the centerline curvature cannot capture the shape.}
    \label{fig:rop}
\end{figure}

\paragraph{Retinopathy of Prematurity}
\label{sec:ROP}
We focus on the Retina \cite{brown2018automated} dataset with ROP images labeled as ``Normal,'' ``Pre-plus'', or ``Plus'', indicating no ROP, intermediate, or severe cases, respectively. Generally, the more dilated (increased vessel thickness) and tortuous (higher curvature) the vessels, the more severe the ROP~\cite{chiang2021international}. Fig.~\ref{fig:rop} presents Pre-Plus and Plus images from top to bottom, with TCF results in the second column and the centerline in the last. The red box regions highlight examples of dilated vessels, where, as discussed in Section~\ref{sec:same_centerline}, the centerline cannot capture the shape in these regions.

\subsection{Three-Dimensional Results}
Here, we include results from (1)  three-dimensional artificial images with various curved structures, (2)  three-dimensional medical images, and (3) an additional analysis on voxel isotropy.  
\subsubsection{Artificial Images}
\begin{figure}[t]
    \centering
    \begin{subfigure}[b]{0.49\textwidth}
        \centering
        \includegraphics[trim={0.8cm 0.85cm 0.8cm 1.0cm },clip,width=0.48\textwidth]{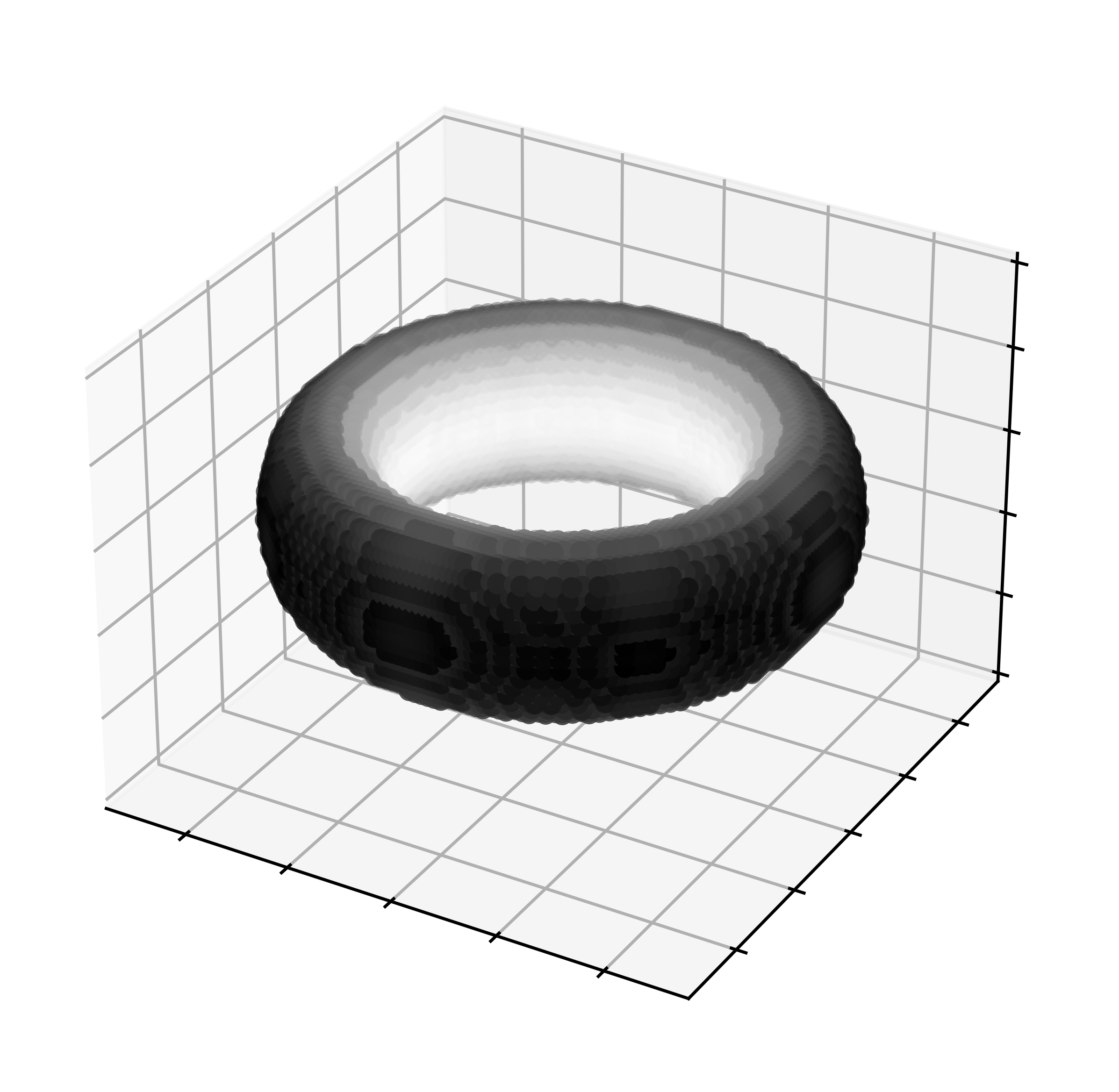}
        \includegraphics[trim={0.8cm 0.85cm 0.8cm 1.0cm },clip,width=0.48\textwidth]{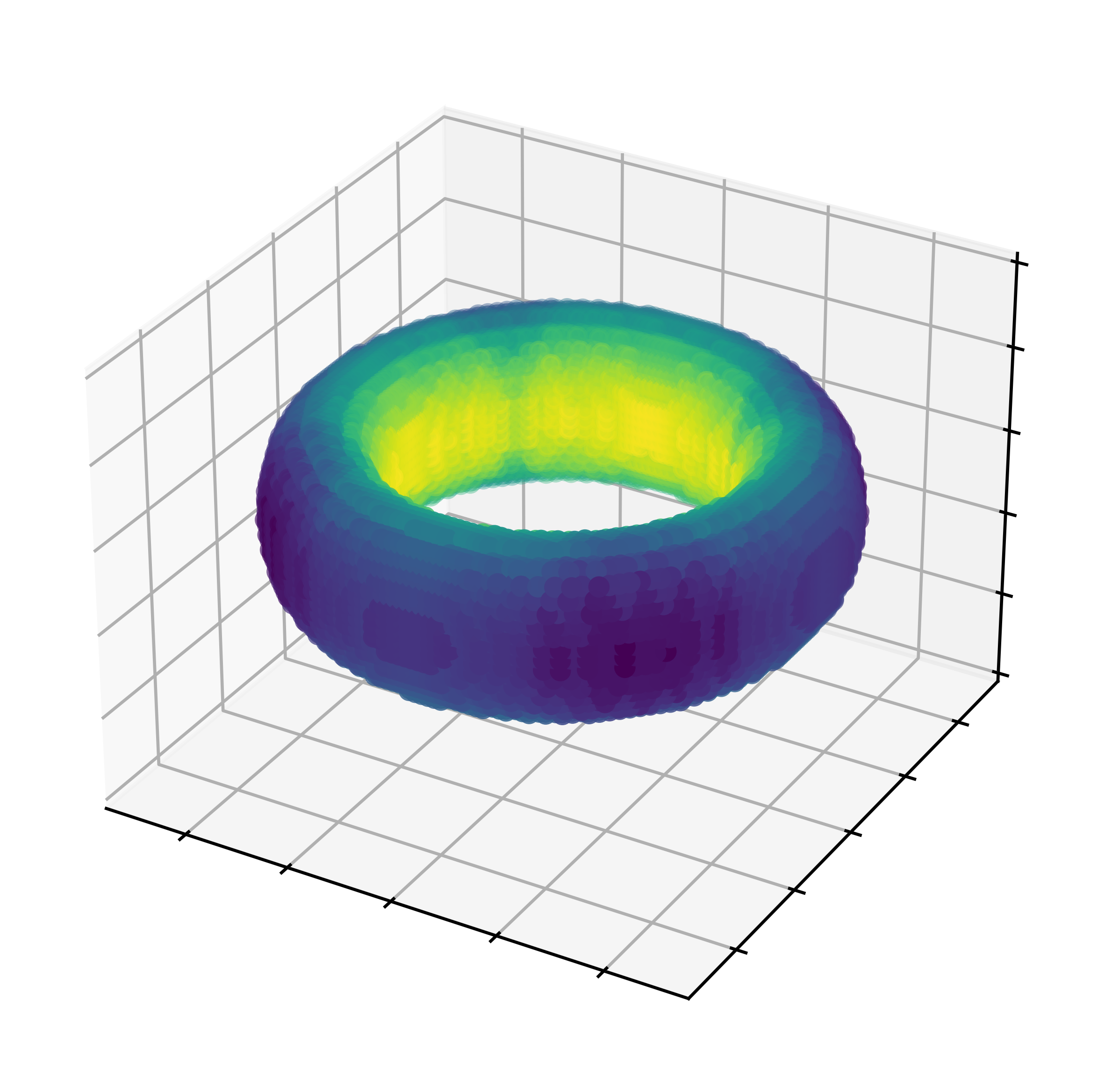}
    \end{subfigure} 
    \begin{subfigure}[b]{0.49\textwidth}
        \centering
        \includegraphics[trim={0.8cm 0.85cm 0.8cm 1.0cm },clip,width=0.48\textwidth]{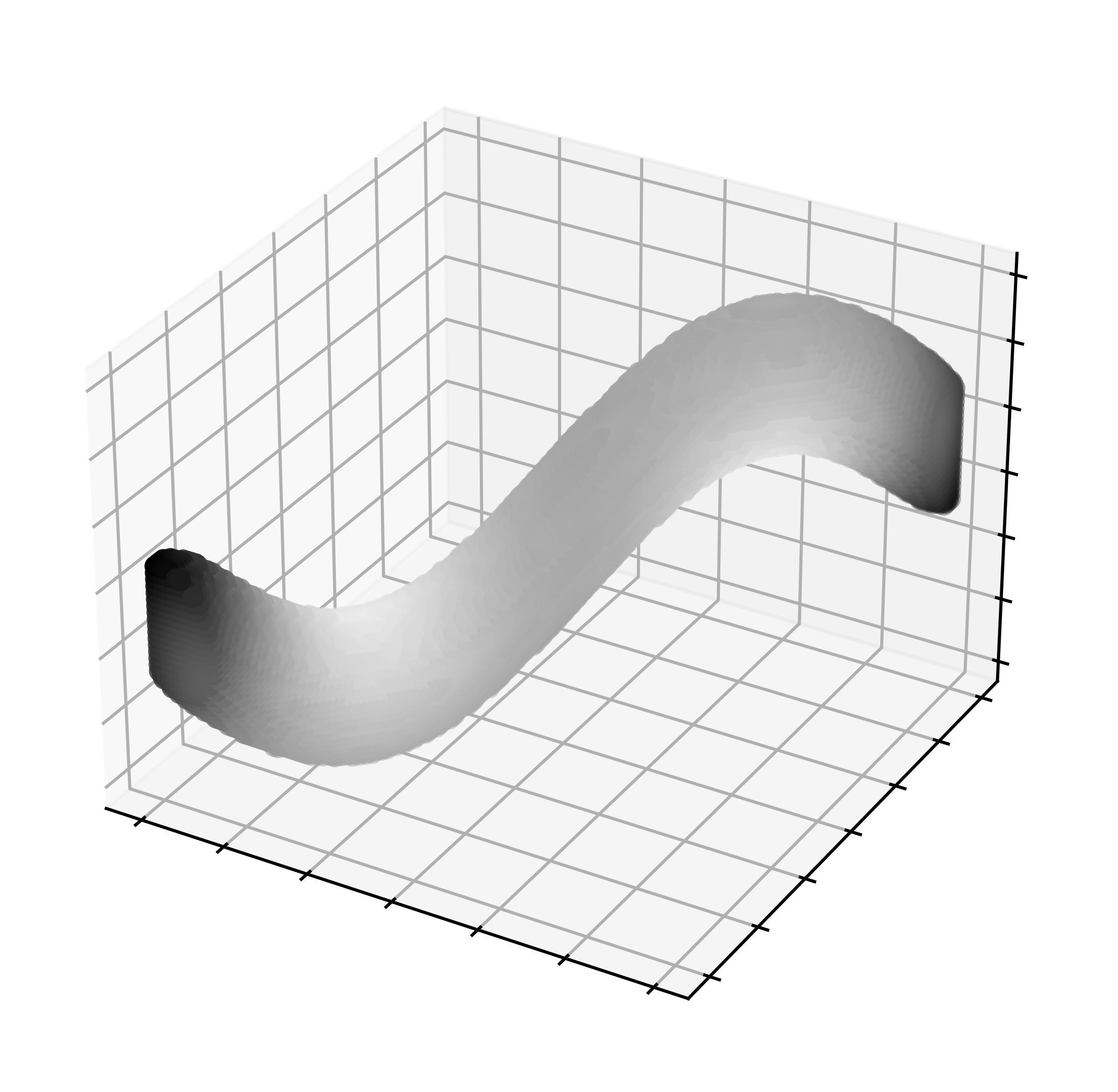}
        \includegraphics[trim={0.8cm 0.85cm 0.8cm 1.0cm },clip,width=0.48\textwidth]{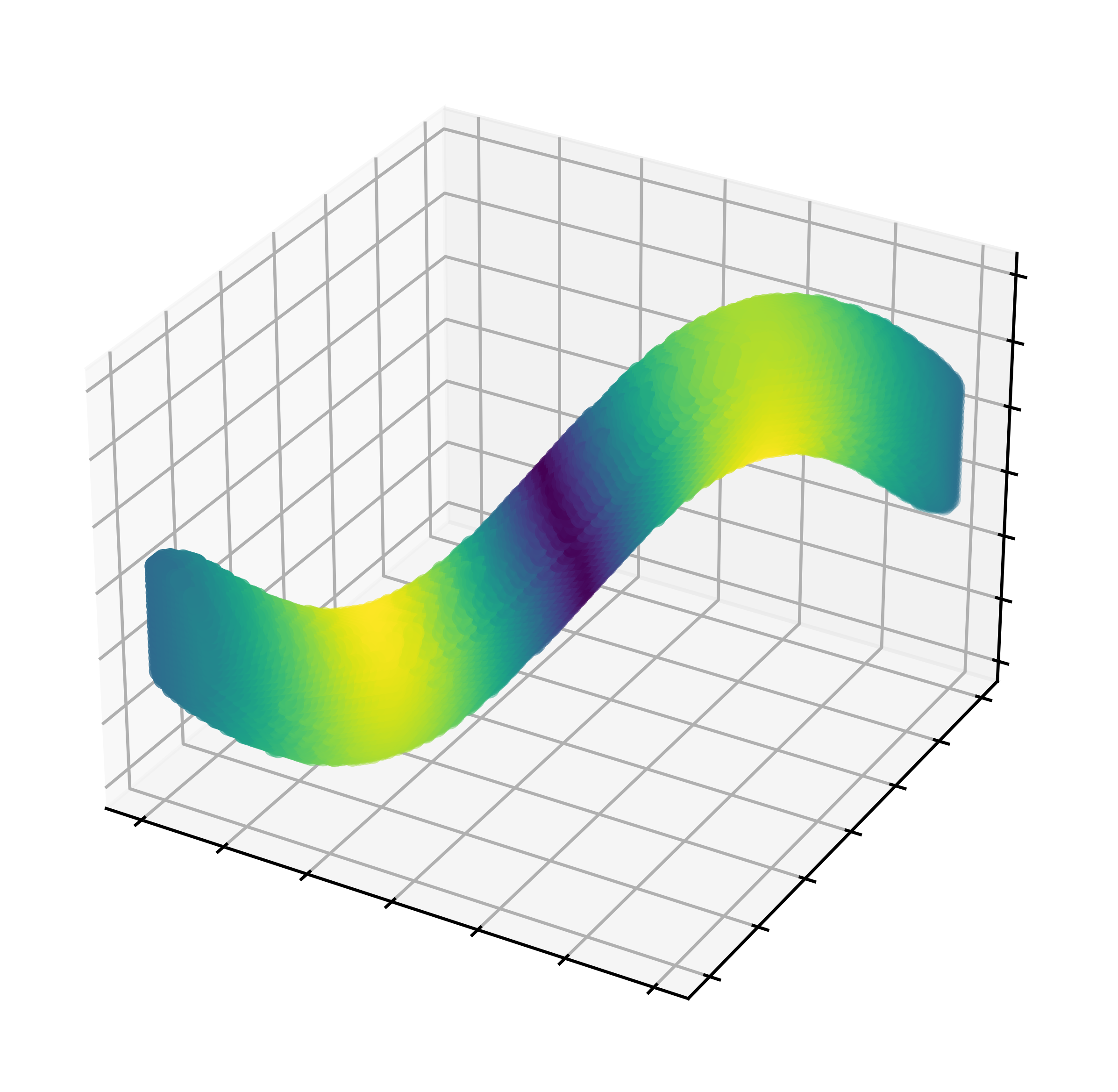}
    \end{subfigure}
    \begin{subfigure}[b]{0.6\textwidth}
        \centering
        \includegraphics[width=0.7\textwidth]{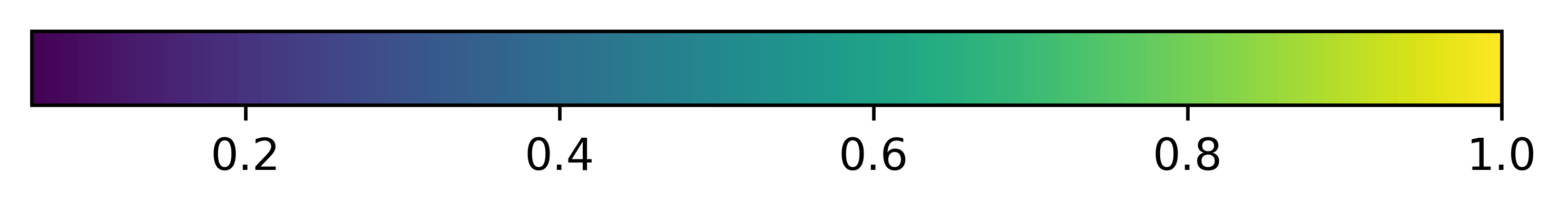} 
    \end{subfigure}
    \caption{On the left: The three-dimensional blurry-ring function image and its TCF result. On the right: The three-dimensional sine function image and its TCF result. At the bottom: The Viridis colormap for the curvature plots.}
    \label{fig:artificial_results_3d}
\end{figure}

We present three-dimensional artificial images: a blurry ring and a sine wave. Fig.~\ref{fig:artificial_results_3d} displays these images with their corresponding curvature images using TCF, and the corresponding colormap. We employ Gaussian kernels given in \eqref{eq:kde}, with the scale matrix $\mathbf{S}_j = \mathbf{I}_3$, where $\mathbf{I}_3$ is the three-dimensional identity matrix, and $w_j = 1$ for $j \in \{1, \ldots, n\}$. Functions are defined by positioning evenly spaced isotropic Gaussian kernels, followed by the creation of images using uniform grids as explained in Section~\ref{sec:artificial_images}.

\paragraph{Image of a Three-Dimensional Blurry Ring}
\label{sec:blurry_ring_3d}

$p_{\text{blurry\_ring\_3d}}(\pointx)= \sum_{j=1}^{n} K_{I} (\pointx-\pointx_{j})$ is the three-dimensional blurry ring function, which is defined by placing equally spaced isotropic Gaussian kernels. Here, $\pointx_j \in \{\pointx = [r\sin{(\theta)},r\cos{(\theta)}, \sin(r)]^T \,|\, \theta \in \mathcal{G}(40,0,2\pi)\}$, and $r$ is 2. Fig.~\ref{fig:artificial_results_3d} presents $p_{\text{blurry\_ring\_3d}}(\pointx)$ and its curvature results on the left. Similar to Section~\ref{sec:blurry_ring}, the curvature is reciprocal to the radius of the ring and TCF gives higher values to the points closer to the center. 

\paragraph{Image of a Three-Dimensional Sine Wave}
We define a sine function and place equally spaced isotropic Gaussian kernels as $p_{\text{sine\_3d}}(\pointx)= \sum_{j=1}^{n} K_{I} (\pointx-\pointx_{j}),$ where $\pointx_j \in \{\pointx = [m,m,\sin{(m)}]^T \,|\, m \in \mathcal{G}(40,-\pi,\pi)\}$. Fig.~\ref{fig:artificial_results_3d} shows $p_{\text{sine\_3d}}(\pointx)$ and its TCF result on the right. 
Pixels on the inner side of the curves have higher curvature than those on the outer side, and pixels near the function's maximum and minimum exhibit higher curvature values than those farther away.




\subsubsection{Medical Images}
This section examines examples from the Computed Tomography Angiography~\cite{ZENG2023102287} and the Coronary Artery Segmentations~\cite{dalvit2023automated} datasets. We employed Gaussian kernels as the interpolation method, as defined in \eqref{eq:kde_knearest}. In three dimensions, $\tilde{\pointx}_j\in \mathbb{R}_+^3$, and $K_{S_j}(\cdot): \mathbb{R}^3 \rightarrow \mathbb{R}$ is the interpolation kernel and we used a scale matrix $\textbf{S}_j = \mathbf{I}_3$. For the Computed Tomography Angiography dataset~\cite{ZENG2023102287}, $k = 20$, and for the Coronary Artery Segmentations dataset~\cite{dalvit2023automated}, $k = 40$. Curvature results were visualized with contrast enhancement using histogram equalization~\cite{jain1989fundamentals}.
\begin{figure}[t]
    \centering
    \begin{subfigure}[b]{\textwidth}
        \centering
        \includegraphics[trim={0.9cm 0.85cm 0.9cm 1cm },clip, width=0.24\textwidth]{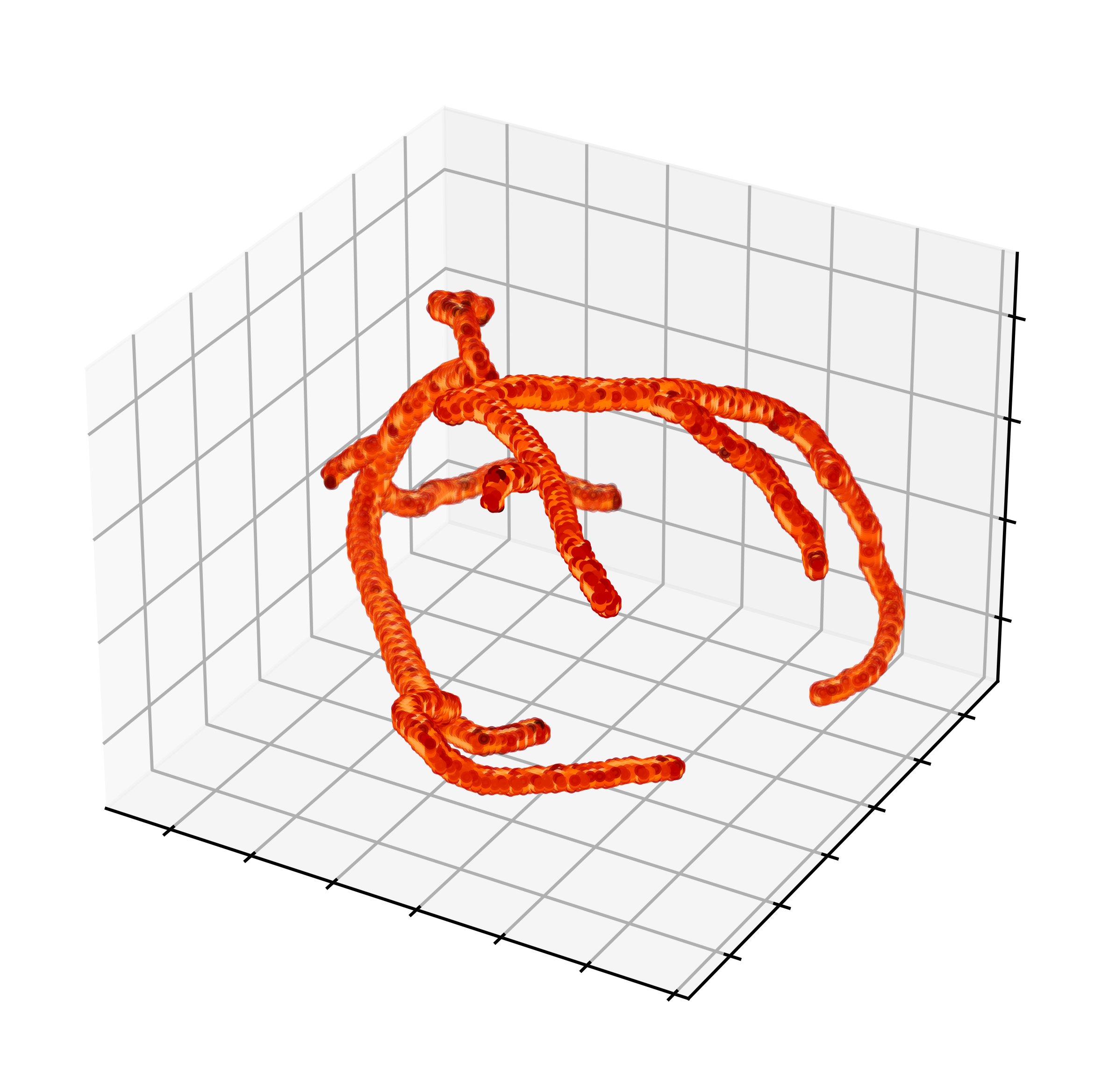}
        \includegraphics[trim={1.3cm 1.7cm 1.3cm 1.6cm},clip, width=0.24\textwidth]{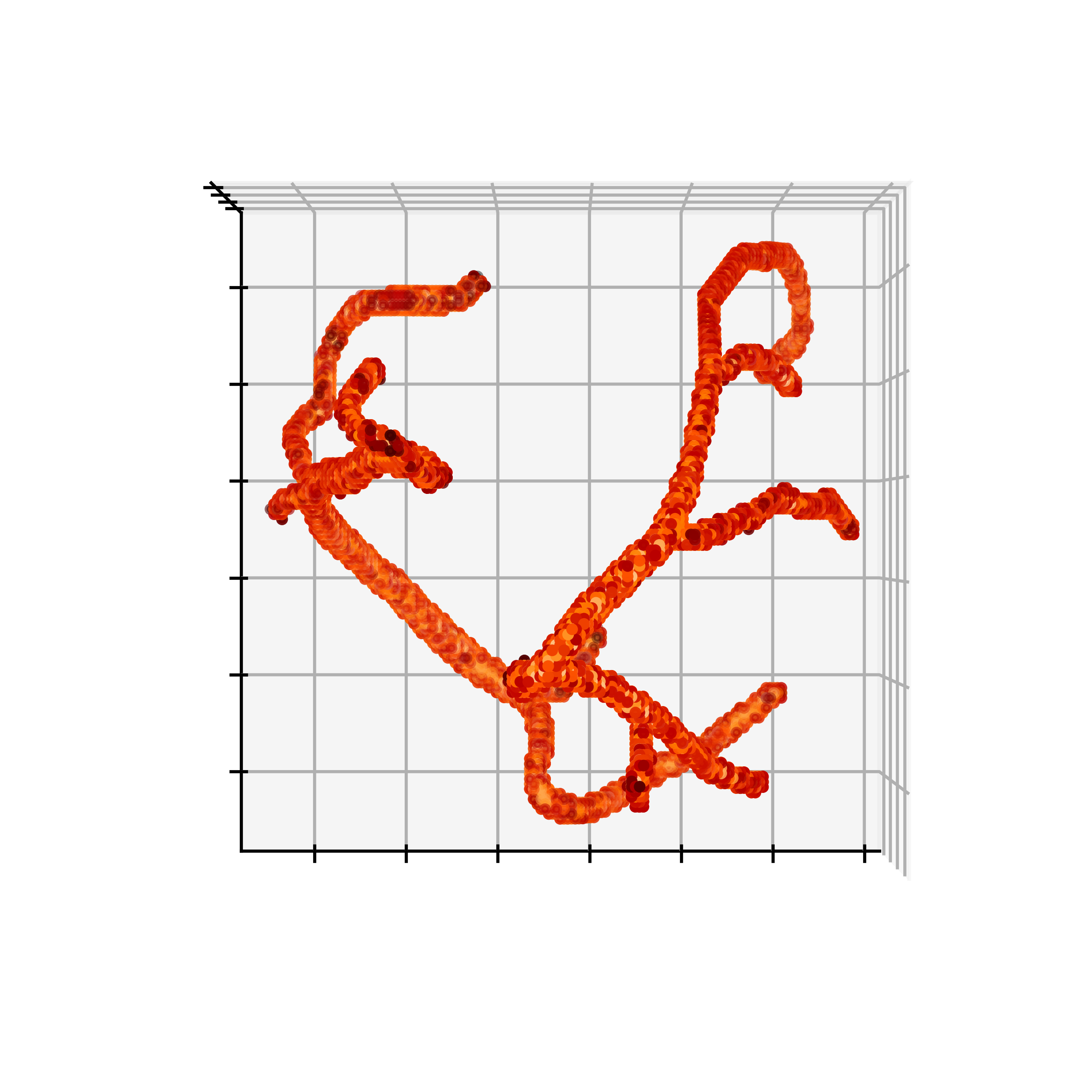}
        \includegraphics[trim={1.3cm 2.5cm 1.3cm 2.8cm},clip, width=0.24\textwidth]{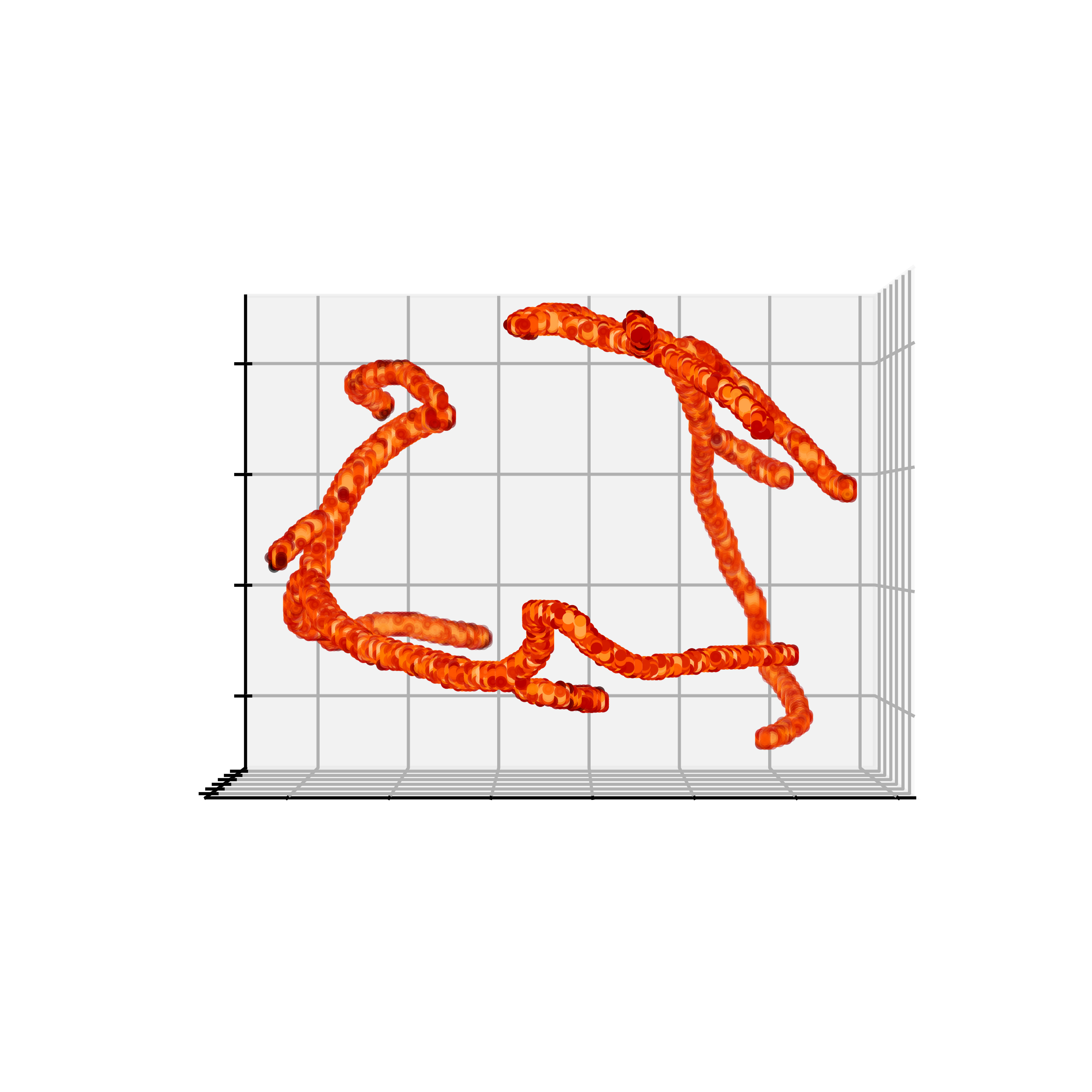}
        \includegraphics[trim={1.3cm 2.5cm 1.3cm 2.5cm},clip, width=0.24\textwidth]{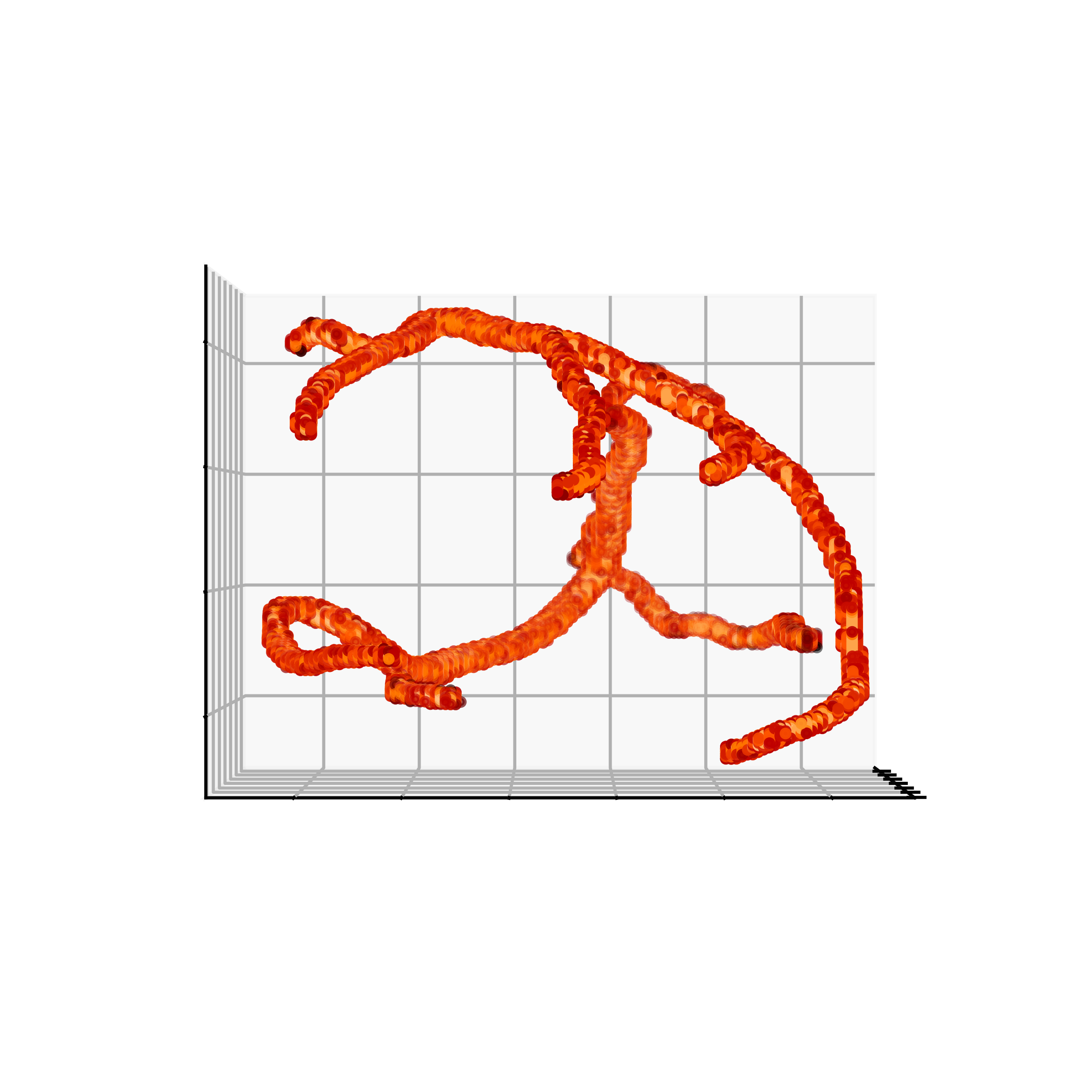} 
    \end{subfigure} 
    \hfill
    \begin{subfigure}[b]{\textwidth}
        \centering
        \includegraphics[trim={0.9cm 0.85cm 0.9cm 1cm },clip, width=0.24\textwidth]{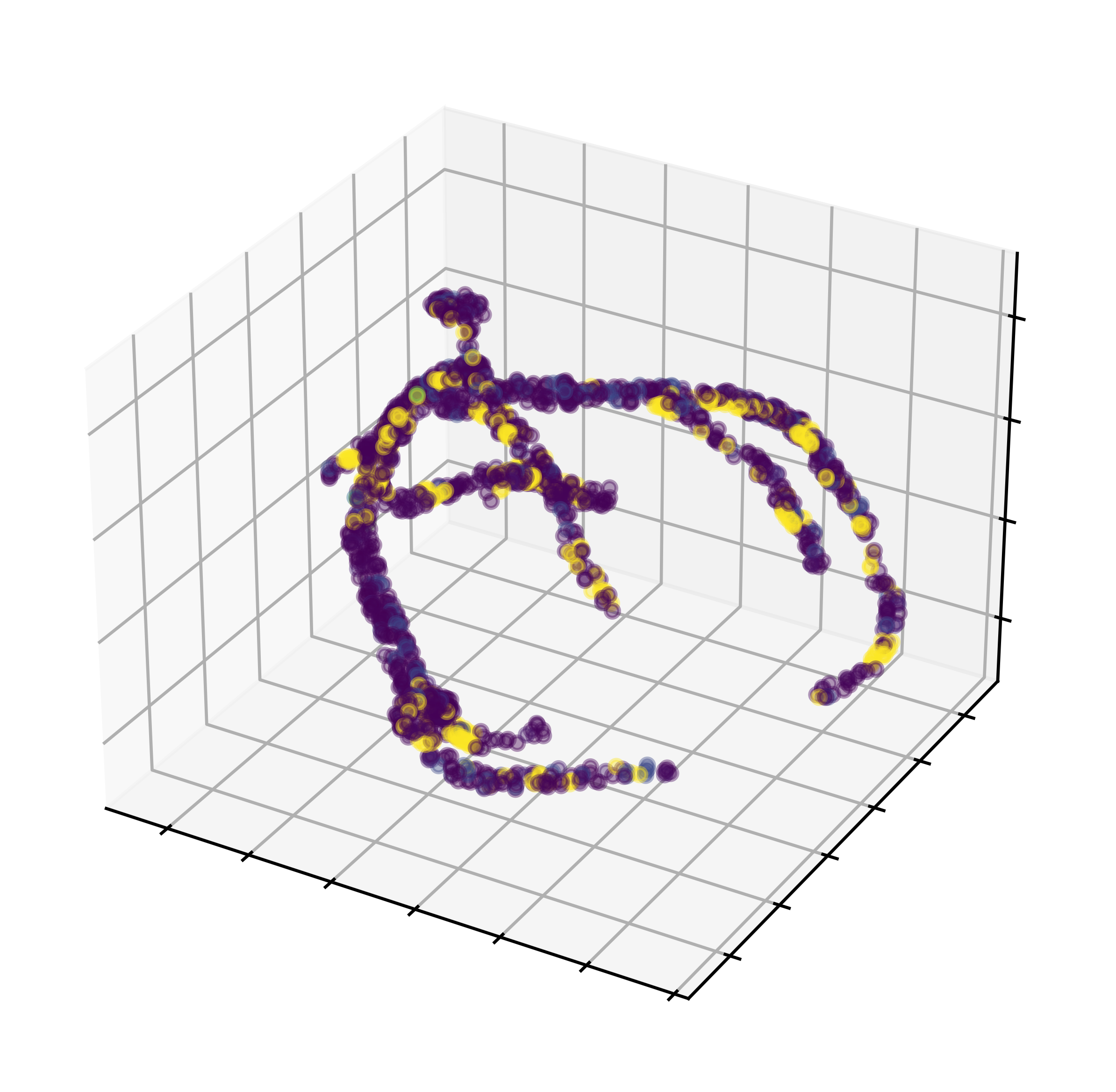}
        \includegraphics[trim={1.3cm 1.7cm 1.3cm 1.6cm},clip, width=0.24\textwidth]{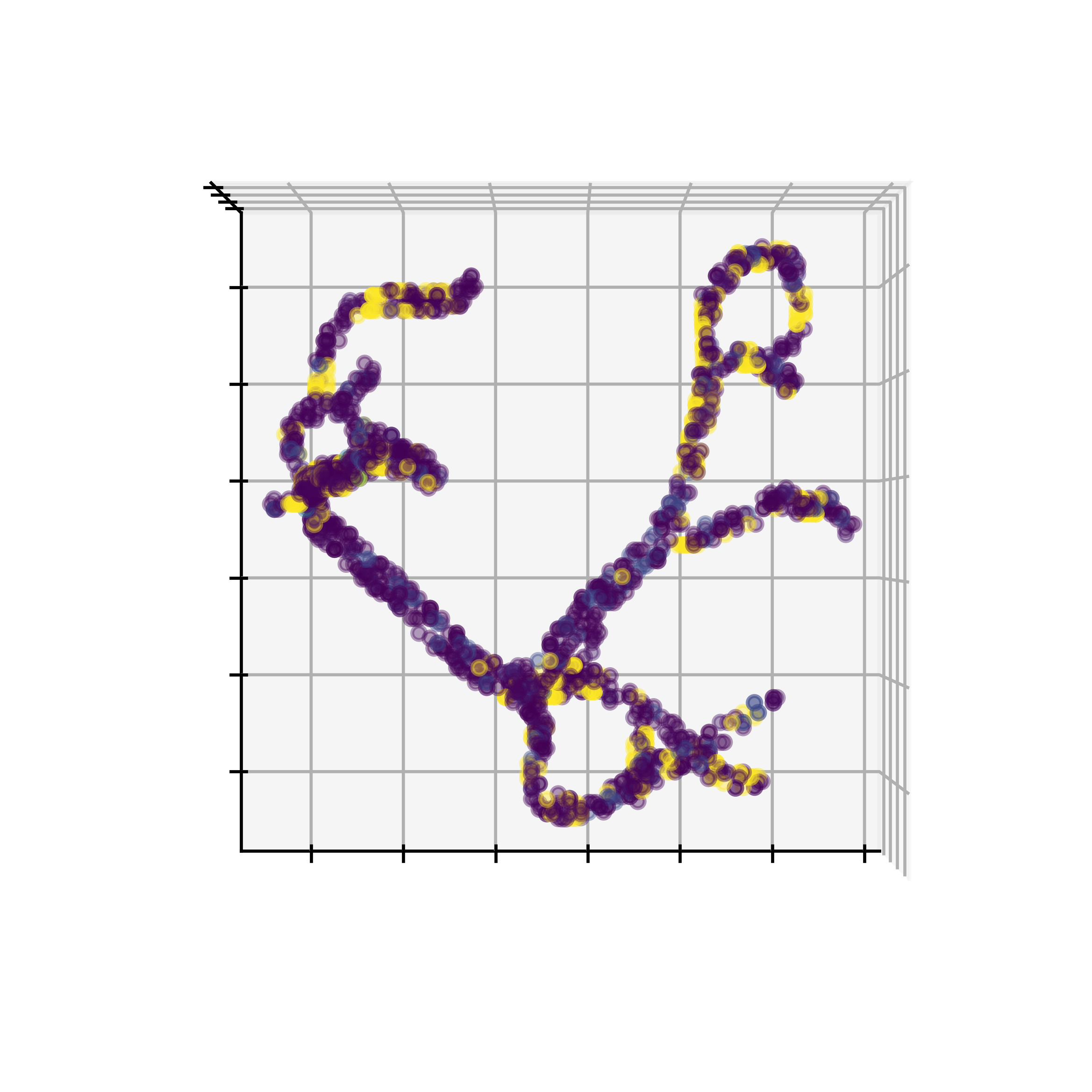}
        \includegraphics[trim={1.3cm 2.5cm 1.3cm 2.8cm},clip, width=0.24\textwidth]{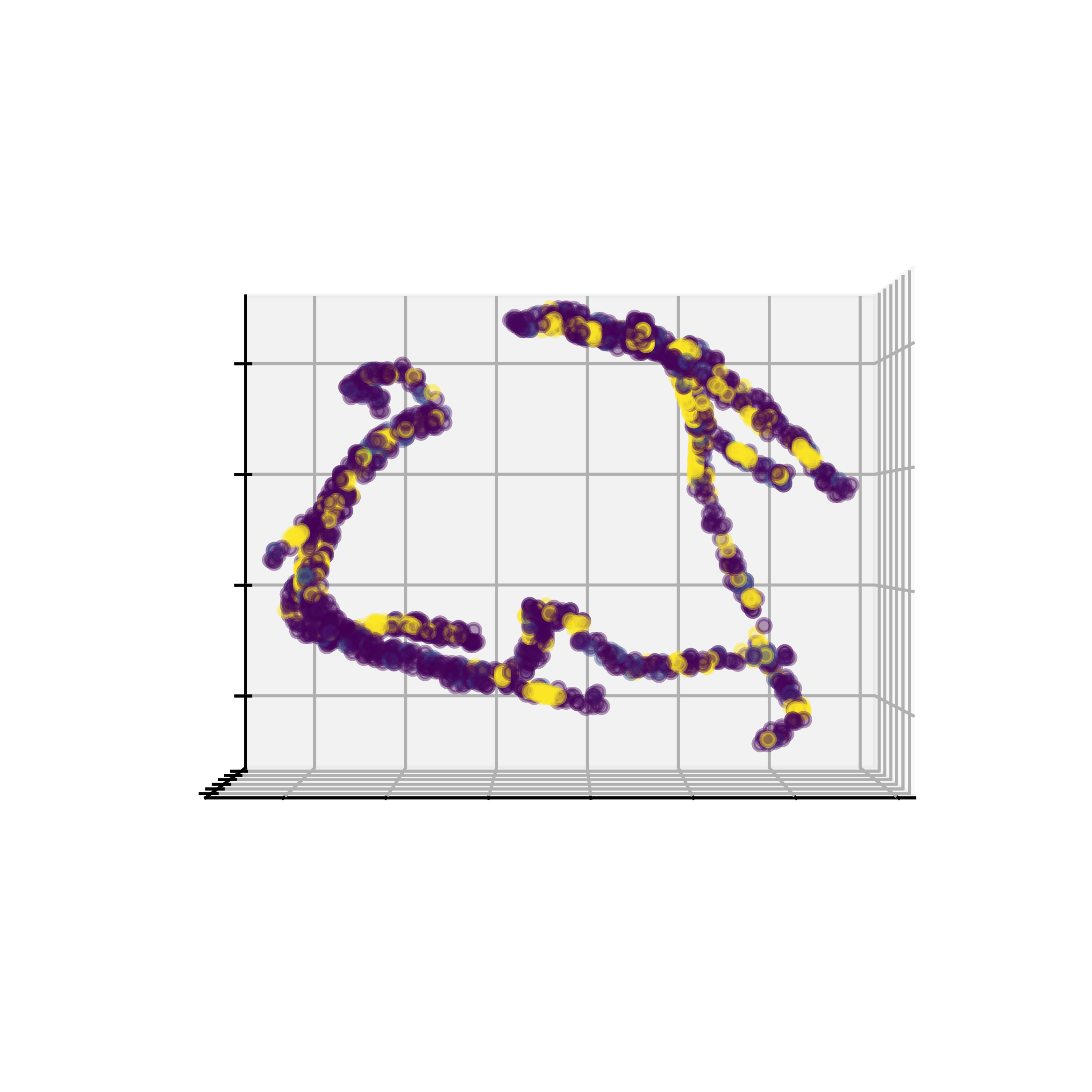} 
        \includegraphics[trim={1.3cm 2.5cm 1.3cm 2.5cm},clip, width=0.24\textwidth]{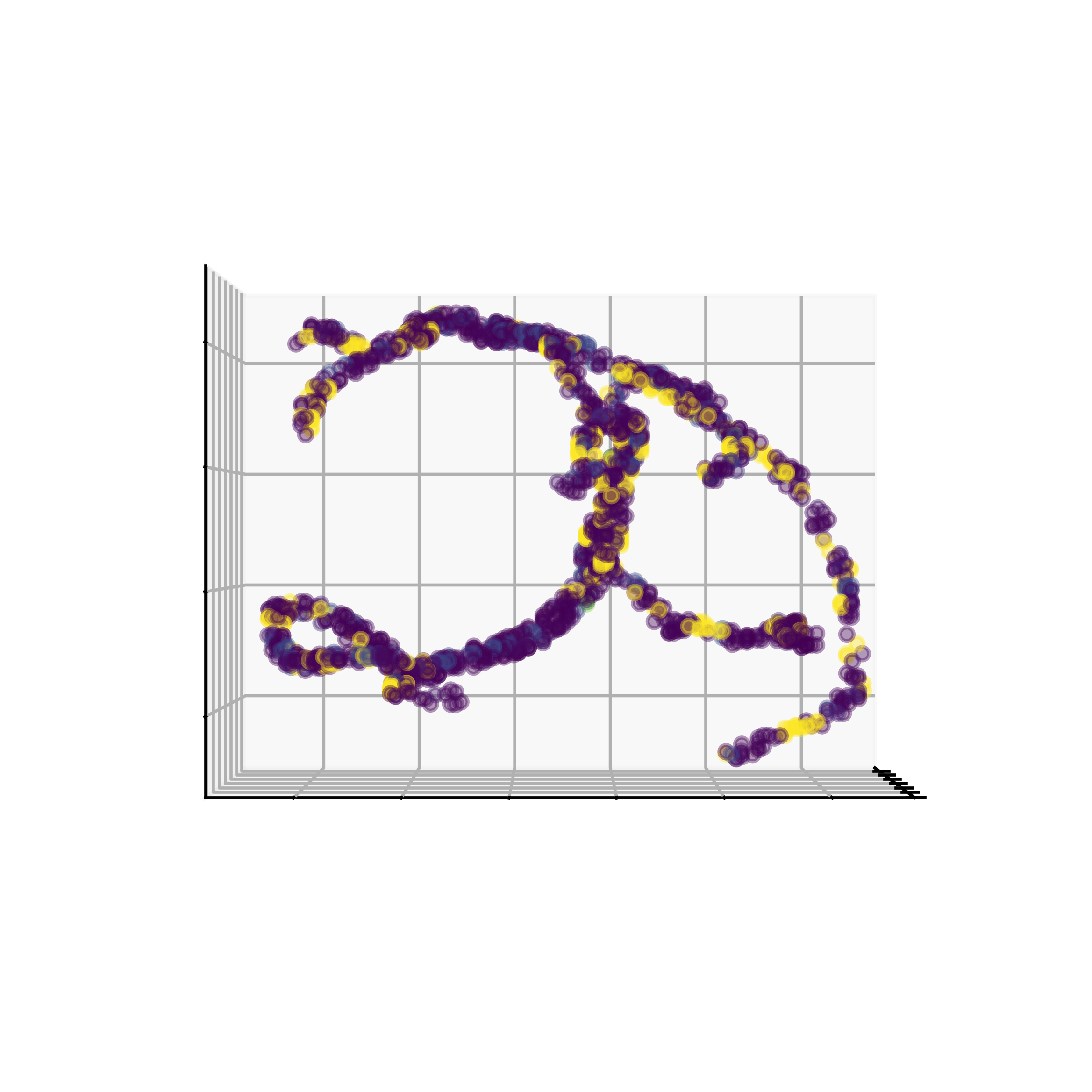} 
    \end{subfigure}
    \hfill
    \begin{subfigure}[b]{0.6\textwidth}
        \centering
        \includegraphics[width=0.7\textwidth]{JMO/figures/colorbar.png} 
    \end{subfigure}
    \caption{First row: The coronary angiography with views from XY, XZ and YZ planes, respectively. Second row: Their TCF results with contrast-enhanced using histogram equalization~\cite{jain1989fundamentals}. Third row: Viridis colormap applied to curvature plots.}
    \label{fig:real_results_3d}
\end{figure}

\paragraph{Computed Tomography Angiography}
\label{sec:real_results_3d_cta}
We calculated the curvature of a segmented coronary artery image from a three-dimensional computed tomography angiography (CTA) scan, downsampling the image by a factor of 2. Fig.~\ref{fig:real_results_3d} shows the interpolated image on the first row, and the curvature results with TCF on the second row. The second, third, and fourth columns display the views in the XY, XZ, and YZ planes, respectively.
Viridis colormap is used for curvature results. Similar to Section~\ref{sec:real_images_angiography}, high curvature values at the branching points and twisted vessels are observed in this example, as expected. 
\paragraph{Coronary Artery}
The Coronary Artery Segmentations dataset~\cite{dalvit2023automated} includes both segmented coronary arteries and their reference centerlines. We selected a sample coronary vessel segment from the dataset, downsampled it by a factor of 2, and then applied interpolation to calculate the TCF.

\begin{figure}[t]
    \centering
    \begin{subfigure}[b]{0.99\textwidth}
        \centering
        \includegraphics[trim={0.9cm 0.75cm 0.9cm 0.9cm },clip, width=0.24\textwidth]{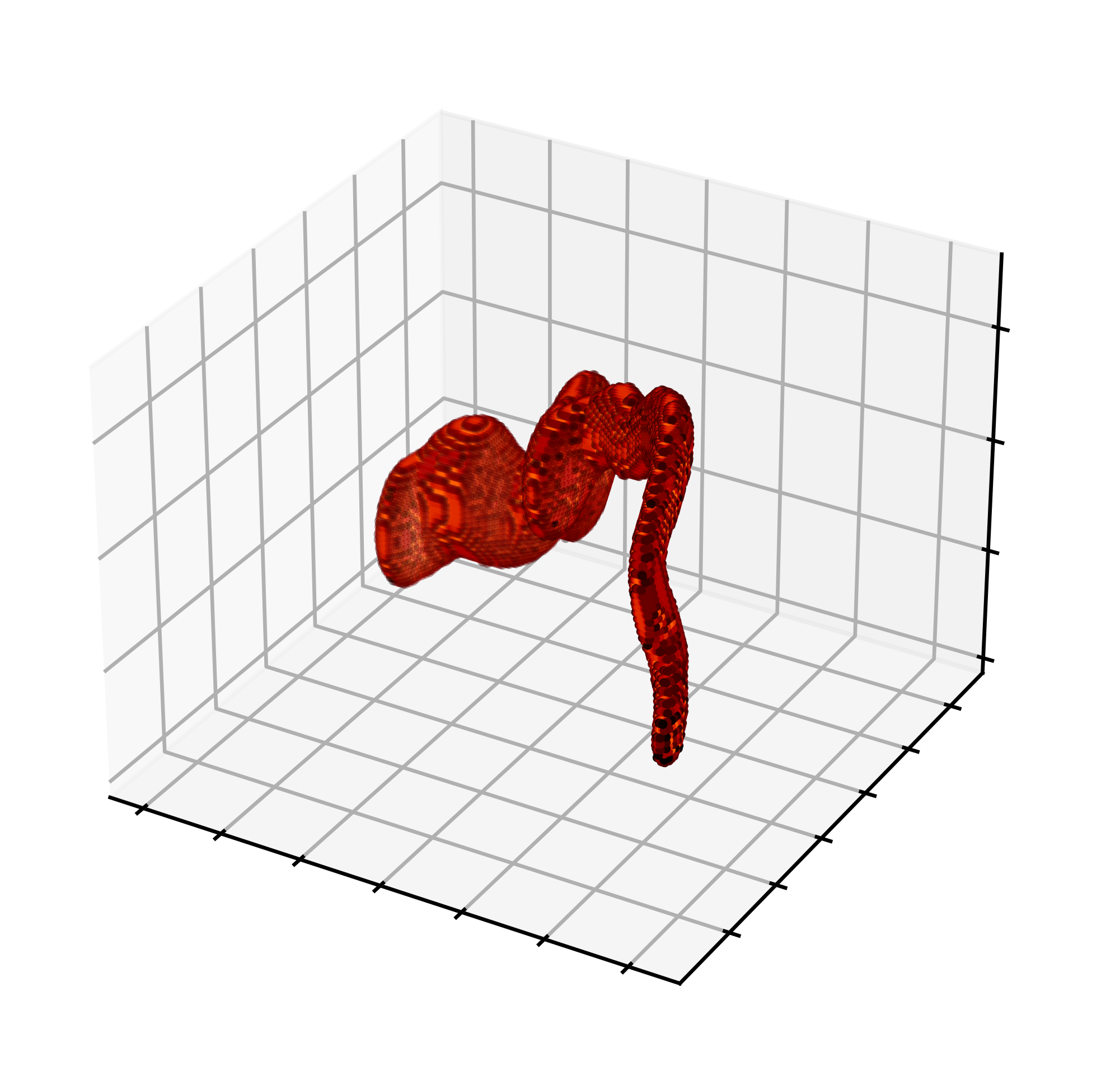}
        \includegraphics[trim={1.3cm 1.3cm 1.3cm 1.3cm },clip, width=0.24\textwidth]{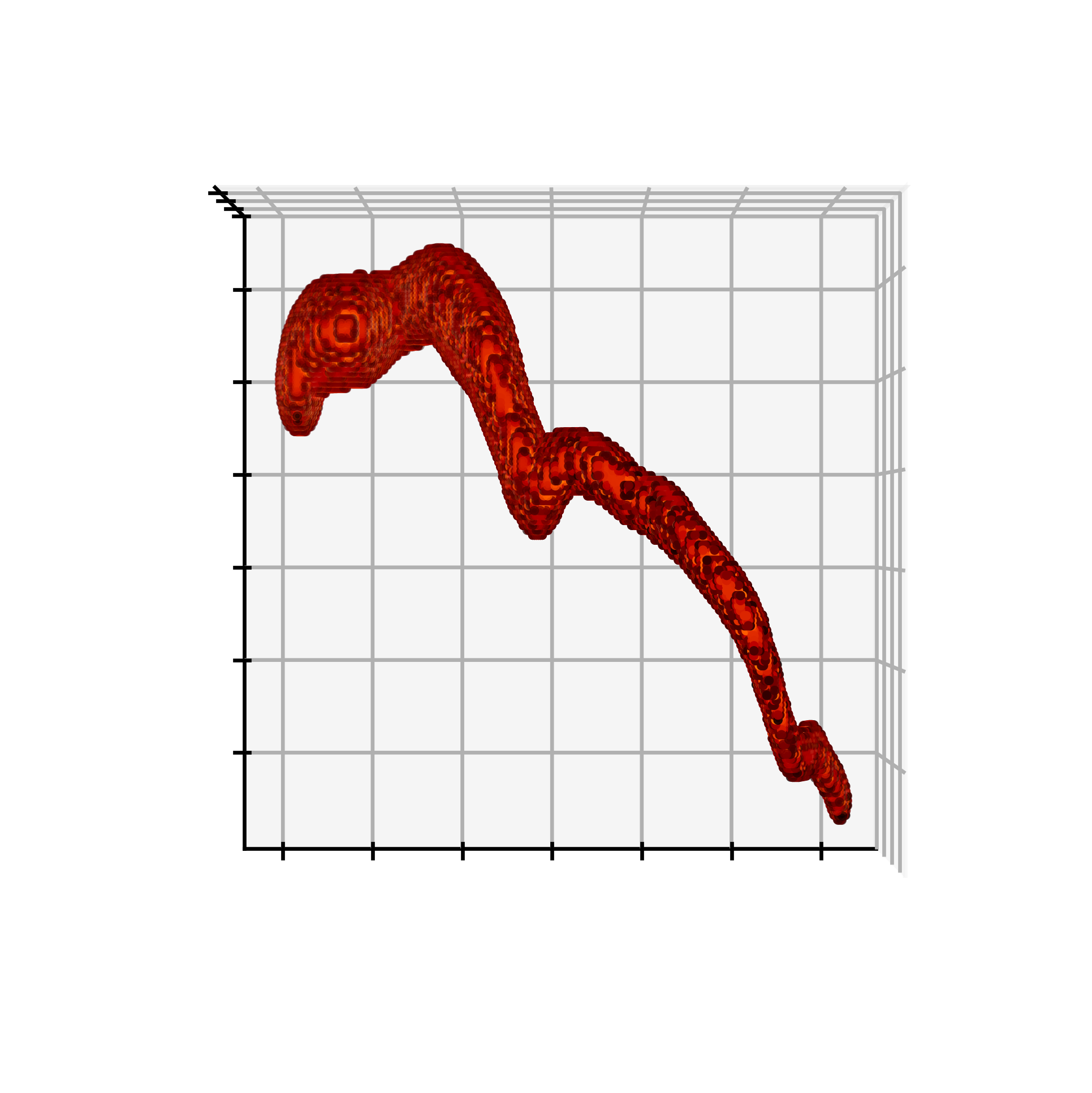}
        \includegraphics[trim={1.3cm 1.7cm 1.3cm 2cm},clip, width=0.24\textwidth]{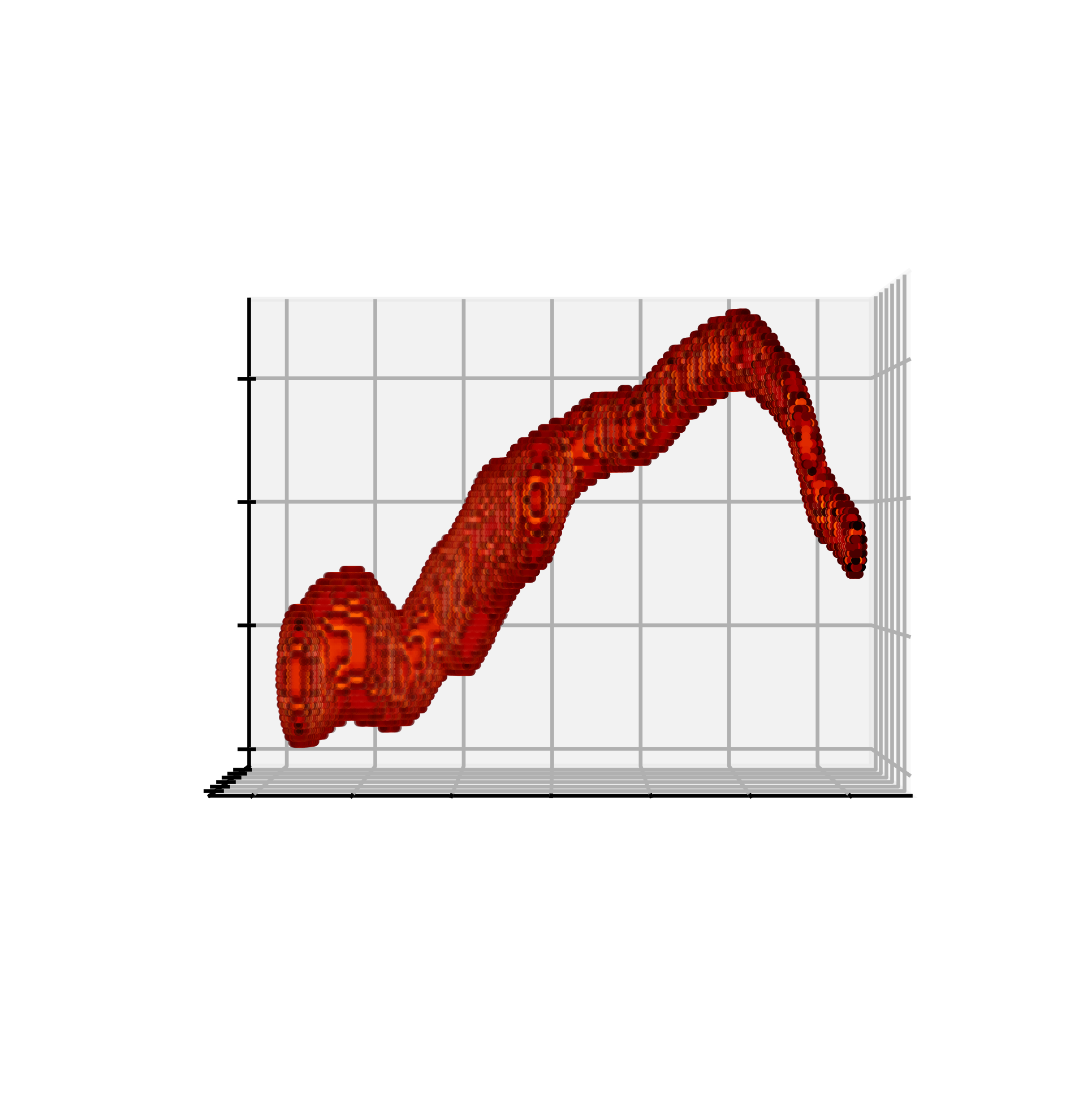}
        \includegraphics[trim={1.3cm 1.7cm 1.3cm 2cm},clip, width=0.24\textwidth]{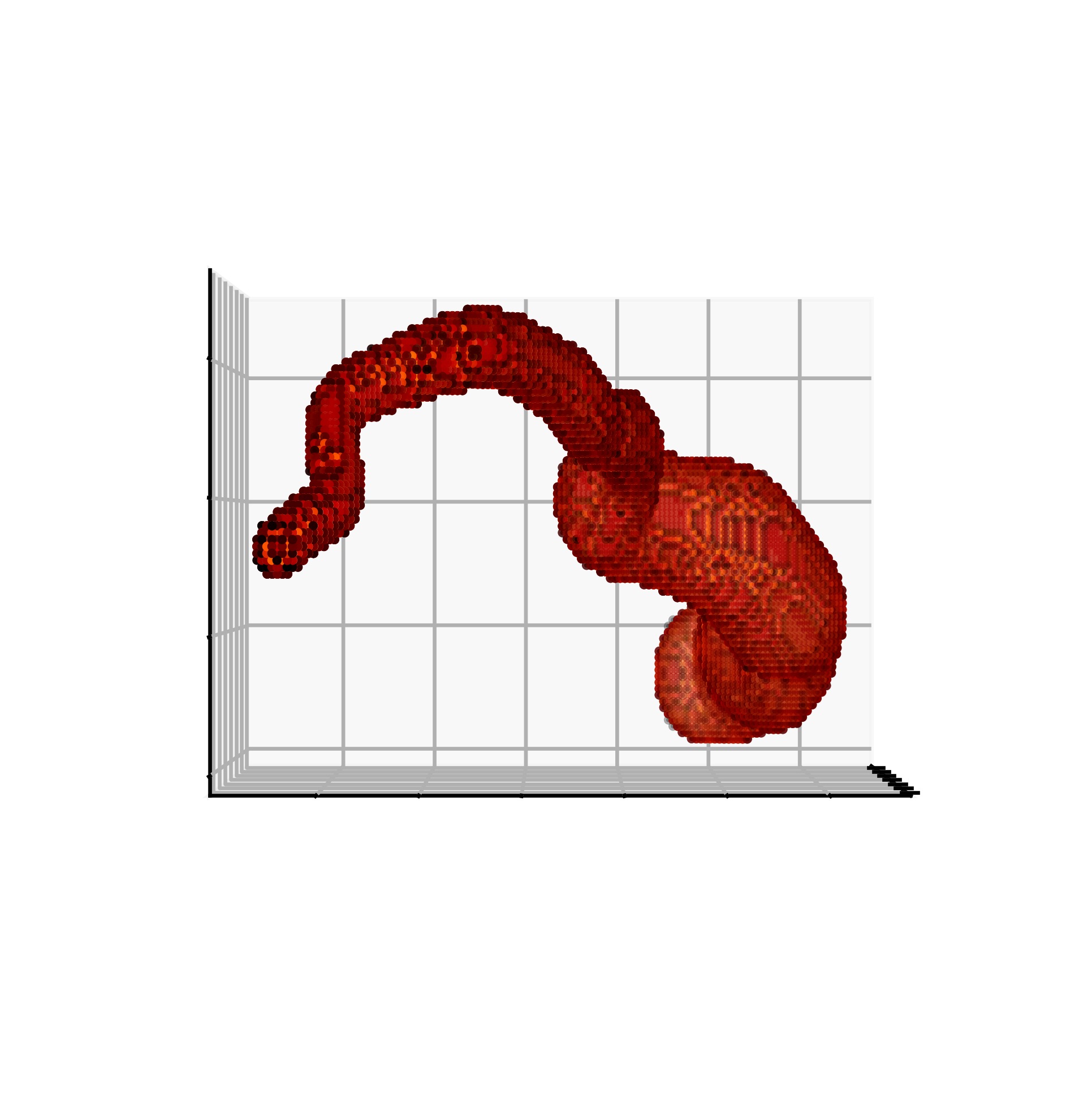} 
    \end{subfigure} 
    \hfill
    \begin{subfigure}[b]{0.99\textwidth}
        \centering
        \includegraphics[trim={0.9cm 0.75cm 0.9cm 0.9cm },clip, width=0.24\textwidth]{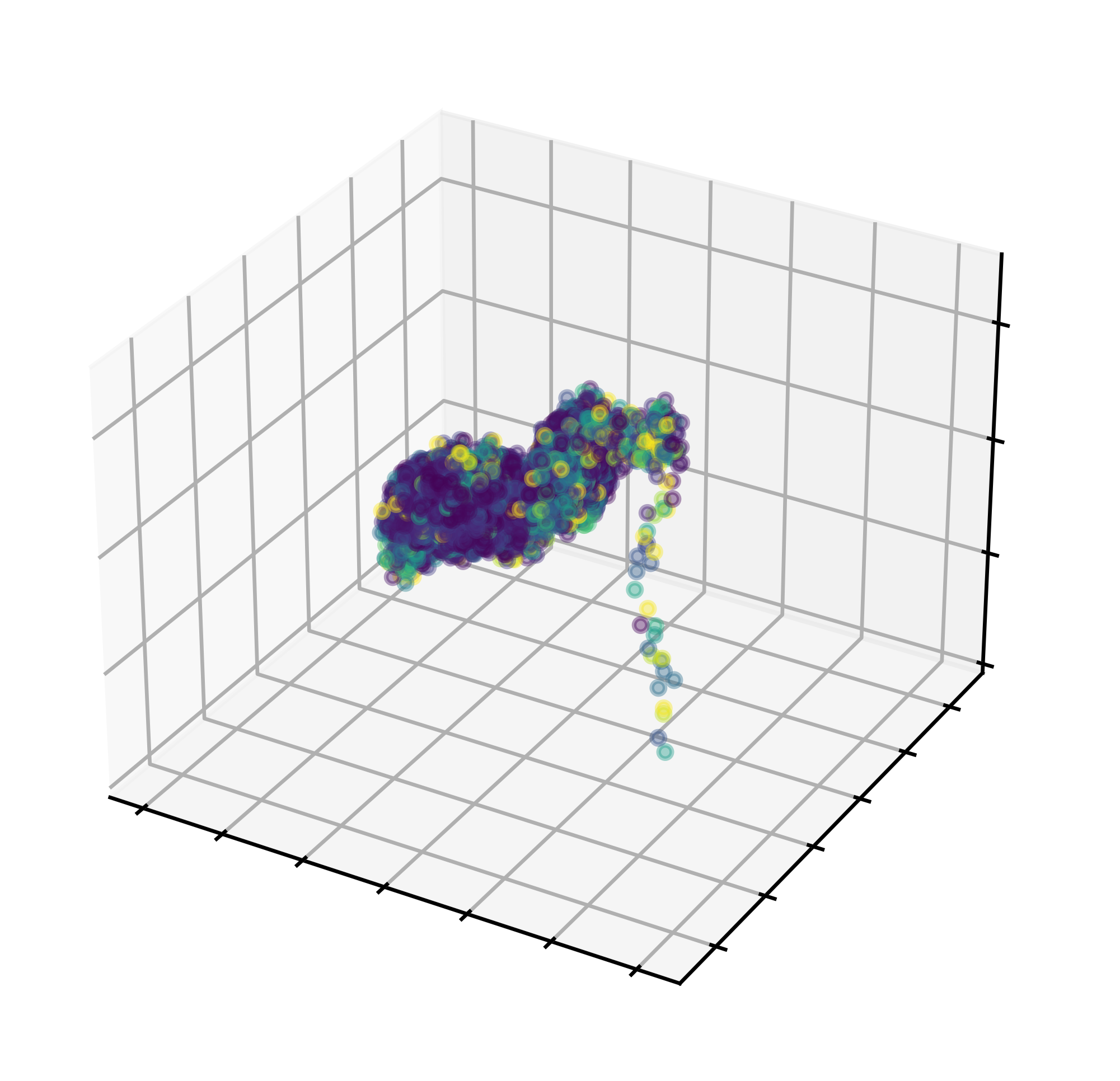}
        \includegraphics[trim={1.3cm 1.3cm 1.3cm 1.3cm },clip, width=0.24\textwidth]{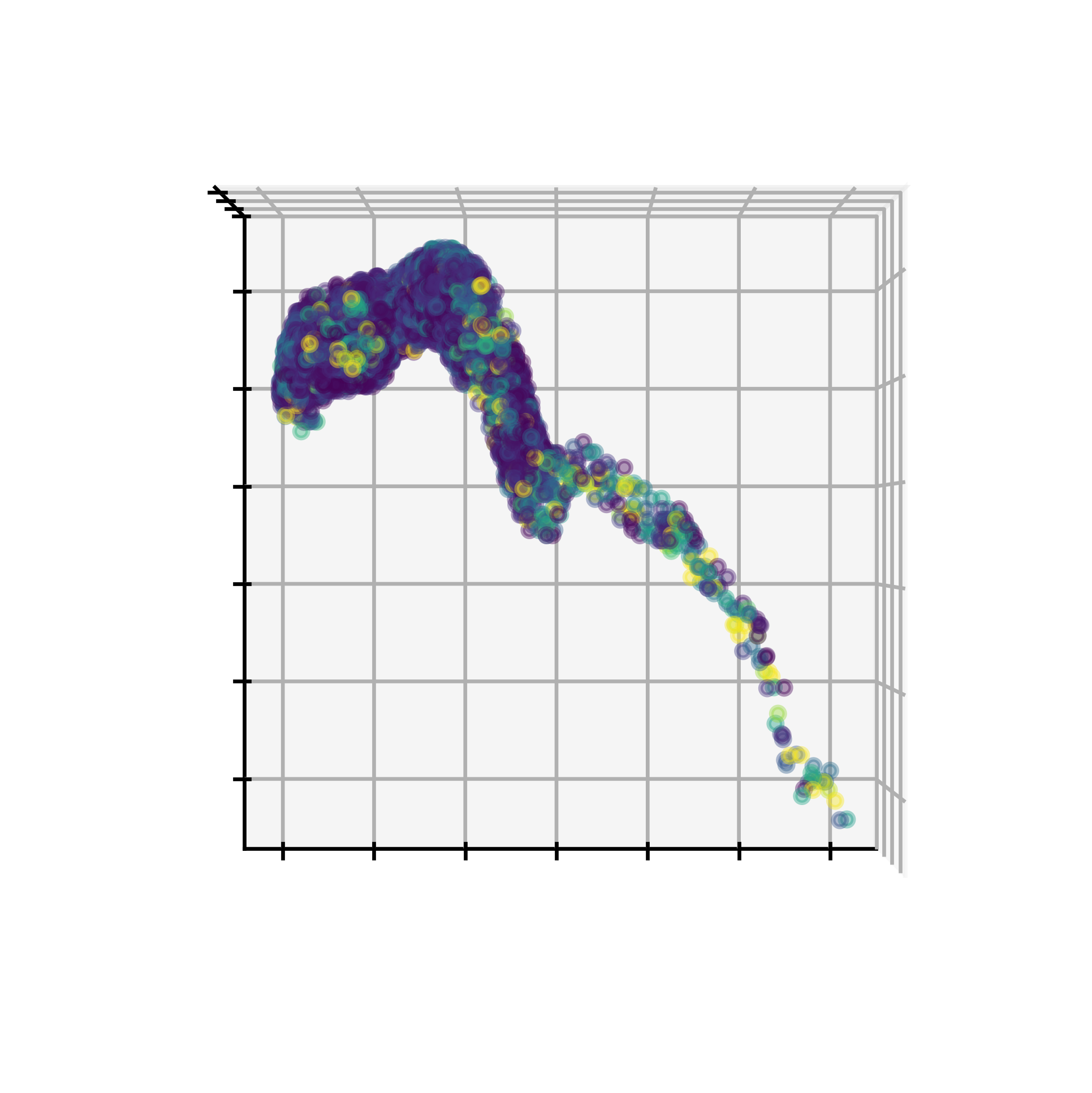}
        \includegraphics[trim={1.5cm 1.7cm 1.5cm 2cm},clip, width=0.24\textwidth]{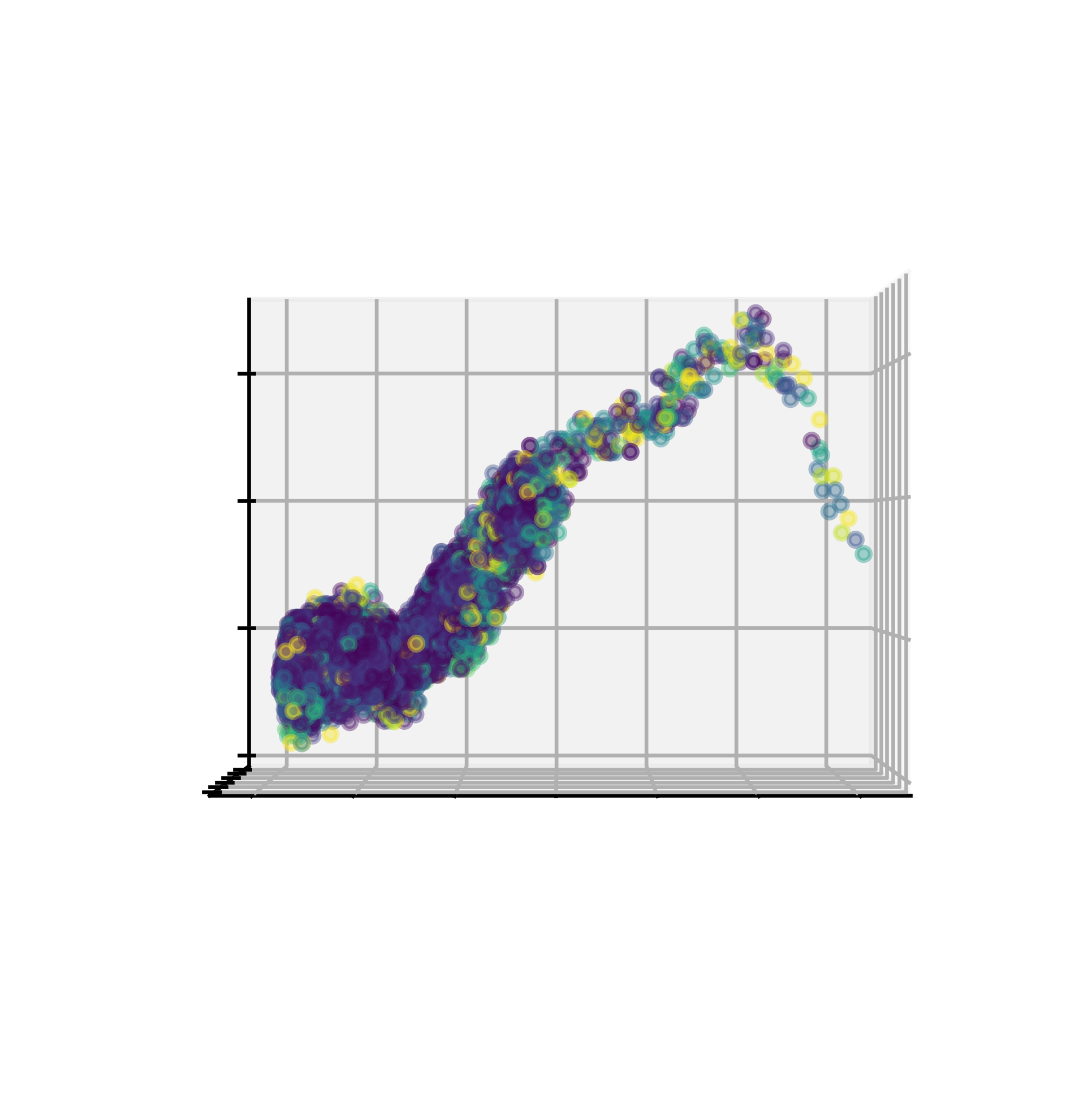} 
        \includegraphics[trim={1.3cm 1.7cm 1.3cm 2cm},clip, width=0.24\textwidth]{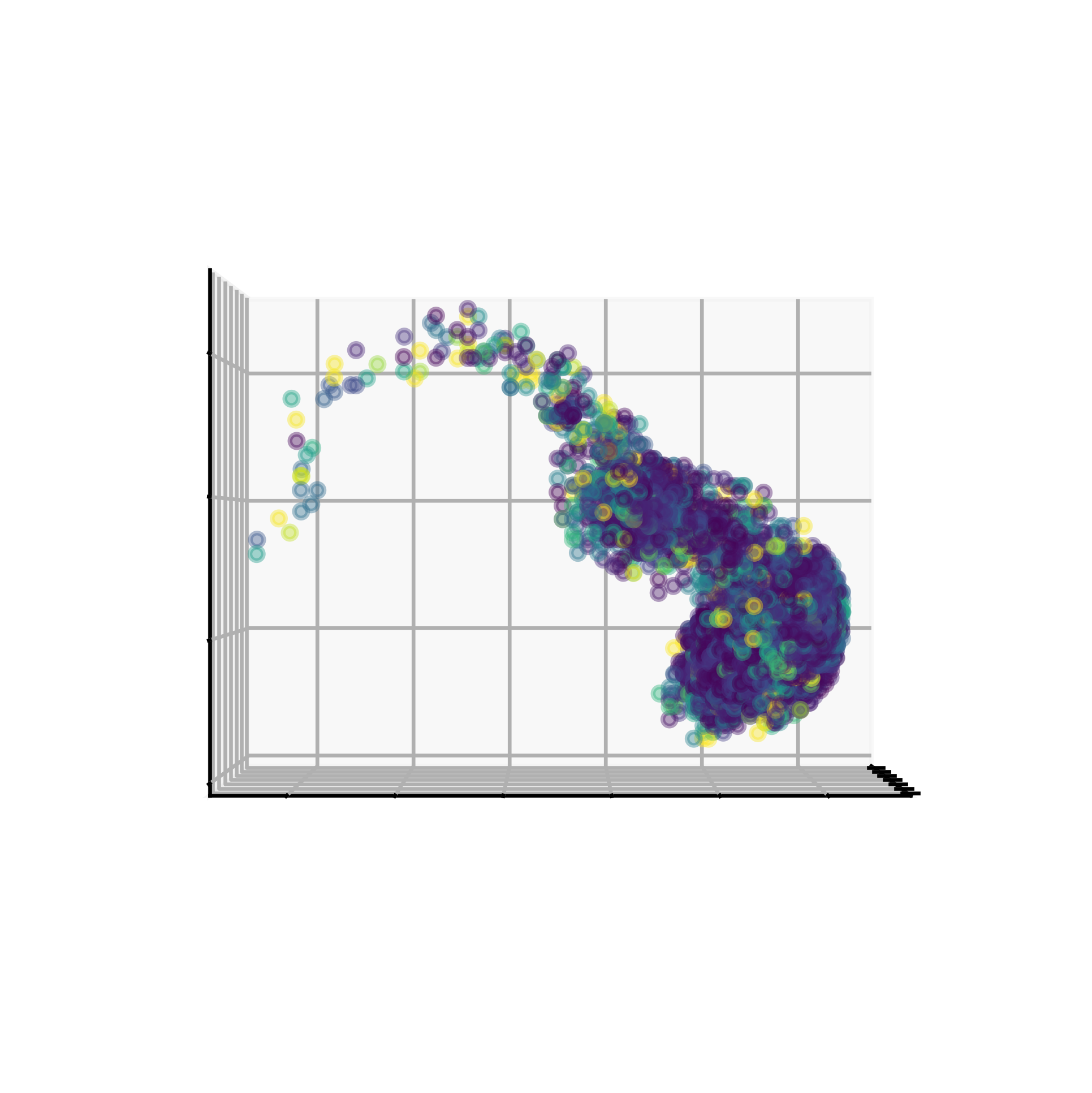} 
    \end{subfigure}
    \hfill
    \begin{subfigure}[b]{0.99\textwidth}
        \centering        \includegraphics[trim={0.9cm 0.75cm 0.9cm 0.9cm },clip, width=0.24\textwidth]{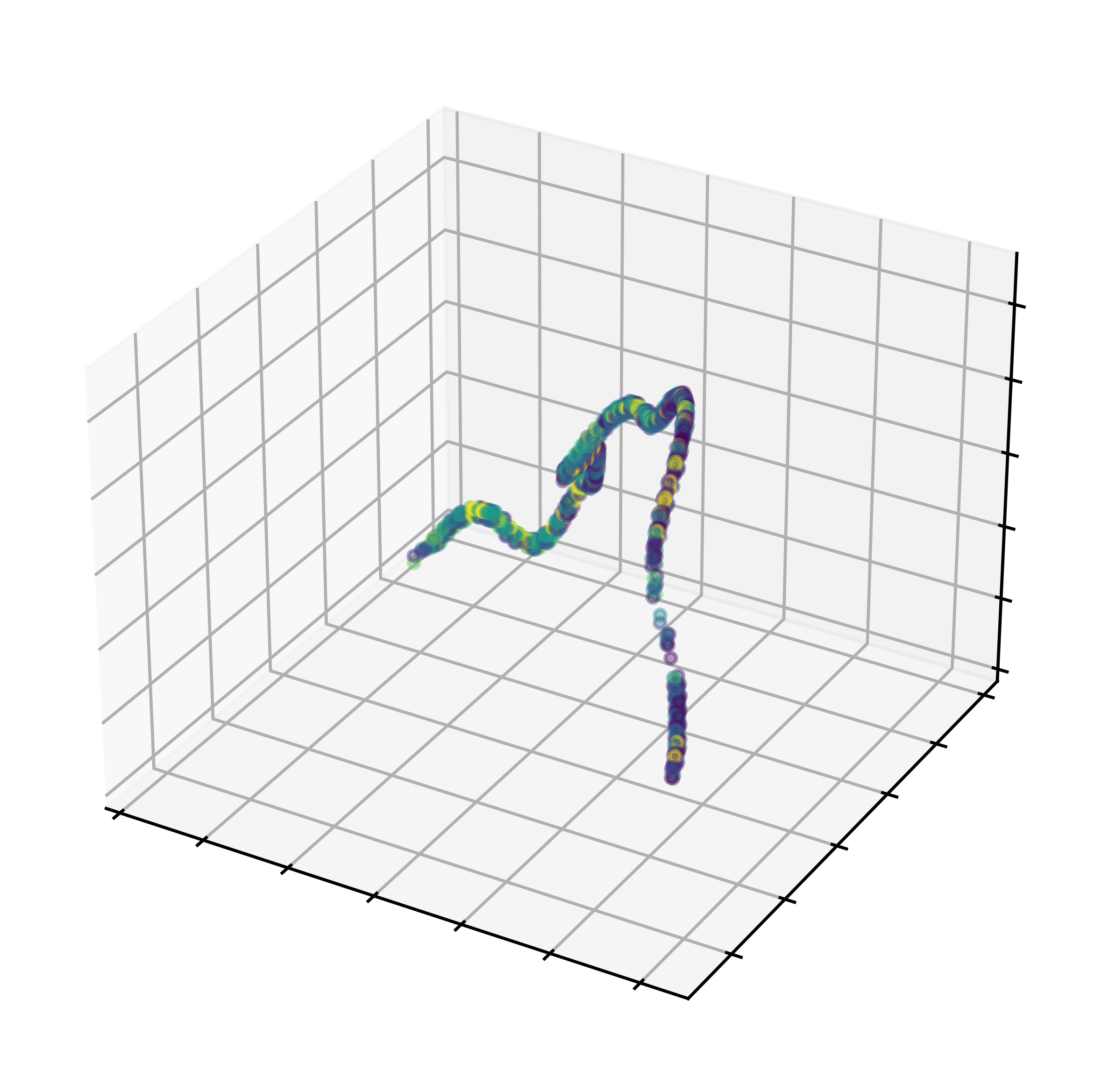}
        \includegraphics[trim={1.3cm 1.3cm 1.3cm 1.3cm },clip, width=0.24\textwidth]{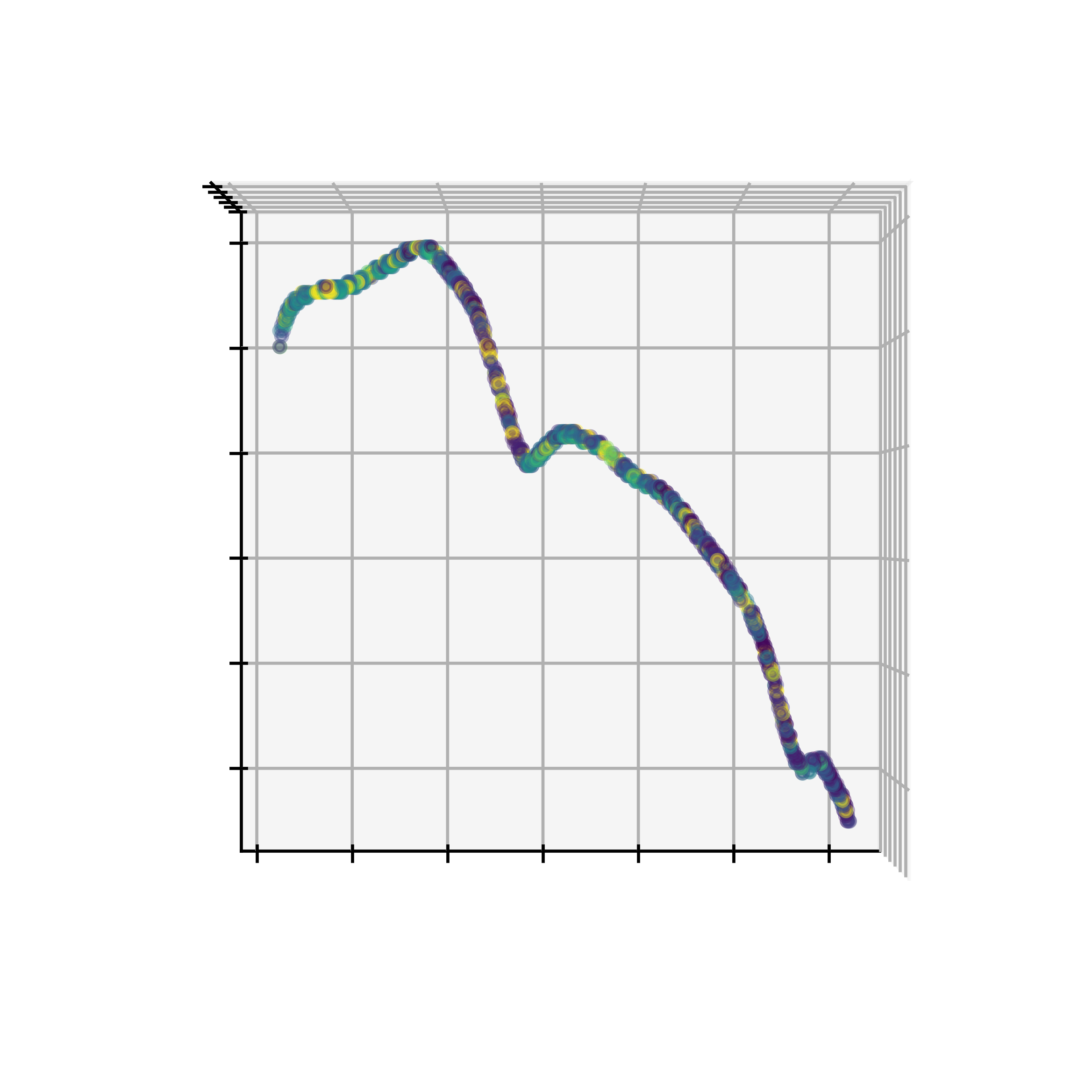}
        \includegraphics[trim={1.3cm 1.7cm 1.3cm 2cm},clip, width=0.24\textwidth]{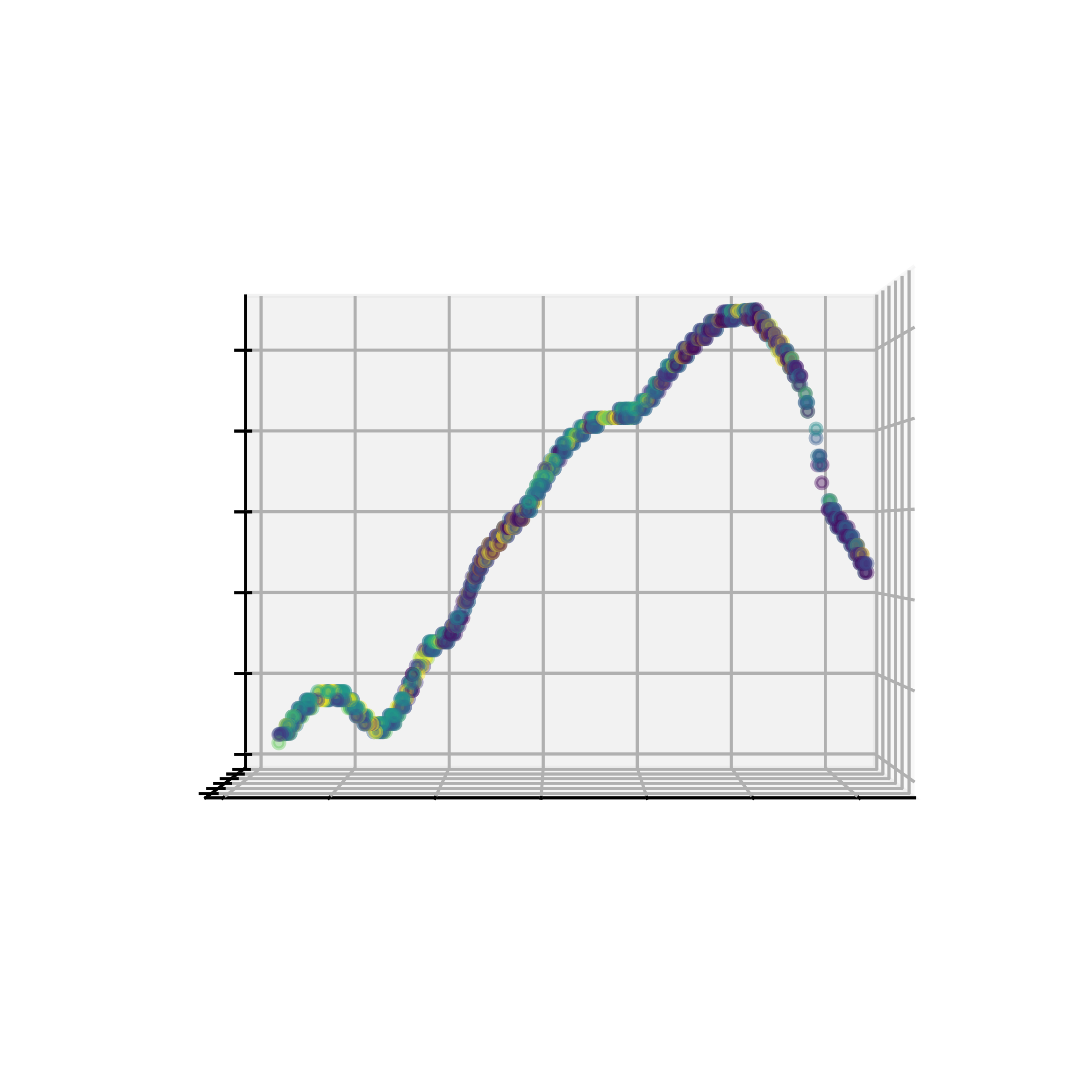}
        \includegraphics[trim={1.3cm 1.7cm 1.3cm 2cm},clip, width=0.24\textwidth]{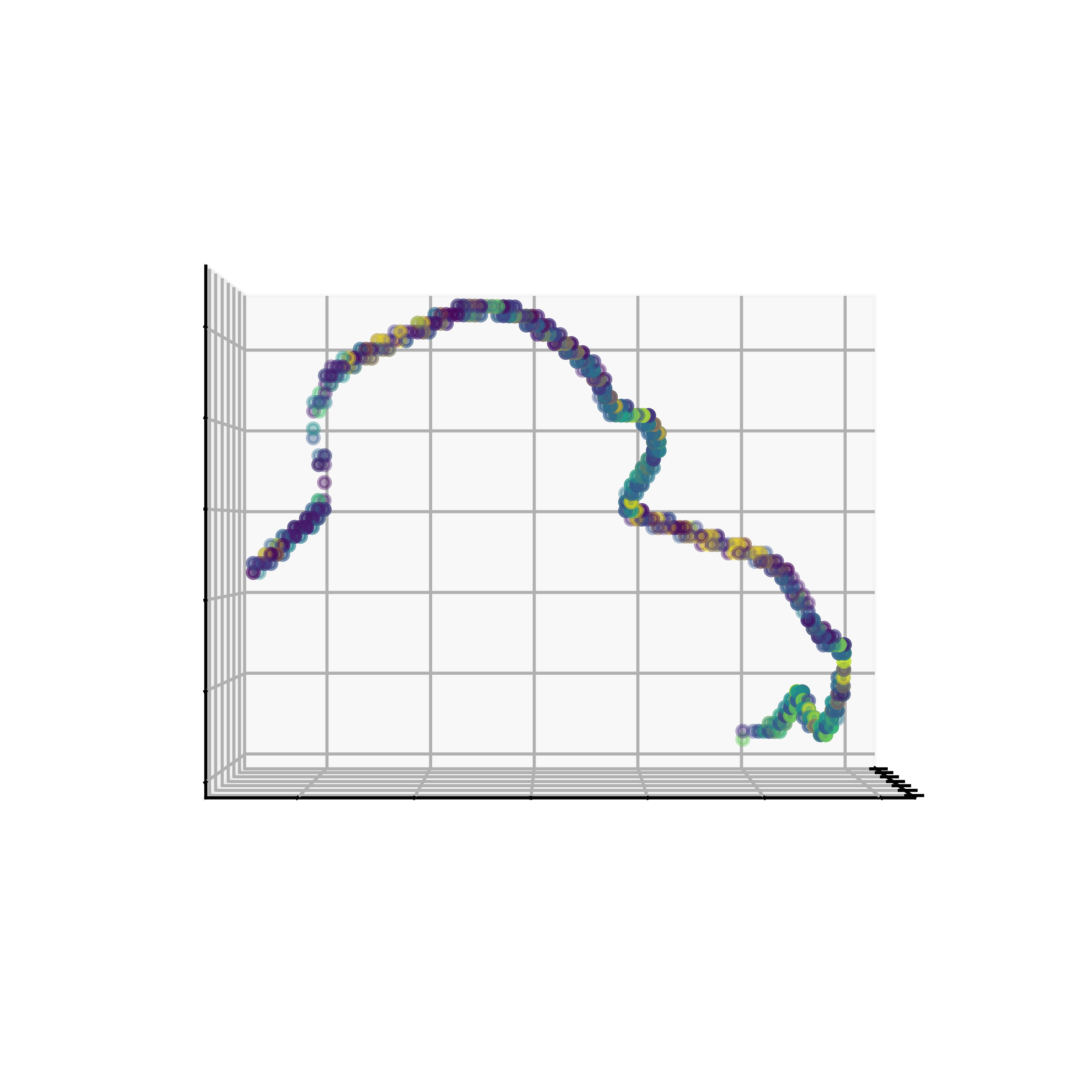} 
    \end{subfigure}     
    \begin{subfigure}[b]{0.6\textwidth}
        \centering
        \includegraphics[width=0.7\textwidth]{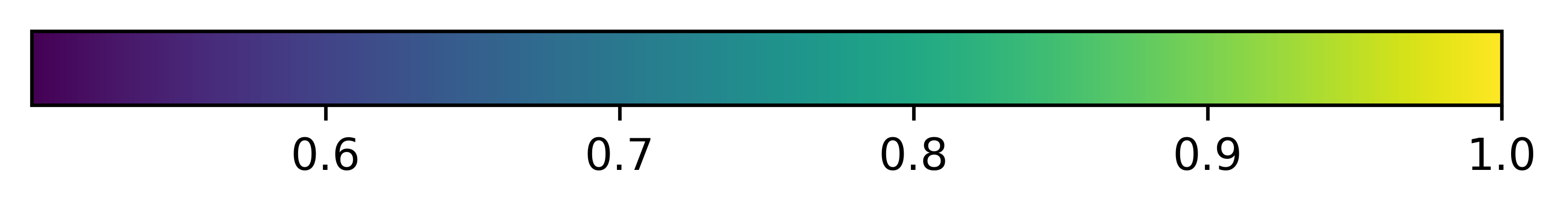} 
    \end{subfigure}
    \caption{First row: The coronary artery segment with views from XY, XZ and YZ planes, respectively. Second row: Their TCF results. Third row: The centerline curvatures using the implicit planar curvature method~\cite{GOLDMAN2005632}. Curvature contrast is enhanced with histogram equalization\cite{jain1989fundamentals} with the viridis colormap applied to curvature plots.}
    \label{fig:real_results_3d_coronary}
\end{figure}

Fig.~\ref{fig:real_results_3d_coronary} shows the interpolated vessel segment on the first row, the TCF results on the second row and the centerline curvatures computed using the implicit planar curvature method~\cite{GOLDMAN2005632} on the third row. The second, third, and fourth columns show views from the XY, XZ, and YZ planes. Curvature is visualized using the Viridis colormap. This dataset includes arteries with varying vessel widths, which centerlines alone cannot capture. For example, in Fig.~\ref{fig:real_results_3d_coronary}, high curvature values are expected in the bottom right of third row last column view from the centerline results. However, when we examine the image itself on the first row, this region appears to be an artery segment with a large width and no sharp curvature changes. TCF results show that this region has low curvature values, which the centerline cannot detect. Similar to Section~\ref{sec:real_results_3d_cta}, high curvature values are detected at the twisted vessel (in the bottom-right region of the second column of Fig.~\ref{fig:real_results_3d_coronary}), as expected.

\subsubsection{Additional Analyses}
\begin{figure}[t]
    \centering
    \begin{subfigure}[b]{0.49\textwidth}
        \centering
        \includegraphics[trim={0.95cm 0.75cm 0.9cm 0.9cm },clip,width=0.48\textwidth]{JMO/figures/circle3d_intp.png}
        \includegraphics[trim={0.95cm 0.75cm 0.9cm 0.9cm },clip,width=0.48\textwidth]{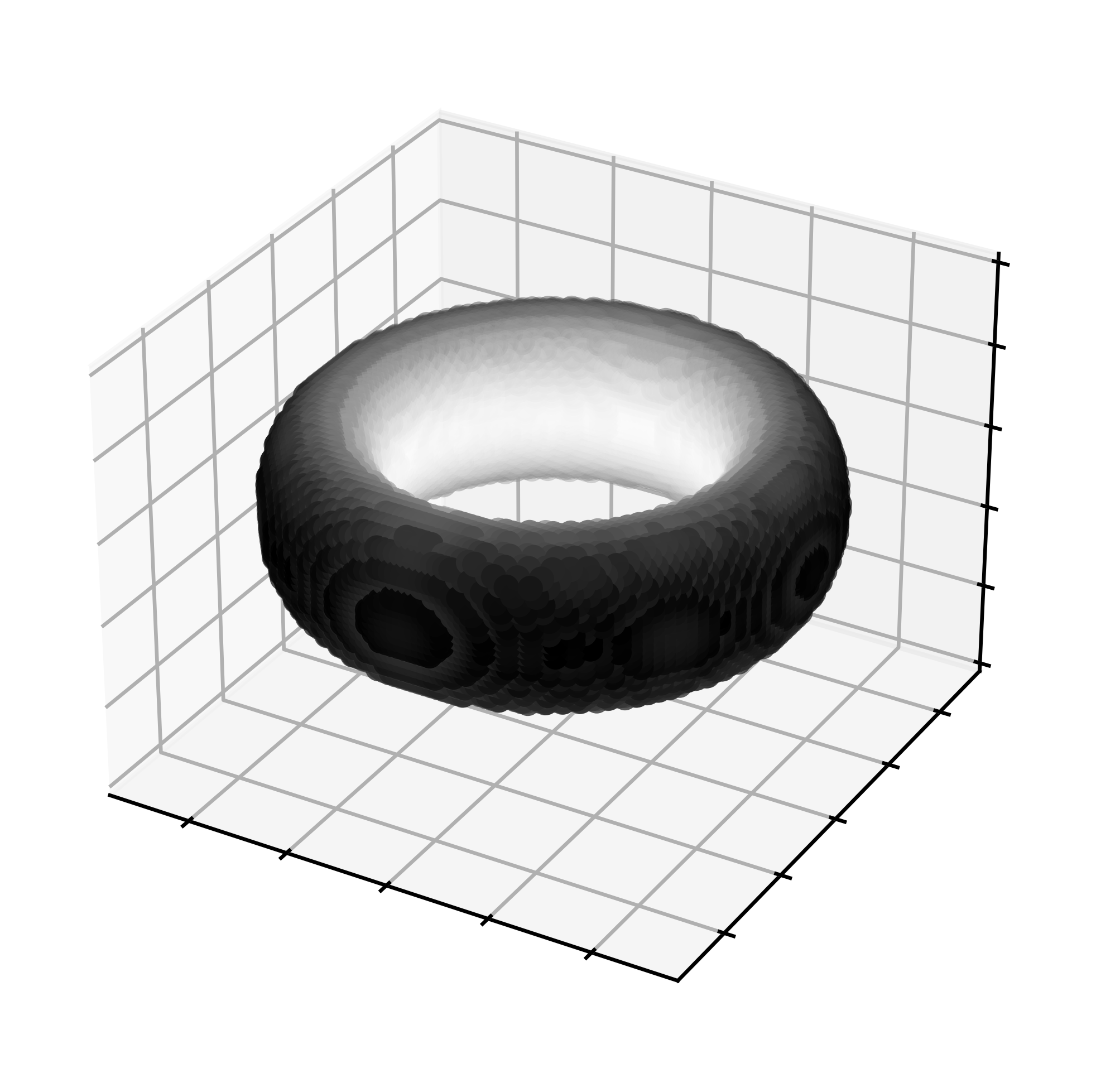}
    \end{subfigure} 
    \begin{subfigure}[b]{0.49\textwidth}
        \centering
        \includegraphics[trim={0.95cm 0.75cm 0.9cm 0.9cm },clip,width=0.48\textwidth]{JMO/figures/circle3d_curv.png}
        \includegraphics[trim={0.95cm 0.75cm 0.9cm 0.9cm },clip,width=0.48\textwidth]{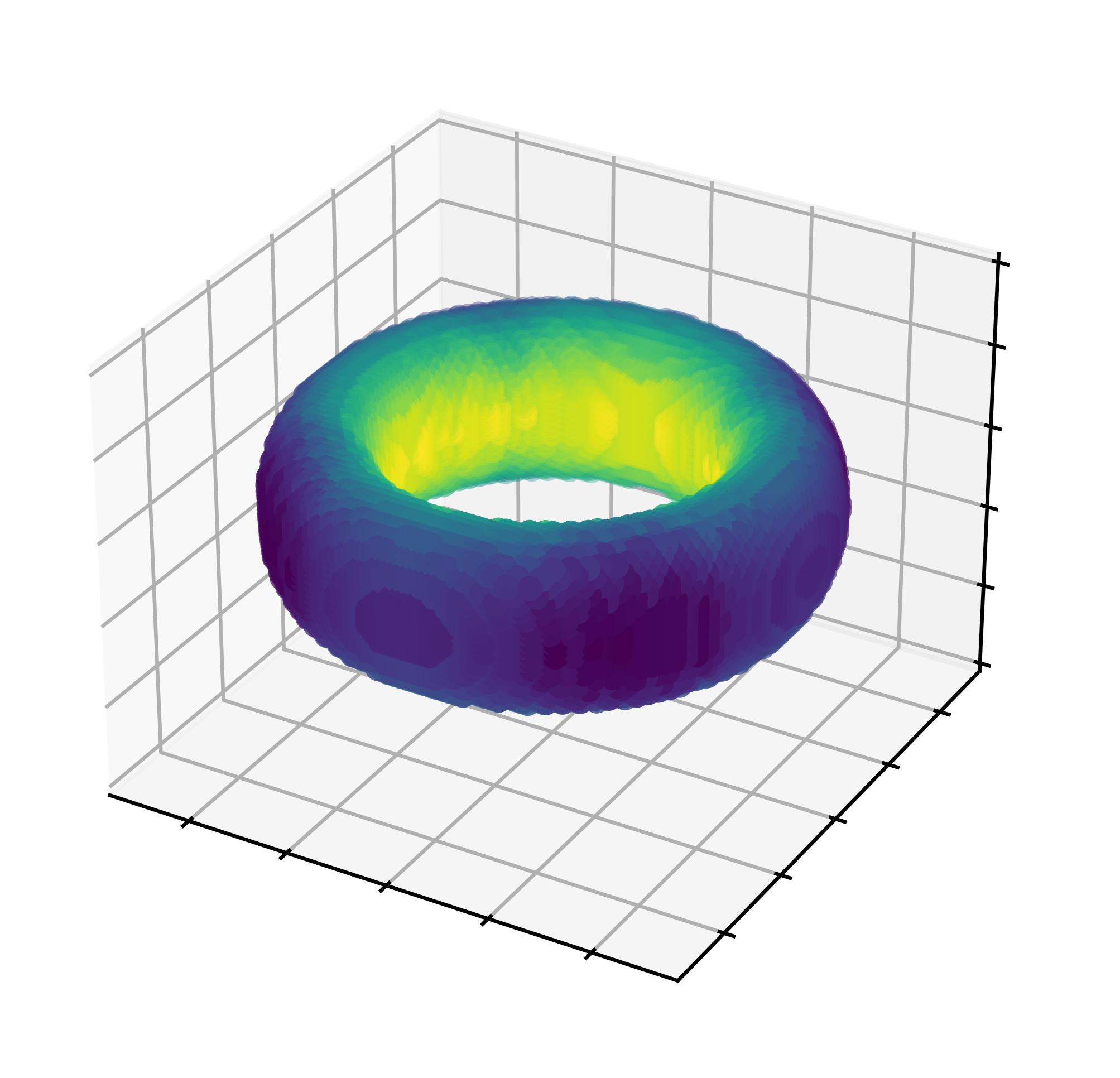}
    \end{subfigure}
    \begin{subfigure}[b]{0.6\textwidth}
        \centering
        \includegraphics[width=0.7\textwidth]{JMO/figures/colorbar.png} 
    \end{subfigure}
    \caption{On the left: The three-dimensional images of the blurry-ring function with isotropic and anisotropic voxels, respectively. On the right: their TCF results. The Viridis colormap at the bottom is used for curvature plots. The mean absolute difference between the ground truth and TCF is $2.493\times 10^{-3}$ for the isotropic and $2.495 \times 10^{-3}$ for the anisotropic case.
}
    \label{fig:voxel_anisotropy}
\end{figure}

We analyze TCF on images with isotropic and anisotropic voxel sizes. The ground-truth curvature is calculated as $1/r$, where $r$ is the distance of the points from the ring's center. We compute the mean absolute difference between the ground-truth results and TCF.
\paragraph{Voxel Anisotropy}
We use $p_{\text{blurry\_ringring\_3d}}(\pointx)$ to show how voxel anisotropy affects TCF results. For the isotropic case, the voxel sizes are $0.6 \times 0.6 \times 0.6$. For the anisotropic case, they are $0.6 \times 0.8 \times 0.4$. Fig.~\ref{fig:voxel_anisotropy} shows the image of $p_{\text{blurry\_ringring\_3d}}(\pointx)$ for isotropic and anisotropic voxels on the left, with their TCF results on the right. Both cases perform similarly, with mean absolute differences of $2.493\times 10^{-3}$ and $2.495 \times 10^{-3}$, respectively. The ground truth curvature values range between 0.37 and 0.59 at the points in the image.

%% file: JMO/tex/conclusion.tex
\label{sec:conclusion}
This paper defines local tubular curvature and presents a methodology, Tubular Curvature Filter (TCF), to calculate local curvature of tubular objects using interpolated intensity images. TCF implicitly calculates the tubular curvature locally for a bundle of parallel curves that extend along the tube. TCF does not require the extraction of centerlines or curves. To our knowledge, TCF is the first of its kind, in the sense that it focuses on local curvature calculations for curves that traverse space along the tubular object, and achieves implicit calculations by determining the acceleration vectors for these curves using the eigenvectors of Hessian matrices of the intensity function. TCF exhibits sensitivity to varying curvature levels in a tube along the cross-section. The results on artificial and medical images demonstrate that this method can highlight tortuous regions in vessels and detect curvature differences between points on the inner and outer sides of curved structures, unlike centerline-based methods. 

TCF is particularly useful in medical fields that require vasculature curvature analysis from images. Biological vascular structures often have non-uniform diameters and may exhibit various conformations in both normal physiology and disease states. Measuring the curvature of anatomic structures is a common task for clinicians and is increasingly performed by automated techniques. In diseases such as Retinopathy of Prematurity (ROP), a leading cause of childhood blindness, the dilation and tortuosity of the retinal blood vessels are key components of disease classification; the more dilated and tortuous the vessels, the more severe the ROP. This has led the quantification of ``vascular severity'' to become a key biomarker target for image analysis techniques. One application of this technique would be to apply TCF to better understand the relationship between local retinal vessel tortuosity and disease progression. TCF has an advantage over deep learning techniques in that it is interpretable and can be applied both locally and globally within medical images and across image types, whereas many convolutional neural networks may face generalizability challenges.

%% file: JMO/tex/appendix_derivation.tex
\label{appendix:derivation_of_definition1}

Given a point $\pointx \in \realnumbers^2 $, let $\functionf(\mathbf{x})$ be the logarithm of the interpolation function $\interpolatefunc(\pointx)$. Note that $\pointx$ is a point on the curve $s$, i.e., $\pointx(s)$. As explained in Section~\ref{sec:tub_curv}, we drop the $(s)$ notation for simplicity. Let us define $\interpolatefunc(\pointx)$ and $\functionf(\mathbf{x})$ as:  
\begin{equation}
    \begin{aligned}
    \label{eq:interpolation_function}
    \interpolatefunc(\pointx)= \sum_{j=1}^{n} w_{j}K_{\textbf{S}_j} (\pointx-\pointx_j)
    \text{  and  } \functionf(\pointx)  &=  \log (\interpolatefunc(\pointx)).
    \end{aligned}
\end{equation}
$K_{S_j}(\cdot): \realnumbers^{2} \rightarrow \realnumbers$ is the interpolation kernel with scale matrix $\textbf{S}_j$,  $\pointx$ and $w_j\in\mathbb{R}_+$, $j\in\{1,\ldots, n\}$, are the location and intensity of a point indexed by $j \in \{ 1,\dots,n \}$ with $n$ pixels. 

Consider $\nabla$ to be a column derivative operator. Let $\gradiant(\pointx)$ be the gradient of $\functionf(\mathbf{x})$, $\mathbf{g}_{p}(\pointx)$ be the gradient of $\interpolatefunc(\pointx)$, $\hessian(\pointx)$ be the Hessian of $\functionf(\mathbf{x})$ and $\hessian_{p}(\pointx)$ be the hessian of $\interpolatefunc(\pointx)$ . Dropping ($\pointx$) for brevity, $\gradiant = \nabla \functionf = \nabla \interpolatefunc / \interpolatefunc = \gradiant_p / \interpolatefunc$ and:
    \begin{equation}
    \label{eq:hessianlogp}
        \begin{aligned}
        \hessian 
        = p^{-1} \nabla \nabla^T  \interpolatefunc  -
        p^{-2} \nabla \interpolatefunc \nabla^T  \interpolatefunc 
        = p^{-1} \hessian_{p} -
        \gradiant \gradiant^T.
        \end{aligned}    
    \end{equation}
If $\hessian_p$ has eigenvectors that are equal or close to each other for a point under a ridge with a constant value, then the second gradient term in \eqref{eq:hessianlogp} aligns one of the eigenvectors of $\hessian$ to be orthogonal to the ridge and the other to be parallel. We will use $ \partial\hessian / \partial x_i $ for \eqref{eq:derivationT1} where $x_i$ is the i-th dimension of $\mathbf{x}$. 
    \begin{equation}
        \begin{aligned}
        \frac{ \partial \hessian } {\partial x_i} =& \frac{\frac{ \partial \nabla\nabla^T \interpolatefunc}{\partial x_i}\interpolatefunc- \nabla\nabla^T\interpolatefunc\frac{\partial \interpolatefunc}{\partial x_i} }{\interpolatefunc^2} \\
        &-\frac{(\frac{\partial\nabla\interpolatefunc}{\partial x_i}\nabla^T\interpolatefunc+\nabla\interpolatefunc\frac{\partial\nabla^T \interpolatefunc}{\partial x_i})\interpolatefunc^2}{\interpolatefunc^4} + \frac{2\interpolatefunc\nabla\nabla^T\interpolatefunc\frac{\partial\interpolatefunc}{\partial x_i}}{\interpolatefunc^4} \\
        =& \frac{1}{\interpolatefunc}\frac{ \partial \nabla\nabla^T \interpolatefunc}{\partial x_i} -\frac{1}{\interpolatefunc^2}(\nabla\nabla^T \interpolatefunc\frac{\partial \interpolatefunc}{\partial x_i} +
        \frac{\partial\nabla\interpolatefunc}{\partial x_i}\nabla^T\interpolatefunc \\& +\nabla\interpolatefunc\frac{\partial\nabla^T \interpolatefunc}{\partial x_i})+ \frac{2}{\interpolatefunc^3}\nabla\interpolatefunc\nabla^T\interpolatefunc\frac{\partial\interpolatefunc}{\partial x_i}.
        \end{aligned}
    \end{equation}
Following the steps from (\ref{eq:hessian})-(\ref{eq:matrix_m}), Hessian and its derivatives are obtained to calculate the acceleration vector at $\pointx$, where $\{(\lambda_1 (\pointx), \eigenvec_1(\pointx)),(\lambda_2(\pointx), \eigenvec_2(\pointx)))\} $ are the eigenvalue-eigenvector pairs of $\hessian(\pointx)$, and $|\lambda_1 (\pointx)| > |\lambda_2(\pointx)|$. We define $\eigenvec_i$ as $\eigenvec_i = \begin{bmatrix}\text{q}_{i1} &\text{q}_{i2}\end{bmatrix}^T$ for two-dimensional.
$ \hessian(\pointx) $ is decomposed into its eigenvectors in \eqref{eq:hessian}, where $\bigQ$ is the $2\times2$ matrix whose ith column is the eigenvector $\eigenvec_i$, and $\bigLambda$ is the diagonal matrix whose diagonal elements are the corresponding eigenvalues. 
To obtain ${\partial \eigenvec_1(\pointx)}/{\partial \pointx} $ differentiating the Hessian eigendecomposition, and introducing $\tensor_i(\pointx)$, $\textbf{V}_i(\pointx)$ and $\bigL_i(\pointx)$ as in \eqref{eq:tensor_i_derivation_2d}-\eqref{eq:l_i_derivation_2d}, $\tensor_i$ for $i=1$ is derived as:
    \begin{equation} 
        \label{eq:derivationT1}
        \begin{aligned}
        \tensor_1 = \sum_{j=1}^{2}\lambda_j\frac{\partial \eigenvec_j}{\partial x_1}\eigenvec_j^T
        + l_{1j}\eigenvec_j\eigenvec_j^T 
        +\! \lambda_j\eigenvec_j\frac{\partial \eigenvec_j}{\partial x_1}^T\!.
        \end{aligned}
    \end{equation}
Linear equations for $i\!=\!1$ from \eqref{eq:matrix_form_both} are derived in \eqref{eq:teqs}-\eqref{eq:linear_eigs}. From \eqref{eq:derivationT1}, we have three equations which are defined for $i\!=\!1$:

\begin{equation} 
\begin{aligned}
\label{eq:teqs}
\text{t}_{111} &= \lambda_1\text{v}_{111}\text{q}_{11}+ \lambda_2\text{v}_{112}\text{q}_{21}+
l_{11}\text{q}_{11}\text{q}_{11}\\&+l_{12}\text{q}_{21}\text{q}_{21}
+  \lambda_1\text{q}_{11}\text{v}_{111}+ \lambda_2\text{q}_{21} \text{v}_{112}, \\
\text{t}_{112} &= \lambda_1\text{v}_{111}\text{q}_{12}+ \lambda_2\text{v}_{112}\text{q}_{22}
+l_{11}\text{q}_{11}\text{q}_{12}\\&+l_{12}\text{q}_{21}\text{q}_{22}
+  \lambda_1\text{q}_{11}\text{v}_{121}+ \lambda_2\text{q}_{21} \text{v}_{122}, \\
\text{t}_{122} &= \lambda_1\text{v}_{121}\text{q}_{12}+ \lambda_2\text{v}_{122}\text{q}_{22}
+l_{11}\text{q}_{12}\text{q}_{12}\\&+l_{12}\text{q}_{22}\text{q}_{22}
+  \lambda_1\text{q}_{12}\text{v}_{121}+ \lambda_2\text{q}_{22} \text{v}_{122}.
\end{aligned}
\end{equation}
Two properties of eigenvectors are: they are unit vectors and perpendicular to each other. Using these properties, we derived three linear equations in \eqref{eq:linear_eigs}.
\begin{equation}
\label{eq:linear_eigs}
\begin{aligned}
0&= \text{q}_{11}\text{v}_{111}+\text{q}_{12}\text{v}_{121}, \\
0&= \text{q}_{21}\text{v}_{112}+\text{q}_{22}\text{v}_{122}, \\
0= \text{q}_{21}\text{v}_{111}&+\text{q}_{22}\text{v}_{121} + \text{q}_{11}\text{v}_{112}+\text{q}_{12}\text{v}_{122}.
\end{aligned}
\end{equation}
We have 6 equations and 12 unknowns. Writing the set of linear equations, we obtain \eqref{eq:matrix_form_both}-\eqref{eq:matrix_m}. The elements of $\textbf{V}(\pointx)$ and $\bigL(\pointx)$ are found via a linear solver. $\partial \eigenvec_1(\pointx) / \partial \pointx$ are calculated from $\textbf{V}(\pointx)$ to obtain the acceleration vector at $\pointx$.

%% file: JMO/tex/appendix_n_dimensions.tex
\label{appendix:N-dimensional}
Similar to the 2-dimensional case, the curvature in N-dimensional space can be obtained from the acceleration of the \emph{first} eigen vector of the $N \times N$ Hessian matrix. Let $ \pointx \in \realnumbers^N $ and $\functionf(\pointx) : \realnumbers^{N} \rightarrow \realnumbers$ as before with $w_{j}$, $\pointx_j$ indicating image intensity and location for hypervoxel $j \in \{ 1,\dots,n \}$. Interpolate using a suitable kernel $K_{S_j}(\cdot): \realnumbers^{N} \rightarrow \realnumbers$ to obtain $functionf(\pointx)=ln\interpolatefunc(\pointx)$ as in~\eqref{eq:interpolation_function}. At $\pointx$, $\eigenvec_1$ of the Hessian is the unit velocity vector of the curve $s$, which is the parallel curve to the principle curve of the function. With this as the unit velocity vector, acceleration is obtained as its directional derivative yielding the local curvature value. Following \textbf{Definition 1}, direction derivatives $\eigenvec_{1}$, i.e., $
\begin{bmatrix}
\partial \textbf{q}_{1}/{\partial x_1} \cdots \partial \textbf{q}_{1}/{\partial x_N} \end{bmatrix}$ must be determined.
Following (\ref{eq:hessian})-(\ref{eq:last_prependicular}), Hessian and its derivatives are identified for the N-dimensional case. 

\subsection{Three-Dimensional Input}
\label{appendix:three-dimensional}

Following the steps in Eqs (\ref{eq:hessian})-(\ref{eq:matrix_m}), Hessian and its derivatives are identified to determine the acceleration vector at $ \pointx \in \realnumbers^3 $, where $\{(\lambda_1 (\pointx), \eigenvec_1(\pointx)),(\lambda_2(\pointx), \eigenvec_2(\pointx)),(\lambda_3(\pointx), \eigenvec_3(\pointx))\} $ are the eigenvalue-eigenvector pairs of $\hessian(\pointx)$, and $|\lambda_1 (\pointx)| > |\lambda_2(\pointx)|> |\lambda_3(\pointx)|$. Define $\eigenvec_i$ as $\eigenvec_i = \begin{bmatrix}\text{q}_{i1} &\text{q}_{i2} &\text{q}_{i3}\end{bmatrix}^T$.
Decompose $ \hessian(\pointx) $ as in \eqref{eq:hessian}, where $\bigQ$ is now a $3\times3$ matrix whose ith column is the eigenvector $\eigenvec_i$, and $\bigLambda$ is the diagonal matrix that contains the corresponding eigenvalues:
    \begin{align}
        \label{eq:hessian_decomposed_3d}
        \hessian(\pointx)=\begin{bmatrix}\eigenvec_1, \eigenvec_2, \eigenvec_3\end{bmatrix}
        \begin{bmatrix}\lambda_1 &0 &0\\ 0 &\lambda_2 &0\\ 0 &0 &\lambda_3\end{bmatrix}
        \begin{bmatrix}\eigenvec_1, \eigenvec_2,\eigenvec_3\end{bmatrix}^T,
    \end{align}
Calculate ${\partial \eigenvec_1(\pointx)}/{\partial \pointx} $ by differentiating the eigendecomposition of $\hessian(\pointx)$ and define $\tensor_i(\pointx)$, $\textbf{V}_i(\pointx)$ and $\bigL_i(\pointx)$ as:
    \begin{equation}
    \begin{aligned}
        \tensor_i(\pointx) =& \frac{\partial \hessian(\pointx)}{ \partial x_i} = \begin{bmatrix}
        \text{t}_\text{i11} & \text{t}_\text{i12}& \text{t}_\text{i13} \\ 
        \text{t}_\text{i21} &\text{t}_\text{i22}& \text{t}_\text{i23} \\ 
        \text{t}_\text{i31} &\text{t}_\text{i32}& \text{t}_\text{i33} 
        \end{bmatrix}, \\
        \textbf{V}_i(\pointx) 
        =& \begin{bmatrix}
        \frac{\partial \eigenvec_1}{\partial x_i}, \frac{\partial \eigenvec_2}{\partial x_i}, \frac{\partial \eigenvec_3}{\partial x_i}
        \end{bmatrix}
        = \begin{bmatrix}
        \text{v}_\text{i11} & \text{v}_\text{i12} & \text{v}_\text{i13} \\ 
        \text{v}_\text{i21} &\text{v}_\text{i22} &\text{v}_\text{i23} \\
        \text{v}_\text{i31} &\text{v}_\text{i32} &\text{v}_\text{i33} 
        \end{bmatrix}, \\
        \bigL_{i}(\pointx)=&
        \begin{bmatrix}
        \frac{\partial \lambda_{1}}{\partial x_i} &0 & 0\\
        0 &\frac{\partial \lambda_{2}}{\partial x_i} & 0 \\
        0 & 0 &\frac{\partial \lambda_{3}}{\partial x_i}
        \end{bmatrix}
        =
        \begin{bmatrix}
        l_\text{i1} &0 &0 \\ 
        0 &l_\text{i2} &0 \\
        0 &0 &l_\text{i3}
        \end{bmatrix}.
        \end{aligned}
    \end{equation}
$\tensor_i$ is symmetric due to thrice-differentiable interpolation, and calculated as in Eq.~\eqref{eq:derivationT1} with $j=1,2,3$. As in Appendix~\ref{appendix:derivation_of_definition1}, one can derive linear equations to obtain \eqref{eq:matrix_form_both_3d}. By utilizing two properties of the eigenvectors are explained in Appendix~\ref{appendix:derivation_of_definition1}; $\textbf{M}$ is defined in \eqref{eq:matrix_m_3d}. The elements of $\textbf{V}(\pointx)$ and $\bigL(\pointx)$ are determined by solving the linear equations to obtain $\partial \eigenvec_1(\pointx) / \partial \pointx$ from $\textbf{V}(\pointx)$. The acceleration at $\pointx$ is then computed using (\ref{eq:acceleration_eq}).
    \begin{equation}
        \label{eq:matrix_form_both_3d}
        \begin{bmatrix}
        \text{t}_{111}\\
        \text{t}_{112}\\
        \text{t}_{113}\\
        \text{t}_{122}\\
        \text{t}_{123}\\
        \text{t}_{133}\\
        0\\
        0\\
        0\\
        0\\
        0\\
        0\\
        \end{bmatrix}
        \!= \textbf{M} 
        \begin{bmatrix}
        \text{v}_{111}\\
        \text{v}_{112}\\
        \text{v}_{113}\\
        l_{11}\\
        l_{12}\\
        l_{13}\\
        \text{v}_{121}\\
        \text{v}_{122}\\
        \text{v}_{123}\\
        \text{v}_{131}\\
        \text{v}_{132}\\
        \text{v}_{133}
        \end{bmatrix} 
        \!\textrm{, }\!
        \begin{bmatrix}
        \text{t}_{211}\\
        \text{t}_{212}\\
        \text{t}_{213}\\
        \text{t}_{222}\\
        \text{t}_{223}\\
        \text{t}_{233}\\
        0\\
        0\\
        0\\
        0\\
        0\\
        0\\
        \end{bmatrix}
        \!= \textbf{M}
        \begin{bmatrix}
        \text{v}_{211}\\
        \text{v}_{212}\\
        \text{v}_{213}\\
        l_{21}\\
        l_{22}\\
        l_{23}\\
        \text{v}_{221}\\
        \text{v}_{222}\\
        \text{v}_{223}\\
        \text{v}_{231}\\
        \text{v}_{232}\\
        \text{v}_{233}
        \end{bmatrix}
         \!\textrm{,}\!
         \begin{bmatrix}
        \text{t}_{311}\\
        \text{t}_{312}\\
        \text{t}_{313}\\
        \text{t}_{322}\\
        \text{t}_{323}\\
        \text{t}_{333}\\
        0\\
        0\\
        0\\
        0\\
        0\\
        0\\
        \end{bmatrix}
        \!= \textbf{M}
        \begin{bmatrix}
        \text{v}_{311}\\
        \text{v}_{312}\\
        \text{v}_{313}\\
        l_{31}\\
        l_{32}\\
        l_{33}\\
        \text{v}_{321}\\
        \text{v}_{322}\\
        \text{v}_{323}\\
        \text{v}_{331}\\
        \text{v}_{332}\\
        \text{v}_{333}
        \end{bmatrix},
    \end{equation}

    \begin{equation}
        \begin{aligned}
            \label{eq:matrix_m_3d}
            \!\textbf{M}\! = \resizebox{0.87\textwidth}{!}{\setlength{\arraycolsep}{1.2pt}$\begin{bmatrix}
                2\lambda_1\text{q}_{11} &2\lambda_2\text{q}_{21} &2\lambda_3\text{q}_{31} &\text{q}_{11}^2 &\text{q}_{21}^2 &\text{q}_{31}^2  &0 &0 &0 &0 &0 &0\\
                \lambda_1\text{q}_{12} & \lambda_2\text{q}_{22} & \lambda_3\text{q}_{32} &\text{q}_{11}\text{q}_{12} &\text{q}_{21}\text{q}_{22} &\text{q}_{31}\text{q}_{32} &\lambda_1\text{q}_{11} & \lambda_2\text{q}_{21}& \lambda_3\text{q}_{31}&0 &0 &0\\
                \lambda_1\text{q}_{13} & \lambda_2\text{q}_{23} & \lambda_3\text{q}_{33} &\text{q}_{11}\text{q}_{13} &\text{q}_{21}\text{q}_{23} &\text{q}_{31}\text{q}_{33} &0 &0 &0 &\lambda_1\text{q}_{11} & \lambda_2\text{q}_{21}& \lambda_3\text{q}_{31}\\
                0 &0&0 &\text{q}_{12}^2 &\text{q}_{22}^2 &\text{q}_{32}^2 &2\lambda_1\text{q}_{12} &2\lambda_2\text{q}_{22}&2\lambda_3\text{q}_{32} &0 &0&0\\
                0 &0&0 &\text{q}_{12}\text{q}_{13} &\text{q}_{22}\text{q}_{23} &\text{q}_{32}\text{q}_{33} &\lambda_1\text{q}_{13} &\lambda_2\text{q}_{23}&\lambda_3\text{q}_{33} &\lambda_1\text{q}_{12} &\lambda_2\text{q}_{22}&\lambda_3\text{q}_{32}\\
                0 &0&0 &\text{q}_{13}^2 &\text{q}_{23}^2 &\text{q}_{33}^2 &0 &0&0&2\lambda_1\text{q}_{13} &2\lambda_2\text{q}_{23}&2\lambda_3\text{q}_{33}\\
                \text{q}_{11} &0 &0 &0 &0 &0&\text{q}_{12}&0 &0&\text{q}_{13}&0 &0\\
                0 &\text{q}_{21}&0 &0 &0 &0 &0 &\text{q}_{22}&0 &0&\text{q}_{23}&0\\
                0 & 0 &\text{q}_{31}&0 &0 &0 &0 &0 &\text{q}_{32}&0 &0&\text{q}_{33}\\
                \text{q}_{21} &\text{q}_{11} &0 &0 &0 &0 &\text{q}_{22} &\text{q}_{12} &0 &\text{q}_{23} &\text{q}_{13} &0 \\
                \text{q}_{31}  &0 &\text{q}_{11} &0 &0 &0 &\text{q}_{32} &0 &\text{q}_{12} &\text{q}_{33} &0 &\text{q}_{13} \\
                0 &\text{q}_{31} &\text{q}_{21} &0 &0 &0 &0 &\text{q}_{32} &\text{q}_{22} &0 &\text{q}_{33}  &\text{q}_{23} \\
            \end{bmatrix}.$}
        \end{aligned}
    \end{equation}

%% file: ref.bib
@article{ozertem2011locally,
  title={{Locally defined principal curves and surfaces}},
  author={Ozertem, Umut and Erdogmus, Deniz},
  journal={The Journal of Machine Learning Research},
  volume={12},
  pages={1249--1286},
  year={2011},
  publisher={JMLR. org}
}

@inproceedings{RegionDetector,
author = {Deng, Hongli and Zhang, Wei and Mortensen, Eric and Dietterich, Thomas and Shapiro, Linda},
year = {2007},
month = {06},
pages = {},
title = {{Principal curvature-based region detector for object recognition}},
doi = {10.1109/CVPR.2007.382972}
}

@article{thanh2020melanoma,
  title={{Melanoma skin cancer detection method based on adaptive principal curvature, colour normalisation and feature extraction with the ABCD rule}},
  author={Thanh, Dang NH and Prasath, VB Surya and Hien, Nguyen Ngoc and others},
  journal={Journal of Digital Imaging},
  volume={33},
  number={3},
  pages={574},
  year={2020},
  publisher={Springer}
}

@ARTICLE{735812,
  author={Mokhtarian, F. and Suomela, R.},
  journal={IEEE Transactions on Pattern Analysis and Machine Intelligence}, 
  title={{Robust image corner detection through curvature scale space}}, 
  year={1998},
  volume={20},
  number={12},
  pages={1376-1381},
  doi={10.1109/34.735812}}

@article{10.1007/s10851-017-0728-2,
author = {Tang, Yinhang and Li, Huibin and Sun, Xiang and Morvan, Jean-Marie and Chen, Liming},
title = {{Principal curvature measures estimation and application to 3D face recognition}},
year = {2017},
issue_date = {October 2017},
publisher = {Kluwer Academic Publishers},
address = {USA},
volume = {59},
number = {2},
issn = {0924-9907},
doi = {10.1007/s10851-017-0728-2},
journal = {J. Math. Imaging Vis.},
month = {oct},
pages = {211-233},
numpages = {23}
}

@article{ImageReconstructionbyMinimizingCurvatures,
author = {Qiuxiang, Zhong and Yin, Ke and Duan, Yuping},
year = {2021},
month = {01},
pages = {},
title = {{Image reconstruction by minimizing curvatures on image surface}},
volume = {63},
journal = {Journal of Mathematical Imaging and Vision},
doi = {10.1007/s10851-020-00992-3}
}

@inproceedings{ImageDescriptors,
author = {Fischer, Philipp and Brox, Thomas},
year = {2014},
month = {09},
pages = {239-249},
title = {{Image descriptors based on curvature histograms}},
isbn = {978-3-319-11751-5},
doi = {10.1007/978-3-319-11752-2_19}
}

@article{GOLDMAN2005632,
title = {{Curvature formulas for implicit curves and surfaces}},
journal = {Computer Aided Geometric Design},
volume = {22},
number = {7},
pages = {632-658},
year = {2005},
note = {Geometric Modelling and Differential Geometry},
issn = {0167-8396},
doi = {https://doi.org/10.1016/j.cagd.2005.06.005},
author = {Ron Goldman}
}

@article{he2008corner,
  title={{Corner detector based on global and local curvature properties}},
  author={He, Xiaochen and Yung, Nelson Hon Ching},
  journal={Optical engineering},
  volume={47},
  number={5},
  pages={057008},
  year={2008},
  publisher={SPIE}
}

@article{brown2018automated,
  title={{Automated diagnosis of plus disease in retinopathy of prematurity using deep convolutional neural networks}},
  author={Brown, James M and Campbell, J Peter and Beers, Andrew and Chang, Ken and Ostmo, Susan and Chan, RV Paul and Dy, Jennifer and Erdogmus, Deniz and Ioannidis, Stratis and Kalpathy-Cramer, Jayashree and others},
  journal={JAMA ophthalmology},
  volume={136},
  number={7},
  pages={803--810},
  year={2018},
  publisher={American Medical Association}
}

@article{fabrydisease,
author = {Atiskova, Yevgeniya and Wildner, Jan and Spitzer, Martin and Aries, Charlotte and Muschol, Nicole and Dulz, Simon},
year = {2021},
month = {12},
pages = {},
title = {{Retinal vessel tortuosity as a prognostic marker for disease severity in Fabry disease}},
volume = {16},
journal = {Orphanet Journal of Rare Diseases},
doi = {10.1186/s13023-021-02080-0}
}

@article{twistedbloodvessels,
author = {Han, Hai-Chao},
year = {2012},
month = {03},
pages = {185-97},
title = {{Twisted blood vessels: Symptoms, etiology and biomechanical mechanisms}},
volume = {49},
journal = {Journal of vascular research},
doi = {10.1159/000335123}
}

@article{hathout2012vascular,
  title={{Vascular tortuosity: A mathematical modeling perspective}},
  author={Hathout, Leith and Do, Huy M},
  journal={The Journal of Physiological Sciences},
  volume={62},
  pages={133--145},
  year={2012},
  publisher={Springer}
}

@article{choi2008methods,
author = {Choi, Gilwoo and Cheng, Christopher and Wilson, Nathan and Taylor, Charles},
year = {2008},
month = {12},
pages = {14-33},
title = {{Methods for quantifying three-dimensional deformation of arteries due to pulsatile and nonpulsatile forces: Implications for the design of stents and stent grafts}},
volume = {37},
journal = {Annals of biomedical engineering},
doi = {10.1007/s10439-008-9590-0}
}

@Article{app9245507,
AUTHOR = {Cervantes-Sanchez, Fernando and Cruz-Aceves, Ivan and Hernandez-Aguirre, Arturo and Hernandez-Gonzalez, Martha Alicia and Solorio-Meza, Sergio Eduardo},
TITLE = {{Automatic segmentation of coronary arteries in X-ray angiograms using multiscale analysis and artificial neural networks}},
JOURNAL = {Applied Sciences},
VOLUME = {9},
YEAR = {2019},
NUMBER = {24},
ARTICLE-NUMBER = {5507},
ISSN = {2076-3417},
DOI = {10.3390/app9245507}
}

@article{ZENG2023102287,
title = {{ImageCAS: A large-scale dataset and benchmark for coronary artery segmentation based on computed tomography angiography images}},
journal = {Computerized Medical Imaging and Graphics},
pages = {102287},
year = {2023},
issn = {0895-6111},
doi = {https://doi.org/10.1016/j.compmedimag.2023.102287},
author = {An Zeng and Chunbiao Wu and Wen Xie and Jin Hong and Meiping Huang and Jian Zhuang and Shanshan Bi and Dan Pan and Najeeb Ullah and Kaleem Nawaz Khan and Tianchen Wang and Yiyu Shi and Xiaomeng Li and Guisen Lin and Xiaowei Xu}
}

@article{Nc2019OptimizedMP,
  title={{Optimized maximum principal curvatures based segmentation of blood vessels from retinal images}},
  author={Santosh Kumar Nc and Y. Radhika},
  journal={Biomedical Research-tokyo},
  year={2019},
  volume={30},
  pages={308-318}
}

@ARTICLE{9223670,
  author={Lutton, Judith E. and Collier, Sharon and Bretschneider, Till},
  journal={IEEE Transactions on Medical Imaging}, 
  title={{A curvature-enhanced random walker segmentation method for detailed capture of 3D cell surface membranes}}, 
  year={2021},
  volume={40},
  number={2},
  pages={514-526},
  doi={10.1109/TMI.2020.3031029}}

@inproceedings{gerig2004analysis,
  title={{Analysis of brain white matter via fiber tract modeling}},
  author={Gerig, Guido and Gouttard, Sylvain and Corouge, Isabelle},
  booktitle={The 26th Annual International Conference of the IEEE Engineering in Medicine and Biology Society},
  volume={2},
  pages={4421--4424},
  year={2004},
  organization={IEEE}
}

@article{bullitt2003measuring,
  title={{Measuring tortuosity of the intracerebral vasculature from MRA images}},
  author={Bullitt, Elizabeth and Gerig, Guido and Pizer, Stephen M and Lin, Weili and Aylward, Stephen R},
  journal={IEEE transactions on medical imaging},
  volume={22},
  number={9},
  pages={1163--1171},
  year={2003},
  publisher={IEEE}
}

@inproceedings{goldluecke2011introducing,
  title={{Introducing total curvature for image processing}},
  author={Goldluecke, Bastian and Cremers, Daniel},
  booktitle={2011 International Conference on Computer Vision},
  pages={1267--1274},
  year={2011},
  organization={IEEE}
}

@inproceedings{zhong2020minimizing,
  title={{Minimizing discrete total curvature for image processing}},
  author={Zhong, Qiuxiang and Li, Yutong and Yang, Yijie and Duan, Yuping},
  booktitle={Proceedings of the IEEE/CVF Conference on Computer Vision and Pattern Recognition},
  pages={9474--9482},
  year={2020}
}

@inproceedings{chen2016new,
  title={{A new finsler minimal path model with curvature penalization for image segmentation and closed contour detection}},
  author={Chen, Da and Mirebeau, Jean-Marie and Cohen, Laurent D},
  booktitle={Proceedings of the IEEE Conference on Computer Vision and Pattern Recognition},
  pages={355--363},
  year={2016}
}

@article{lee2005noise,
  title={{Noise removal with Gauss curvature-driven diffusion}},
  author={Lee, Suk-Ho and Seo, Jin Keun},
  journal={IEEE Transactions on Image Processing},
  volume={14},
  number={7},
  pages={904--909},
  year={2005},
  publisher={IEEE}
}

@article{el2016contrast,
  title={{Contrast driven elastica for image segmentation}},
  author={El-Zehiry, Noha Youssry and Grady, Leo},
  journal={IEEE Transactions on Image Processing},
  volume={25},
  number={6},
  pages={2508--2518},
  year={2016},
  publisher={IEEE}
}

@article{zhang2019discrete,
  title={{Discrete curvature representations for noise robust image corner detection}},
  author={Zhang, Weichuan and Sun, Changming and Breckon, Toby and Alshammari, Naif},
  journal={IEEE Transactions on Image Processing},
  volume={28},
  number={9},
  pages={4444--4459},
  year={2019},
  publisher={IEEE}
}

@article{kim2006pde,
  title={{PDE-based image restoration: A hybrid model and color image denoising}},
  author={Kim, Seongjai},
  journal={IEEE Transactions on Image Processing},
  volume={15},
  number={5},
  pages={1163--1170},
  year={2006},
  publisher={IEEE}
}

@book{jain1989fundamentals,
  title={{Fundamentals of digital image processing}},
  author={Jain, Anil K},
  year={1989},
  publisher={Prentice-Hall, Inc.}
}

@article{ozertem2007nonparametric,
  title={{Nonparametric snakes}},
  author={Ozertem, Umut and Erdogmus, Deniz},
  journal={IEEE Transactions on Image Processing},
  volume={16},
  number={9},
  pages={2361--2368},
  year={2007},
  publisher={IEEE}
}

@article{javaheri2020lasting,
  title={Lasting organ-level bone mechanoadaptation is unrelated to local strain},
  author={Javaheri, Behzad and Razi, Hajar and Gohin, Stephanie and Wylie, Sebastian and Chang, Yu-Mei and Salmon, Phil and Lee, Peter D and Pitsillides, Andrew A},
  journal={Science Advances},
  volume={6},
  number={10},
  pages={eaax8301},
  year={2020},
  publisher={American Association for the Advancement of Science}
}

@article{ALETTI201677,
title = {A simplified fluid–structure model for arterial flow. Application to retinal hemodynamics},
journal = {Computer Methods in Applied Mechanics and Engineering},
volume = {306},
pages = {77-94},
year = {2016},
issn = {0045-7825},
doi = {https://doi.org/10.1016/j.cma.2016.03.044},
author = {Matteo Aletti and Jean-Frédéric Gerbeau and Damiano Lombardi}
}

@article{koogler2023analysis,
  title={Analysis of vessel tortuosity and its impact on hemodynamics in retinopathy of prematurity},
  author={Koogler, Brent and Young, Benjamin and Campbell, J Peter and Guidoboni, Giovanna},
  journal={Investigative Ophthalmology \& Visual Science},
  volume={64},
  number={8},
  pages={1237--1237},
  year={2023},
  publisher={The Association for Research in Vision and Ophthalmology}
}

@article{santamarina1998computational,
  title={Computational analysis of flow in a curved tube model of the coronary arteries: effects of time-varying curvature},
  author={Santamarina, Aland and Weydahl, Erlend and Siegel, John M and Moore, James E},
  journal={Annals of biomedical engineering},
  volume={26},
  pages={944--954},
  year={1998},
  publisher={Springer}
}

@article{van2014scikit,
  title={scikit-image: image processing in Python},
  author={Van der Walt, Stefan and Sch{\"o}nberger, Johannes L and Nunez-Iglesias, Juan and Boulogne, Fran{\c{c}}ois and Warner, Joshua D and Yager, Neil and Gouillart, Emmanuelle and Yu, Tony},
  journal={PeerJ},
  volume={2},
  pages={e453},
  year={2014},
  publisher={PeerJ Inc.}
}

@article{dalvit2023automated,
  title={Automated Coronary Artery Tracking with a Voronoi-Based 3D Centerline Extraction Algorithm},
  author={Dalvit Carvalho da Silva, Rodrigo and Soltanzadeh, Ramin and Figley, Chase R},
  journal={Journal of Imaging},
  volume={9},
  number={12},
  pages={268},
  year={2023},
  publisher={MDPI}
}

@article{lorigo2001curves,
  title={Curves: Curve evolution for vessel segmentation},
  author={Lorigo, Liana M and Faugeras, Olivier D and Grimson, W Eric L and Keriven, Renaud and Kikinis, Ron and Nabavi, Arya and Westin, C-F},
  journal={Medical image analysis},
  volume={5},
  number={3},
  pages={195--206},
  year={2001},
  publisher={Elsevier}
}

@article{law2009efficient,
  title={Efficient implementation for spherical flux computation and its application to vascular segmentation},
  author={Law, Max WK and Chung, Albert CS},
  journal={IEEE transactions on image processing},
  volume={18},
  number={3},
  pages={596--612},
  year={2009},
  publisher={IEEE}
}

@inproceedings{law2008three,
  title={Three dimensional curvilinear structure detection using optimally oriented flux},
  author={Law, Max WK and Chung, Albert CS},
  booktitle={ECCV 2008: 10th European Conference on Computer Vision, Marseille, France, October 12-18, 2008, Proceedings, Part IV 10},
  pages={368--382},
  year={2008},
  organization={Springer}
}

@inproceedings{law2013gradient,
  title={Gradient competition anisotropy for centerline extraction and segmentation of spinal cords},
  author={Law, Max WK and Garvin, Gregory J and Tummala, Sudhakar and Tay, KengYeow and Leung, Andrew E and Li, Shuo},
  booktitle={Information Processing in Medical Imaging: 23rd International Conference, IPMI 2013, Asilomar, CA, USA, June 28--July 3, 2013. Proceedings 23},
  pages={49--61},
  year={2013},
  organization={Springer}
}

@book{montiel2009curves,
  title={Curves and surfaces},
  author={Montiel, Sebasti{\'a}n and Ros, Antonio},
  volume={69},
  year={2009},
  publisher={American Mathematical Soc.}
}

@inproceedings{law2010oriented,
  title={An oriented flux symmetry based active contour model for three dimensional vessel segmentation},
  author={Law, Max WK and Chung, Albert CS},
  booktitle={ECCV 2010: 11th European Conference on Computer Vision, Heraklion, Crete, Greece, September 5-11, 2010, Proceedings, Part III 11},
  pages={720--734},
  year={2010},
  organization={Springer}
}

@article{gilbert2001childhood,
  title={Childhood blindness in the context of VISION 2020: the right to sight},
  author={Gilbert, Clare and Foster, Allen},
  journal={Bulletin of the World Health Organization},
  volume={79},
  number={3},
  pages={227--232},
  year={2001},
  publisher={SciELO Public Health}
}

@article{chiang2021international,
  title={International classification of retinopathy of prematurity},
  author={Chiang, Michael F and Quinn, Graham E and Fielder, Alistair R and Ostmo, Susan R and Chan, RV Paul and Berrocal, Audina and Binenbaum, Gil and Blair, Michael and Campbell, J Peter and Capone Jr, Antonio and others},
  journal={Ophthalmology},
  volume={128},
  number={10},
  pages={e51--e68},
  year={2021},
  publisher={Elsevier}
}

@inproceedings{ge2024automatic,
  title={Automatic bony structure segmentation and curvature estimation on ultrasound cervical spine images-a feasibility study},
  author={Ge, Songhan and Tian, Haoyuan and Zhang, Wei and Zheng, Rui},
  booktitle={Journal of Physics: Conference Series},
  volume={2822},
  number={1},
  pages={012023},
  year={2024},
  organization={IOP Publishing}
}

@inproceedings{ernst2019cnn,
  title={A CNN-based framework for statistical assessment of spinal shape and curvature in whole-body MRI images of large populations},
  author={Ernst, Philipp and Hille, Georg and Hansen, Christian and T{\"o}nnies, Klaus and Rak, Marko},
  booktitle={Medical Image Computing and Computer Assisted Intervention--MICCAI 2019: 22nd International Conference, Shenzhen, China, October 13--17, 2019, Proceedings, Part IV 22},
  pages={3--11},
  year={2019},
  organization={Springer}
}

@inproceedings{liu2020spinal,
  title={Spinal curve assessment of idiopathic scoliosis with a small dataset via a multi-scale keypoint estimation approach},
  author={Liu, Tianyu and Yang, Yukang and Wang, Yu and Sun, Ming and Fan, Wenhui and Wu, Cheng and Bunger, Cody},
  booktitle={Adjunct proceedings of the 2020 ACM international joint conference on pervasive and ubiquitous computing and proceedings of the 2020 aCM international symposium on wearable computers},
  pages={665--670},
  year={2020}
}

@inproceedings{zhou2015assessment,
  title={Assessment of scoliosis using 3-D ultrasound volume projection imaging with automatic spine curvature detection},
  author={Zhou, Guang-Quan and Zheng, Yong-Ping},
  booktitle={2015 IEEE International Ultrasonics Symposium (IUS)},
  pages={1--4},
  year={2015},
  organization={IEEE}
}

@article{vogel2022robust,
  title={Robust centerline prediction for accurate vessel wall visualization of intracranial vessels in multi-contrast 3D MRI data},
  author={Vogel, Patrick and Kampf, Thomas and Guggenberger, Konstanze and Raithel, Esther and Forman, Christoph and Meckel, Stephan and Ludwig, Ute and Krafft, Axel J and Hennig, J{\"u}rgen and Bley, Thorsten A},
  journal={arXiv preprint arXiv:2209.01422},
  year={2022}
}

@inproceedings{wang2024b,
  title={B-spine: Learning B-spline curve representation for robust and interpretable spinal curvature estimation},
  author={Wang, Hao and Song, Qiang and Yin, Ruofeng and Ma, Rui},
  booktitle={Proceedings of the AAAI Conference on Artificial Intelligence},
  volume={38},
  number={6},
  pages={5381--5389},
  year={2024}
}
